\documentclass[fleqn,usenatbib]{mnras}

\usepackage[T1]{fontenc}
\usepackage{ae,aecompl}

\usepackage{graphicx}	
\usepackage{amsmath}	
\usepackage{amssymb}	
\usepackage{comment}
\usepackage{footmisc}
\usepackage{subcaption}
\usepackage{placeins}
\usepackage{pdflscape}
\usepackage{mdwlist}

\usepackage{tikz}
\usetikzlibrary{shapes.geometric, arrows}
\tikzstyle{startstop} = [rectangle, rounded corners, minimum width=3cm, minimum height=1cm,text centered, text width=3cm, draw=black, fill=red!30]
\tikzstyle{process} = [rectangle, minimum width=3cm, minimum height=1cm, text centered, text width=3cm, draw=black, fill=blue!30]
\tikzstyle{decision} = [diamond, minimum width=3cm, minimum height=1cm, text centered, text width=3cm, draw=black, fill=green!30]
\tikzstyle{arrow} = [->, >=latex', shorten >=1pt, thick]
\tikzstyle{line} = [-, >=latex', shorten >=1pt, thick]
\tikzstyle{label} = [text width=2.5cm, text centered] 

\usepackage{newtxtext,newtxmath}

\newcommand{\pmra}{$\mu{}_{\alpha\cos\delta}$}
\newcommand{\pmdec}{$\mu{}_{\delta}$}
\newcommand{\masyr}{mas~yr$^{-1}$}
\defcitealias{LCS18}{Paper 1}

\title[VIRAC2]{VIRAC2: NIR Astrometry and Time Series Photometry for 500M+ Stars from the VVV and VVVX Surveys}

\author[L. C. Smith et al.]{
Leigh C. Smith$^1$\thanks{lsmith@ast.cam.ac.uk},
Philip W. Lucas$^2$,
Sergey E. Koposov$^{3,1,4}$, 
Carlos Gonzalez-Fernandez$^{1}$,
\newauthor
Javier Alonso-Garc\'{i}a$^{5,6}$,
Dante Minniti$^{7,8}$,
Jason L. Sanders$^{9}$,
Luigi R. Bedin$^{10}$,
\newauthor
Vasily Belokurov$^1$,
N. Wyn Evans$^1$,
Maren Hempel$^{7,11}$,
Valentin D. Ivanov$^{12}$,
\newauthor
Radostin G. Kurtev$^{13,14}$,
and Roberto K. Saito$^{15}$
\\
$^{1}$Institute of Astronomy, University of Cambridge, Madingley Rd, Cambridge CB3 0HA, UK\\
$^{2}$Centre for Astrophysics Research, University of Hertfordshire, College Lane, Hatfield AL10 9AB, UK\\
$^{3}$Institute for Astronomy, University of Edinburgh, Royal Observatory, Blackford Hill, Edinburgh EH9 3HJ, UK\\
$^{4}$Kavli Institute for Cosmology, University of Cambridge, Madingley Road, Cambridge CB3 0HA, UK\\
$^{5}$Centro de Astronom\'{i}a (CITEVA), Universidad de Antofagasta, Av. Angamos 601, Antofagasta, Chile\\
$^{6}$Millennium Institute of Astrophysics, Nuncio Monse\~nor Sotero Sanz 100, Of. 104, Providencia, Santiago, Chile\\
$^{7}$Instituto de Astrofísica, Dep. de Ciencias Físicas, Facultad de Ciencias Exactas, Universidad Andres Bello, Av. Fernández Concha 700, Santiago, Chile\\
$^{8}$Vatican Observatory, Specola Vaticana, V-00120, Vatican City, Vatican City State\\
$^{9}$ Department of Physics and Astronomy, University College London, London WC1E 6BT, UK\\
$^{10}$INAF - Osservatorio Astronomico di Padova, Vicolo dell'Osservatorio 5,  Padova I-35122, Italy\\
$^{11}$Max-Planck Institute for Astronomy, Königstuhl 17, 69117 Heidelberg, Germany\\
$^{12}$European Southern Observatory, Karl-Schwarzschild-Str. 2, 85748, Garching bei M\"unchen, Germany\\
$^{13}$Instituto de F\'isica y Astronom\'ia, Universidad de Valpara\'iso, Av. Gran Breta\~na 1111, Playa Ancha, Valpara\'iso, Chile\\
$^{14}$The Millennium Institute of Astrophysics (MAS), Av. Vicu\~na Mackenna 4860, 782-0436 Macul, Santiago, Chile\\
$^{15}$Departamento de F\'isica, Universidade Federal de Santa Catarina, Trindade 88040-900, Florian\'opolis, Brazil\\
}

\date{Accepted 2024 Dec 18. Received YYY; in original form 2024 July 5}

\pubyear{2024}

\begin{document}
\label{firstpage}
\pagerange{\pageref{firstpage}--\pageref{lastpage}}
\maketitle

\begin{abstract}
We present VIRAC2, a catalogue of positions, proper motions, parallaxes and $Z$, $Y$, $J$, $H$, and $K_s$ near-infrared photometric time series of 545\,346\,537 unique stars. The catalogue is based on a point spread function fitting reduction of nearly a decade of VISTA VVV and VVVX images, which cover $560~{\rm deg}^2$ of the Southern Galactic plane and bulge. The catalogue is complete at the $>90$ per cent level for $11<K_s~{\rm mag}<16$ sources, but extends to $K_s\approx{}17.5$~mag in most fields. Astrometric performance for $11<K_s~{\rm mag}<14$ sources is typically $\approx{}0.37~{\rm mas~yr}^{-1}$ per dimension for proper motion, and $1~{\rm mas}$ for parallax. At $K_s=16$ the equivalent values are around $1.5~{\rm mas~yr}^{-1}$ and $5~{\rm mas}$. These uncertainties are validated against Gaia DR3 and Hubble Space Telescope astrometry. The complete catalogues are available via the ESO archive.
We perform an initial search of the catalogue for nearby ultracool dwarf candidates. In total we find 26 new sources whose parallaxes place them within 50 parsecs of the Sun. Among them we find two high-confidence T dwarfs and a number of other sources that appear to lie close to the L/T transition.
\end{abstract}




\section{Introduction}

The VISTA Variables in the Via Lactea (VVV; \citealt{minniti10}) is a near-infrared multi-epoch photometric survey conducted using the VISTA telescope at the Paranal Observatory, Chile. It observed 560 deg$^2$ of the bulge and southern disc of the Milky Way between 2009 and 2015, and comprises roughly a hundred epochs in the $K_s$ bandpass at a typical pointing, with additional epochs in $Z$, $Y$, $J$ and $H$ bandpasses at the beginning and end of the survey.
The recently completed VVV extended survey \citep[VVVX;][]{vvvx} built on the VVV survey by extending the area coverage to 1700~deg$^2$ in the $J$, $H$ and $K_s$ bandpasses. It also included many additional epochs for the VVV survey area, extending the time baseline by more than a factor of two.

The bulge and disc comprise the vast majority of the resolvable stellar content of the Milky Way; fertile ground for a great number of subfields of stellar and galactic astronomy. One obvious use of the VVV and VVVX time series data covering this region is the measurement of stellar proper motions. In this context, the datasets remain useful even in the Gaia era due to their ability to survey deeper into regions of high Galactic extinction (see e.g. \citealt{LCS18}, figure 4). Several previous works have made proper motion measurements within subsections of the VVV data (e.g. \citealp{Libralato15,CR17,griggio24}). 

The VVV Infrared Astrometric Catalogue (VIRAC v1 hereafter; \citealt{LCS18}) capitalised on the original VVV survey, providing proper motion measurements from the time sequence observations of some 300 million unique sources over 5 years. Additionally, for around 7000 objects that exhibited large proper motions it provided parallax measurements at $>5\sigma$. At the time, the lack of suitable astrometric reference catalogues meant that astrometric calibration had to be performed in a \textit{relative} sense, i.e. measured motions of individual stars were relative to those of the nearby field. This was a major drawback of VIRAC v1, e.g. for the purposes of large scale studies of the motions of Milky Way stars. Shortly after VIRAC v1 was published, the second data release of the Gaia survey became available. Some authors \citep{clarke19, sanders19} capitalised on this by performing their own recalibration of VIRAC v1 (designated VIRAC v1.1 in the \citealt{sanders19} case) in order to study the kinematic properties of the Galactic bar, though these catalogue versions were never made available publicly.

VIRAC v1 was built from the aperture photometric catalogues provided by the Cambridge Astronomical Survey Unit (CASU). These data products are superb for observations with limited blending, but do suffer from source confusion in regions of significant stellar crowding. The VVV and VVVX surveys cover regions of the Milky Way with the highest stellar densities, in infrared bandpasses that are less impacted by the effects of interstellar reddening, and hence are subject to significant source confusion. Point source profile fitting photometry is usually better suited to such observations \citep{daophot}.

\citet{LCS18} described their planned version two of VIRAC, based on profile fitting photometry, and using \textit{Gaia} as an external reference catalogue. This paper describes the VIRAC version 2 (VIRAC2 hereafter) pipeline and catalogue. In addition to the above enhancements, VIRAC2 includes more observations, covering a longer time baseline; it incorporates an image-level astrometric calibration algorithm that used Gaia DR3 (and its reduced systematic errors relative to Gaia DR2) as an external reference catalogue; and it benefits from a global photometric calibration algorithm, with a secondary component designed to reduce high-frequency atmosphere-induced photometric scatter.

This paper is organised as follows: in Section \ref{data} we describe the data, source detection, astrometric and photometric calibration algorithms, and the main VIRAC2 pipeline. In Section \ref{catalogues} we describe the catalogues, the steps taken to clean them, validate their contents against external sources, and demonstrate how they may be accessed. In Section \ref{sec:eg_applications} we outline and present the results of searches within the catalogues for new nearby sources such as brown dwarfs and white dwarfs that are either too optically faint for Gaia or otherwise overlooked. Such sources are of interest in order to complete the census of systems in nearby space, enabling a better understanding of star formation and stellar evolution. These searches add to previous VVV-based searches for brown dwarfs and high proper motion stars \citep{beamin13, ivanov13, smith15, kurtev17, LCS18} and earlier searches in the Galactic plane based on other data sets, e.g. \citet{looper07, folkes07, folkes12, phan-bao08, lucas10, burningham11, smith14} and many more. We show that ground-based near infrared searches continue to be valuable in the era of {\it Gaia} \citep{prusti16} and the Wide Field Survey Explorer (WISE, \citealt{wright10}) despite the leading role now played by those two all-sky survey missions.

\section{Data and Methods}\label{data}

The data flow for VIRAC2, from reduced images to astrometrically and photometrically calibrated source lists and time series, comprised multiple fairly distinct components. In order, they were: source detection, astrometric calibration, stellar time series production and mean astrometry fitting, and photometric calibration.

The VIRAC2 pipeline went through a number of design iterations during development. One version, internally designated VIRAC2$\beta{}$, produced a completed catalogue that was used in a number of published articles \citep[e.g.][to name a few]{husseiniova21, agarcia21, minniti21, wit08, molnar22, sanders22, sandersmira22, kaczmarek22, pramirez22, sormani22, sanders24, minnitti24, lucas24, luna24, Kaczmarek24, Nieuwmunster24}. In addition to those works, it was used at various points during the final catalogue production, primarily the photometric calibration component (see Section \ref{photcal}), and as a seed catalogue for the source list of the main pipeline (see Section \ref{initmatch}).
Since this earlier pipeline and catalogue influenced the final versions, for completeness we describe the main differences between the VIRAC2$\beta{}$ and final VIRAC2 pipeline versions in Appendix \ref{app:virac2b}.

\subsection{Data}

The recently retired VISTA Infrared Camera (VIRCAM) was the largest near-infrared detector array ever used for astronomy, with sixteen 2048$\times$2048 pixel arrays. VISTA and VIRCAM image a total area of 0.6 deg$^{2}$ at each pointing position or `pawprint'. Detectors are arranged in a 4$\times$4 grid with spacings of 90\% of a detector width in the $Y$ dimension and 42.5\% of a detector width in the $X$ dimension. The conventional VIRCAM tiling pattern consists of six of these pawprints (three offset in $X$ and two in $Y$) that fill a VIRCAM `tile' when stacked. VIRCAM tiles are approximately $1.4^{\circ}\times{}1.1^{\circ}$. Most positions in a VIRCAM tile are observed twice due to the $\approx{}50$\% pawprint overlap in the $X$ direction.
VISTA and VIRCAM are described by \citet{vista}. CASU provide pipeline data reduction and calibration of the photometry and astrometry via the VISTA Data Flow Pipeline \citep{vdfs}, see also  \citet{lewis10} and \href{http://casu.ast.cam.ac.uk/surveys-projects/vista/technical}{http://casu.ast.cam.ac.uk/surveys-projects/vista/technical}.

From CASU we acquired 179\,403 VVV, and 18\,020 VVVX observations (pawprints) that cover the VVV area. Pawprints, rather than tile stacks, should be used for precise astrometric work \citep{AG18}. The process of stacking to produce tiles complicates the point spread function (PSF) and background estimation. The observations spanned 30th January 2010 to 1st September 2019 and passed our basic quality control cuts -- images with incomplete FITS headers, seeing $>2.0$\arcsec{}, source count $<$ 20\,000, or sky level as determined by \textsc{imcore}\footnote{\citet{irwin85}; \url{http://casu.ast.cam.ac.uk/surveys-projects/software-release/imcore}} $>2000$ counts per second were rejected.

\subsection{Source Detection and Preliminary Processing}

Source detection and photometry was performed in an automated manner using a version of \textsc{DoPhot} \citep{dophot1, dophot2}, developed to perform PSF photometry extraction on highly crowded photometric images. \citet{AG18} demonstrated that it is capable of detecting a significantly higher proportion of stars in VVV fields. Further modification was undertaken by ourselves in order to extract astrometric uncertainties. \textsc{DoPhot} produced $1.14\times{}10^{11}$ tentative source detections from the 197\,423 images.

\begin{figure}
  \begin{center}
    \includegraphics[width=0.45\textwidth,keepaspectratio]{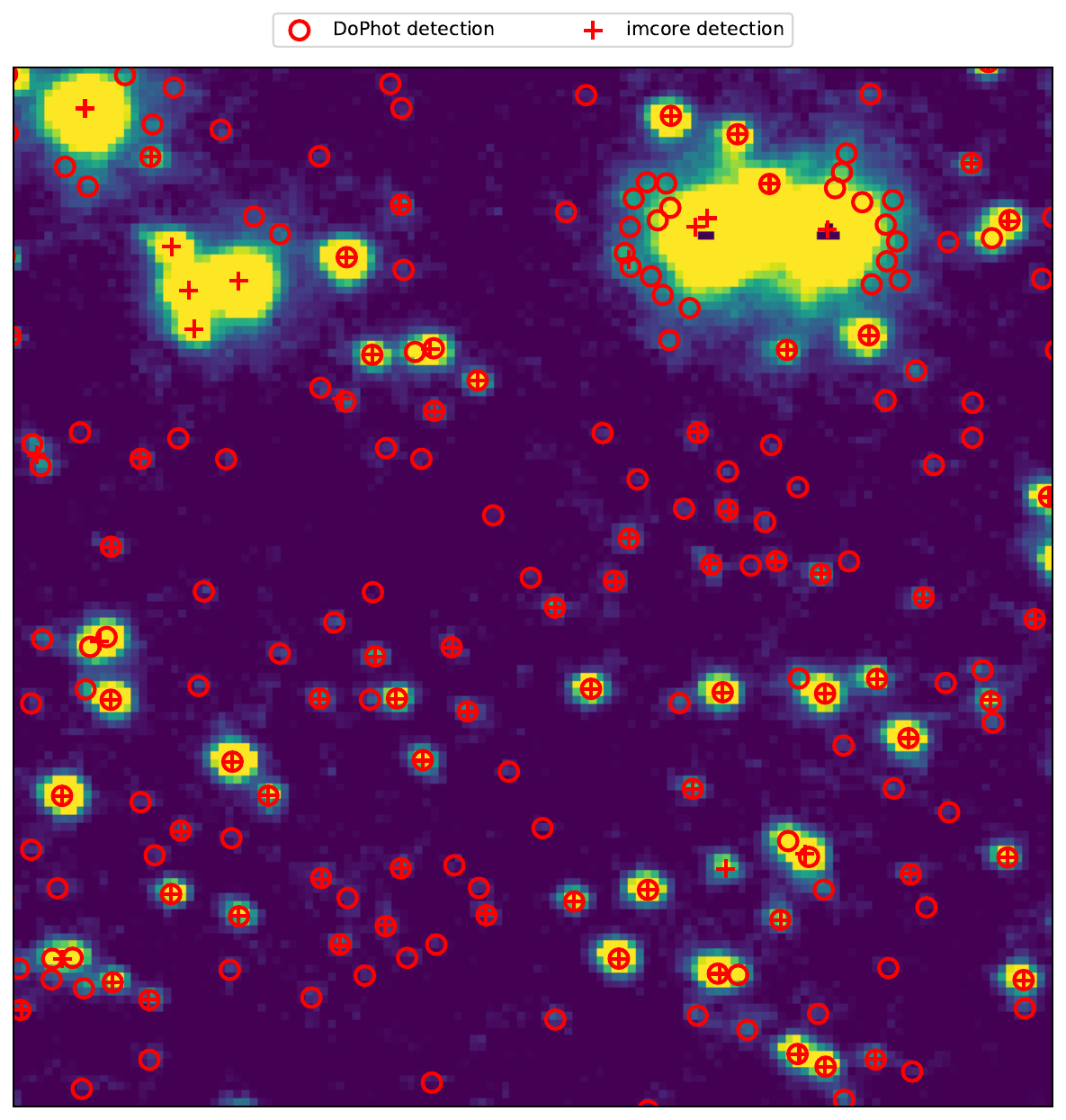}
    \caption{CASU \textsc{imcore} vs \textsc{DoPhot} source detections for an example image cutout, selected to showcase the strengths and weaknesses of each algorithm. \textsc{DoPhot} detections are plotted as red circles and \textsc{imcore} detections are plotted as red pluses. \textsc{DoPhot} is prone to erroneous detection of sources in the wings of bright stars and fails to detect saturated ones, but it is better able to separate blends and has a fainter detection limit. \textsc{imcore} detections tend to map consistently to real stars and saturated ones are recovered more reliably, though it deblends less effectively and has a brighter upper detection limit.}
    \label{imcore_vs_dophot}
  \end{center}
\end{figure}

VIRAC version 1 was based on source catalogues provided by CASU. These were produced by processing the VVV images with their aperture reduction software, \textsc{imcore}. The VVV and VVVX surveys cover the regions of the sky with the highest Galactic stellar density in the near infrared, and as such suffer from significant blending. Since \textsc{DoPhot} uses PSF fitting, it is better suited to analysis of these heavily blended fields. A comparison of the \textsc{imcore} and \textsc{DoPhot} source detection algorithms for an example image cutout is provided in Figure \ref{imcore_vs_dophot}. This cutout covers the region $1088<X<1220$, $1910<Y<2042$ of detector $1$ from exposure v20120621\_00300\_st. It was selected purely to showcase the strengths and weaknesses of each algorithm, although being nearly $5$ degrees from the Galactic centre it is not a particularly dense field in VVV terms (centered on $l,b=356.0,-2.8$). \textsc{DoPhot} has poor saturated star detection performance relative to \textsc{imcore}. Across the top half of the image we see multiple saturated (left) or highly saturated (right) stars that \textsc{DoPhot} fails to detect. This is partly an artificial restriction, since \textsc{DoPhot} is unable to reliable fit the fluxes of saturated objects it typically masks them instead. Additionally, Figure \ref{imcore_vs_dophot} shows that with our run configuration \textsc{DoPhot} erroneously detects sources in the wings of the highly saturated stars. This was also a conscious choice, since close companions to bright stars can be interesting and should ideally be kept. By contrast, \textsc{imcore} quite reliably detects saturated stars and it shows little contamination by false detections. Across the lower half of the cutout there are several instances where blended stars are reliably deblended by \textsc{DoPhot} but not by \textsc{imcore}. It is also evident that \textsc{DoPhot} detects many more stars than \textsc{imcore} at the fainter end, but some of these are erroneous detections of peaks in sky noise. Again, this is partly driven by our run configuration, which was tuned to push the faint limit. We ultimately utilised the multiple epochs to identify and remove the erroneous detections.

CASU provide 2d arrays mapping detector sensitivity (confidence maps) for every VIRCAM image. We recorded the confidence map value of the pixel containing the centroid of each detected source. One dither contributes $\approx{}50$ to confidence, so regions of the detector covered by both dithers have confidence of $\approx{}100$. Sources detected in image regions with confidence $<25$ were rejected. This most frequently occurs for sources in regions where a defective portion of a chip is not covered by the second dither position, or a where a source is covered by a defective region of the chip in both dither positions. We applied the CASU astrometric solution and radial distortion correction to each array in the catalogue to produce equatorial coordinates. The CASU astrometric solution was based on array positions measured by \textsc{imcore}. We did not necessarily expect \textsc{imcore} centroids to agree exactly with the \textsc{DoPhot} centroids, but they were sufficiently close that the astrometric solution was still valid for preliminary processing purposes. At this stage we simply required equatorial coordinates precise enough to identify sources in common with our astrometric reference catalogue.

\subsection{Astrometric calibration of individual images}\label{sec:astrocalib}


As our astrometric reference catalogue we used the projected position of Gaia DR3 sources at the epoch of the VISTA observation, taking into account their proper motions and parallaxes ($\alpha{}\delta{}_{Gaia}$ hereafter). Of the Gaia reference sources we simply required a full 5-parameter astrometric solution and renormalised unit weight error (ruwe) $<1.4$, to reject sources with poorly behaved astrometric solutions \citep[e.g. binary stars;][]{Lindegren18,stassun21,lindegren21}. Of the VISTA detections we required that reference sources were `perfect' stars according to \textsc{DoPhot}, i.e. they were fitted using the full 7 parameter (sky level, flux, position, and shape) model.

While matching between the VISTA and reference catalogues we found in some cases that the CASU astrometric solution was systematically offset from $\alpha{}\delta{}_{Gaia}$ in portions of the detector by up to $300$~mas. As a result, it was necessary to first refine the CASU astrometry by fitting a simple 6 parameter linear transformation matrix to align the two coordinate systems using sources matched within $1$\arcsec{}. After this, to produce a final pool of astrometric reference sources we cross matched the $\alpha{}\delta{}_{Gaia}$ catalogue to the VISTA catalogues with a 0.25\arcsec{} matching radius.

The $\alpha{}\delta{}_{Gaia}$ positions of the reference sources were then TAN projected using a tangent point at the center of the VIRCAM focal plane given in the FITS header of the original VISTA image. The resultant tangent plane coordinates were the $\chi{}\eta{}_{Gaia}$ astrometric reference frame.

We fitted Chebyshev polynomials of varying degree to map reference source VIRCAM array coordinate positions to their $\chi{}\eta{}_{Gaia}$ coordinates. This approach was found to perform well in our many test fields, while also limited over-fitting in regions with relatively few Gaia reference sources. 

For each chip of each observation we measured residuals to least squares fits of increasing degrees of Chebyshev polynomials using 5-fold cross validation. The sequence of polynomial degrees tested was: 3, 4, 5, 6, 7, 8, 10, and 13. At the point that the standardised cross-validated residuals began to deteriorate, the previous value was determined to be optimal to avoid over-fitting and the testing sequence was terminated. Once the optimal value was determined, reference sources with a residual in either dimension greater than $10\sigma$ were removed and a final re-fitting of Chebyshev polynomial coefficients was performed.

Our treatment of positional uncertainties began with a formal propagation of the array coordinate uncertainties reported by \textsc{DoPhot} through the Chebyshev polynomial fitted in the previous stage, to obtain uncertainties in the $\chi{}\eta{}_{Gaia}$ reference frame. Rather than explicitly incorporating the uncertainties on the Chebyshev polynomial coefficients themselves, we elected to fit for calibration uncertainties to be added in quadrature and also a multiplicative scaling factor to be applied to the uncertainties reported by \textsc{DoPhot}. I.e.:
\begin{equation}\label{eqn:errorscale}
    \sigma{}^2_{\rm VISTA} = \sigma{}^2_{\rm cal} + l\,\sigma{}^2_{\textsc{DoPhot}}
\end{equation}
\noindent where $\sigma_{\rm cal}$ and $l$ are the calibration uncertainties and error scaling factor, respectively.

For each dimension, over $N$ equal width magnitude bins we minimised the function:
\begin{equation}\label{eqn:escale_ast}
    \sum_{n=1}^{N}{\Biggl(\ln{\Biggl(
    k\cdot{}{\rm MAD}\Biggl(r\cdot{}\sqrt{\sigma_{\rm Gaia}^2 + \sigma{}^2_{\rm VISTA}}\Biggr)\Biggr)
    \Biggr)^2}}
\end{equation}
\noindent where $k$ is the 1.4826 approximate scaling factor to be applied to the median absolute deviation (MAD) to obtain a reasonable estimate of the standard deviation that is robust against outliers, and $r$ are the separations between the transformed VISTA coordinates and Gaia coordinates in a given dimension. 

The objective was to rescale the uncertainties such that the residual to the coordinate transformation divided by its uncertainty was approximately unit Gaussian. By measuring the spread within sets of equal width magnitude bins we avoided the fainter magnitude bins having undue weight due to their significantly larger source volume. The obtained calibration uncertainty and error scaling factors were applied to the \textsc{DoPhot} positional uncertainties as per Equation \ref{eqn:errorscale} to obtain our final single-epoch positional uncertainties.

\subsection{Catalogue Pipeline}\label{catpipe}

The main catalogue production pipeline can be summarised as an iterative process of two main components: source matching and astrometric fitting. Ultimately, the goal of this iterative procedure was to produce a complete list of sources, having correctly identified their corresponding detections in each VISTA observation. Individual healpixels of resolution 8-10 (approximately 189, 47 or 12 square arcminutes, chosen based on approximate local source density and epoch count) were processed independently, incorporating a small border to include detections of sources straddling the edges of the target healpixel.

\subsubsection{Initial Source Matching}\label{initmatch}

The first step (after data ingestion, etc.) was to perform an initial source matching, producing a complete but highly contaminated (by duplicates, noise, etc.) source list (SL hereafter). The SL was seeded with the VIRAC2$\beta{}$ catalogue, which in turn had been seeded with Gaia DR2 (see Appendix \ref{app:virac2b} for more details).

The positions of SL sources were propagated to the epoch of the first (chronological) VISTA catalogue using their 5-parameter astrometric solutions. The propagated positions were then matched to the VISTA catalogue with a $0.339$\arcsec{} radius (the average VIRCAM pixel size), requiring each match to be the best match in both directions and also within a radius of five times the positional uncertainty. VISTA catalogue row indices of successful matches were recorded for each SL row. The requirement that a match is the best in both directions at this stage implies that multiple SL sources cannot be associated with a single VISTA catalogue detection. Unmatched VISTA catalogue rows were then appended to the SL, and at this stage they were assigned proper motion and error in each dimension equal to the mean and standard deviation of the proper motions of VIRAC2$\beta{}$ seed sources within the same healpixel, and a parallax of $0.0$~mas and uncertainty equal to the standard deviation of the parallaxes of the seed sources. Not taking the local proper motion distribution into account is equivalent to assuming zero proper motion, which is generally incorrect and could become important when crossmatching with large epoch differences. Since the majority of sources in this bandpass, depth and direction are located in the vicinity of the Galactic centre, at $\approx{}8$~kpc, only a small fraction will have parallax greater than a few tenths of a milliarcsecond. An error of even a few milliarcseconds is at the level of a few per cent of the matching radius, which we deemed acceptable, particularly as it is only an initial guess.

This matching procedure then continued for each VISTA observation in chronological order until our complete but contaminated SL was produced. SL rows with fewer than two contributing VISTA detections were removed.

\subsubsection{Astrometric Fitting}\label{astfitting}

With sequences of VISTA detections for each source, we read their $K_s$ astrometric time series and fitted their five astrometric parameters. We used only the $K_s$ bandpass data for astrometry fitting. For sources with fewer than 10 epochs we simply calculated a mean position, epoch and positional uncertainty, and assumed they had the average proper motion of local VIRAC2$\beta{}$ sources and zero parallax (as in Section \ref{initmatch}). For sources with 10 or more epochs we proceeded with a full 5-parameter least squares astrometric solution as described below. We worked in local tangent plane coordinates: the astrometric time series for each source underwent a TAN projection to a $\chi$ and $\eta$ reference system about its mean equatorial position, yielding positional offsets equivalent to $\Delta{}\alpha\cos{\delta}$ and $\Delta{}\delta$.

We measured preliminary residuals to the 5-parameter astrometric model ($\chi_{0}$, $\eta_{0}$, $\omega$, $\mu_{\chi}$, and $\mu_{\eta}$) by least squares fitting the astrometric time series using 5-fold cross-validation, using their inverse variances (see Section \ref{sec:astrocalib}) as weights. Model parameters were fixed at the reference epoch, 2014.0, chosen to be the approximate mid-point of the survey. Observations with $>5\sigma$ 
residuals were removed and if there were still $\ge{}10$ remaining observations then the astrometric model fitting was repeated without cross-validation. The resulting astrometric solution, including its covariance matrix, is recorded and residuals to it are remeasured and recorded. The covariance in the parameters came from a Jacobian approximation to the Hessian of the least squares objective function, provided by the lmdif routine of the \textsc{minpack} library on which \textsc{scipy.optimize.leastsq} relies. The $\chi,\eta$ coordinate system is equivalent to the $\alpha\cos{\delta},\delta$ coordinate system. If sigma clipping reduced the number of remaining observations below $10$ then the source reverted to a mean position and error, as above.

\subsubsection{Iterative Re-Matching}\label{itermatch}

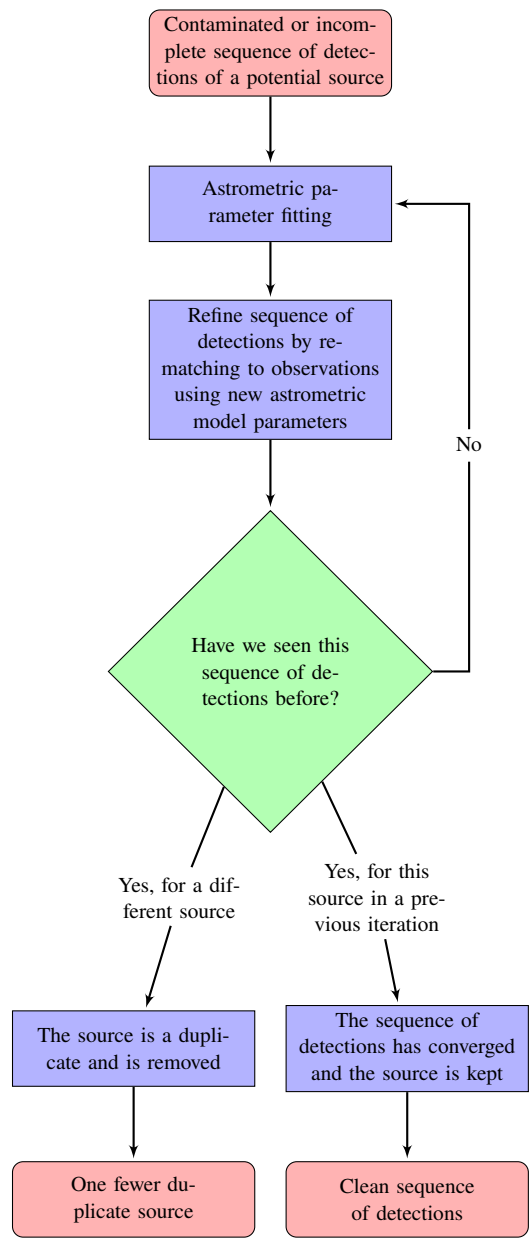
\begin{figure}
\centering
\begin{tikzpicture}[node distance=2cm]

\node (start) [startstop] {Contaminated or incomplete sequence of detections of a potential source};
\node (pro1) [process, below of=start] {Astrometric parameter fitting};
\node (pro2) [process, below of=pro1, yshift=-0.2cm] {Refine sequence of detections by re-matching to observations using new astrometric model parameters};
\node (dec1) [decision, below of=pro2, yshift=-2.0cm] {Have we seen this sequence of detections before?};
\node (pro2a) [process, below of=dec1, yshift=-3.0cm, xshift=1.8cm] {The sequence of detections has converged and the source is kept};
\node (pro2b) [process, below of=dec1, yshift=-3.0cm, xshift=-1.8cm] {The source is a duplicate and is removed};
\node (stopa) [startstop, below of=pro2a] {Clean sequence of detections};
\node (stopb) [startstop, below of=pro2b] {One fewer duplicate source};

\node (lab1) [label, below of=dec1, yshift=-1.0cm, xshift=1.4cm] {Yes, for this source in a previous iteration};
\node (lab2) [label, below of=dec1, yshift=-1.0cm, xshift=-1.2cm] {Yes, for a different source};
\node (lab3) [label, above right of=dec1, yshift=1.6cm, xshift=1.2cm] {No};

\draw [arrow] (start) -- (pro1);
\draw [arrow] (pro1) -- (pro2);
\draw [arrow] (pro2) -- (dec1);
\draw [line] (dec1) -- (lab1);
\draw [arrow] (lab1) -- (pro2a);
\draw [line] (dec1) -- (lab2);
\draw [arrow] (lab2) -- (pro2b);
\draw [line] (dec1.east) -| (lab3);
\draw [arrow] (lab3) |- (pro1.east);
\draw [arrow] (pro2a) -- (stopa);
\draw [arrow] (pro2b) -- (stopb);

\end{tikzpicture}
\caption{An illustration of the iterative process of refinement of the sequence of VISTA detections of a single source.}\label{flow1}
\end{figure}

After initial source matching and fitting of astrometric solutions, subsequent source matching operations were performed slightly differently to those described in Section \ref{initmatch}. Figure \ref{flow1} illustrates the process of SL refinement for a single source. The objective now was to take the complete but contaminated SL, refine the time series for all sources, and reduce the level of contamination. For coordinate propagation we now used the internally determined best-fitting 5-parameter astrometric solutions for sources with $\ge{}10$ detections, and mean field astrometry and average positions for sources with $<10$ detections.

As in Section \ref{initmatch} we propagated the positions of sources to the epoch of each VISTA catalogue, but this time we simply found the closest VISTA catalogue match within our $0.339$\arcsec{} and $5\times$ error matching radius. This now meant that multiple SL entries could incorporate a single VISTA detection (which is a valid treatment of e.g. blended sources). This made identification of the majority of duplicates trivial, since they ultimately converged to the same set of VISTA catalogue detections.

We no longer appended unmatched VISTA catalogue rows to the SL, since our SL at this stage should be as complete as was practical. The sequence of VISTA catalogue detections (which in practice was a sequence of catalogue row indices) for each SL row was hashed (i.e. assigned a pseudo-unique reference integer for a given sequence, for ease of comparison and memory efficiency) and recorded. Duplicates of hashes therefore indicated duplicated sequences of detections and hence duplicated sources. Hashes were compared to all others in the SL and where they had been seen elsewhere the source was considered a duplicate and removed. 

Finally, to determine whether a source had converged we looked for its current hashed list of VISTA catalogue indices among those from all of its previous iterations. If the current hash had been seen previously then the source was considered to have converged. Sources which had not yet converged had their astrometric solutions refitted as described in Section \ref{astfitting} and this re-matching procedure was then repeated until all sources had either been removed as duplicates or had converged.

As mentioned, sources with only a mean position and error used the mean proper motion of nearby sources and a parallax of zero for matching purposes. Once such a source converged we nullified its proper motion and parallax in the output catalogues.

\subsubsection{Additional Duplicate Flagging}
\label{adddupflagging}

At the iterative re-matching stage obvious duplicates were removed by identifying identical lists of VISTA catalogue detections. This strict comparison could leave additional less obvious duplicate sources in the SL. It required a time series difference of only a single VISTA detection to pass the earlier method of checking, which was not uncommon with hundreds to thousands of observations (depending on sky location).

The remaining potential duplicated entries were found by identifying groups of sources within $0.339$\arcsec{}. The source in this group with the most detections was considered the primary source, and all others were flagged as duplicates where they shared more than $20\%$ of their VISTA catalogue detections with other sources in the group. The $20\%$ threshold was necessary to accommodate situations in which a star with e.g. high proper motion happens to be blended with another at the reference epoch (2014.0) and hence was within the $0.339$\arcsec{} threshold, yet was not a genuine duplicate. Such an occurrence is generally improbable for a given source, but among $\sim{}10^9$ sources it does happen (see e.g. \citealt{mcgill19}). Approximately $19\%$ of sources were flagged as probable duplicates.

\subsubsection{Ambiguous Match Flagging}\label{sec:ambmatch}

Our iterative matching routine (Section \ref{itermatch}) allowed multiple sources to associate with single VISTA detections. We identified and flagged these cases in the time series data as ambiguous matches. For primary sources (i.e. those that are not probable duplicates) we flagged detections as ambiguous only where the detection was shared with another primary source. The number of ambiguous detections is provided in the catalogue for each source.

\subsubsection{Observation Counts}\label{obscount}

To determine whether a given source was likely to have been covered by a given observation we first propagated its equatorial position to the epoch of the observation. We then projected the propagated sky coordinates to VIRCAM array coordinates using the skycoord\_to\_pixel function of the Astropy WCS module using the CASU WCS of the original image, and checked this position against a master confidence map.

Our master confidence map is an average of a random selection of confidence maps from late in the survey. We applied a minimum spatial filter, whereby each pixel was assigned the minimum confidence out of itself and its direct neighbour pixels. The application of the minimum spatial filter meant we were effectively requiring all nearby pixels to meet the threshold. This was also necessary to take into account imperfections in the master confidence map due to inherent variation in dither offsets between observations. We simply required a confidence level $>25$ for a given position to be considered covered by a given observation. We used an average confidence map since heavily saturated sources can also produce regions of low confidence on single maps, and we did not want to erroneously reduce the observation counts of saturated stars. Without this consideration we would inflate their ratio of detections to observations, which we have found to be a useful indicator of the reliability of the data for each source (see Section \ref{sec:junk_rejection}).

There is inherent uncertainty in many stages of this procedure and it should therefore be stressed that the provided observation counts are approximate. To illustrate this, we note that $0.6\%$ of sources have more $K_s$ band detections than the number of observations determined by this procedure. Since a source could only be assigned one detection per observation (though a detection could be assigned to multiple sources), this clearly indicates an error in the number of observations. The number of observations of each source is provided in the catalogue, alongside the number of detections.

\subsection{Photometric Calibration}\label{photcal}

The pipeline until this point was largely concerned with the production of a complete catalogue of sources with high quality astrometry and a low level of contamination. This catalogue now allowed us to accurately calibrate the VISTA photometry. We performed this calibration in two stages. The first stage was a coarse survey-wide approach which aimed to anchor the instrumental photometry onto an absolute photometric reference frame. The second stage was a finer calibration at a sub-array level which aimed to further reduce scatter in individual light curves.

We note that the photometric calibration component of the pipeline was produced using the VIRAC2$\beta{}$ version of the catalogue. See Appendix \ref{app:virac2b} for details. The main differences between this preliminary version and the final catalogue version were restricted to astrometric enhancements and minor changes to the time series of individual stars. Considering this, and the considerable computational expense of the photometric calibration, we deemed it unnecessary to rerun this component with the final catalogue version.

Components of the photometric calibration strategy used varying criteria for selection of reference stars. Common source-specific criteria were:
\begin{enumerate}
    \item sources have 5-parameter astrometric solutions and are not flagged as possible duplicates;
    \item sources have no detections that were ambiguously associated with another source;
    \item sources were detected in at least 30\% of their $K_s$ band observations.
\suspend{enumerate}
Common detection-specific criteria were:
\resume{enumerate}
    \item detections are `perfect' according to \textsc{DoPhot} (i.e. flagged as well fitted by the PSF model);
    \item detections have $\chi<5$ as measured by \textsc{DoPhot};
    \item detections have instrumental magnitude error $<0.5$;
    \item detections are not associated with multiple sources;
    \item detections were not outliers at the $5\sigma$ level.
\end{enumerate}
We also incorporated some more specific criteria which will be described in the relevant places, but we will refer back to this list as necessary.

We note that stars that exhibit significant variability represent a tiny fraction of the content of the catalogue, so their impact should be negligible. Even so, each component incorporated some form of outlier removal which should further reduce their impact.

\subsubsection{Primary Calibration}\label{primarycal}

We used an SDSS ubercal \citep{ubercal} inspired approach, whereby spatial- and time-dependent magnitude offsets were fitted for such that they reduced the photometric residuals between successive observations of the same stars. This approach was ideally suited to the VISTA/VIRCAM observing strategy as there is a near guarantee of at least two observations of a given star on different parts of the focal plane, due to the VIRCAM pawprint pattern. Isolated reference stars which were also observed by the two-micron all sky survey (2MASS, \citealt{2mass}) were used to anchor to the photometric system defined by \citet{GF18} (see in particular their equations 5-9).

The functional form of our initial calibration from instrumental to VISTA magnitudes was:
\begin{equation*}
    m_{\rm VISTA} = m + ZP + F(c,t) + I(xy)
\end{equation*}
\noindent where:
\begin{description}
    \item $m_{\rm VISTA}$ is the magnitude of a source in the VISTA system, presumed known for 2MASS reference sources and unknown otherwise.
    \item $m$ is the instrumental magnitude of the source in a given detection.
\end{description}
\noindent and the calibration coefficients were:
\begin{description}
    \item $ZP$ - the overall zero point offset for the relevant bandpass.
    \item $F(c,t)$ - an offset for each chip ($c$) in each observation ($t$).
    \item $I(xy)$ - an offset dependent on position on the chip ($xy$) for a given chip. This is commonly referred to as the illumination correction map, and corrects the detector systematics shown in section 7.3 of \citet{GF18}. Each detector is subdivided into $128\times{}128$ spatial bins and we fitted for a magnitude offset in each bin. One $I(xy)$ map is measured per bandpass. We tested the use of bilinear interpolation for $I(xy)$, but found negligible improvement over simply using the nearest neighbour.
\end{description}

To solve for the calibration coefficients we used the \mbox{\textsc{L-BFGS-B}} minimisation algorithm through the minimize function of the scipy.optimize module. The contribution to the objective function of a 2MASS reference source (i.e. a source with a presumed known $m_{\rm VISTA}$) with $n$ (chronologically ordered) observations was:
\begin{equation}
    g = \sum_{i=1}^{n}{}  \left(\frac{m_{\rm VISTA}-m_i - (ZP+F+I)_i}{\sqrt{\sigma_{m_{\rm VISTA}}^2+\sigma_{m_i}^2}}\right)^2
\end{equation}
\noindent and the contribution to the objective function for a non-2MASS reference source with $n$ observations was:
\begin{equation}
    h = \sum_{i=1}^{n}  \left(\frac{m_{j}-m_i - ((F+I)_{i} - (F+I)_{i-1})}{\sqrt{\sigma_{m_{i-1}}^2+\sigma_{m_i}^2}}\right)^2
\end{equation}
\noindent where:
\begin{equation*}
    j = \begin{cases}
    N, & \text{if $i=1$.}\\
    i-1, & \text{otherwise.}
    \end{cases}
\end{equation*}

The overall objective function then was simply:
\begin{equation}
    f = \left(\sum{g} + \sum{h}\right) N^{-1}
\end{equation}
\noindent where $N$ is the total number of detections across all reference sources.

A reference source pool was selected based on sources that met all of criteria i, ii, and iii, and that had two or more detections in the relevant bandpass that met criteria iv, v, and vi.

From within the reference source pool described above we identified zero point anchors as those that had 2MASS counterparts that were within $0.5$\arcsec{} with `AAA' photometric quality flags. They also had to be fainter than $12$th magnitude in all 2MASS bands to avoid saturation in the VISTA images, and must have no other VIRAC2$\beta{}$ sources within $2$\arcsec{} that might be unresolved in 2MASS.
An additional requirement for 2MASS reference sources was placed on their \citet{sfd98} $E(B-V)$, the threshold of which depended on the bandpass being calibrated: $<0.1$ for $Z$ and $Y$, $<0.8$ for $J$ and $H$, and $<1.0$ for $K_s$.

The statistics provided in Table \ref{ipc_stats} indicate the scale of the minimisation problem. The design matrices were very large but they were also extremely sparse. The majority of design matrix rows had only $4$ nonzero elements out of $\approx{}$350\,000 ($\approx{}$3\,000\,000 in $K_s$), and hence the problem was amenable to sparse matrix techniques. 

\begin{table}
\centering
\caption{Initial photometric calibration statistics. The number of coefficients, the total number of detections across all reference sources, the number of elements in the design matrix, and the number of sources which are $ZP$ anchors.}
\label{ipc_stats}
\begin{tabular}{|c|r|l|c|r|}
\hline
  \multicolumn{1}{|c|}{band} &
  \multicolumn{1}{c|}{coefficients} &
  \multicolumn{1}{c|}{$N$} &
  \multicolumn{1}{c|}{$\mathbf{X}$ size} &
  \multicolumn{1}{c|}{$ZP$ anchors} \\
\hline  
 $Z$   & 343\,025   & $7.32\times{}10^8$    & $2.51\times{}10^{14}$ & 1\,374     \\
 $Y$   & 344\,241   & $8.00\times{}10^8$    & $2.75\times{}10^{14}$ & 1\,378     \\
 $J$   & 374\,561   & $1.60\times{}10^9$    & $6.01\times{}10^{14}$ & 3\,005\,274 \\
 $H$   & 356\,193   & $1.24\times{}10^9$    & $4.41\times{}10^{14}$ & 3\,003\,461 \\
 $K_s$ & 3\,051\,649 & $2.53\times{}10^{10}$ & $7.71\times{}10^{16}$ & 3\,439\,410 \\
\hline\end{tabular}
\end{table}

We ran the minimisation over 30 iterations. Successive iterations had stricter relative error and absolute difference requirements of the reference sources and lower $\frac{\Delta{f}}{f}$ termination tolerance. The first iteration rejected $>10\sigma$ and $>1.0$ mag residuals and terminated the minimiser at $\frac{\Delta{f}}{f}\leq{}10^{-5}$ and those limits decreased linearly to the last iteration which rejected $>3\sigma$ and $>0.1$ mag residuals and terminated the minimiser at $\frac{\Delta{f}}{f}\leq{}10^{-6}$. This minimisation routine was run twice. The first run used the mean $m_{\rm VISTA}-m_i$ for detections of 2MASS reference sources as the starting zero point offset and $0$ for all remaining coefficients, and its purpose was to obtain an approximate illumination map. A second run was then performed using the output illumination map and zero point offsets from the first run as starting values. The starting value of $F(c,t)$ was $2.5\log{\frac{T_{e}}{T_{eb}}}$ where $T_{e}$ is the exposure time of the given observation and $T_{eb}$ is the mode of the exposure time distribution of observations in the lower bulge in the relevant bandpass. The lower bulge region is significant as this is where the majority of zero point offset anchors were located due to the low extinction requirement of 2MASS reference sources. This normalisation between exposure times was most necessary for the $Z$, $Y$, $J$, and $H$ bands, where the default exposure time is different between the disk and bulge and the overlap between them in terms of numbers of reference sources is relatively small. In the K$_s$ bandpass the modal exposure time is $4s$ for all fields so the second run was not so crucial, but we performed it anyway out of an abundance of caution.

\begin{figure*}
  \begin{center}
    \includegraphics[width=\textwidth,keepaspectratio]{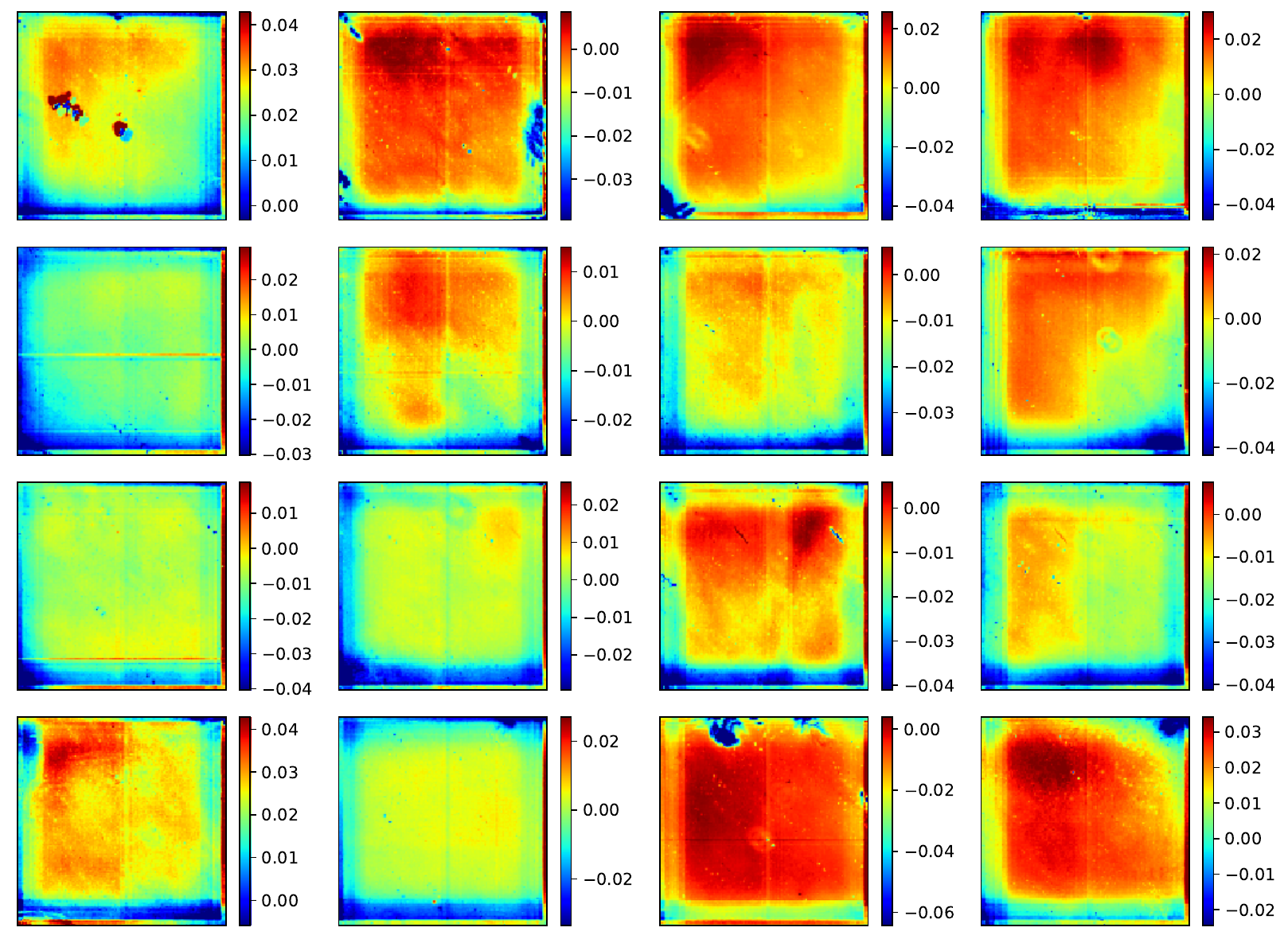}
    \caption{The fitted $K_s$ band $I(xy)$ illumination map. Detector numbers increase from left to right then top to bottom, e.g. the top left is detector 1, the top right is detector 4 and the bottom right is detector 16. Detector $X$ coordinates increase from left to right, and $Y$ coordinates increase from bottom to top. On the colour axis are the magnitude offsets, clipped at their $1$st and $99$th percentiles for clarity.}
    \label{illmap}
  \end{center}
\end{figure*}

Figure \ref{illmap} shows the resultant $K_s$ band $I(xy)$ map. The CASU web pages\footnote{\href{http://casu.ast.cam.ac.uk/surveys-projects/vista/technical/known-issues}{http://casu.ast.cam.ac.uk/surveys-projects/vista/technical/known-issues}} describe some of the defects which are apparent in the map.

\subsubsection{Secondary Calibration}

\begin{figure*}
  \begin{center}
    \includegraphics[width=.45\textwidth,keepaspectratio]{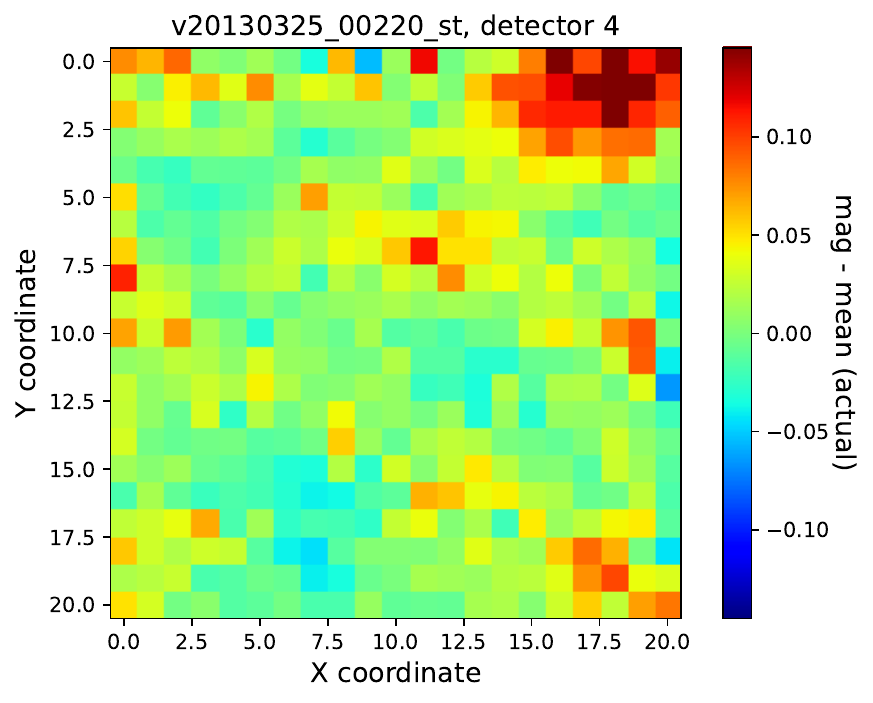}
    \includegraphics[width=.45\textwidth,keepaspectratio]{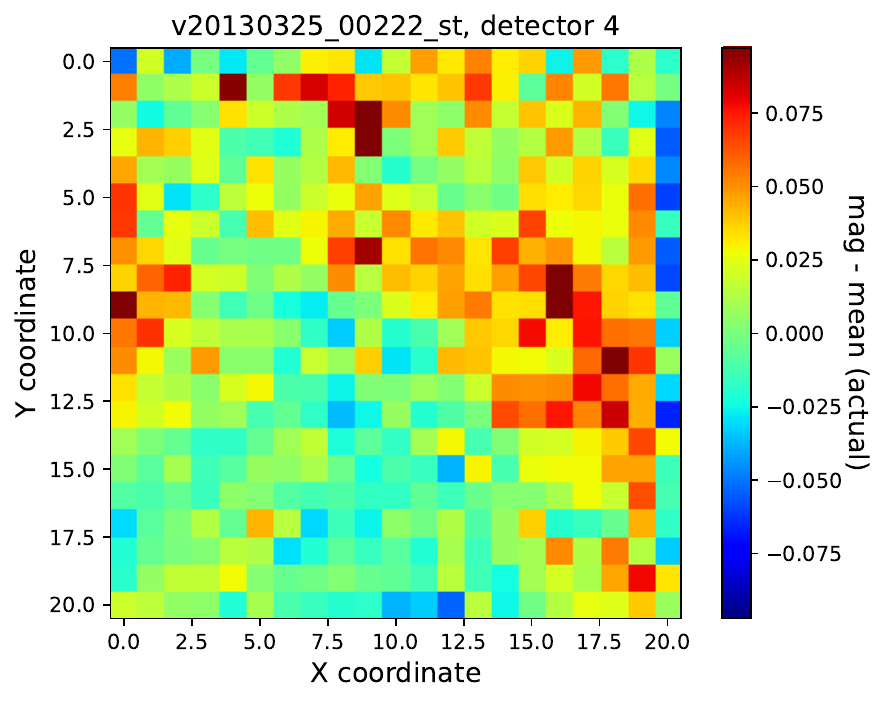}\\
    \includegraphics[width=.45\textwidth,keepaspectratio]{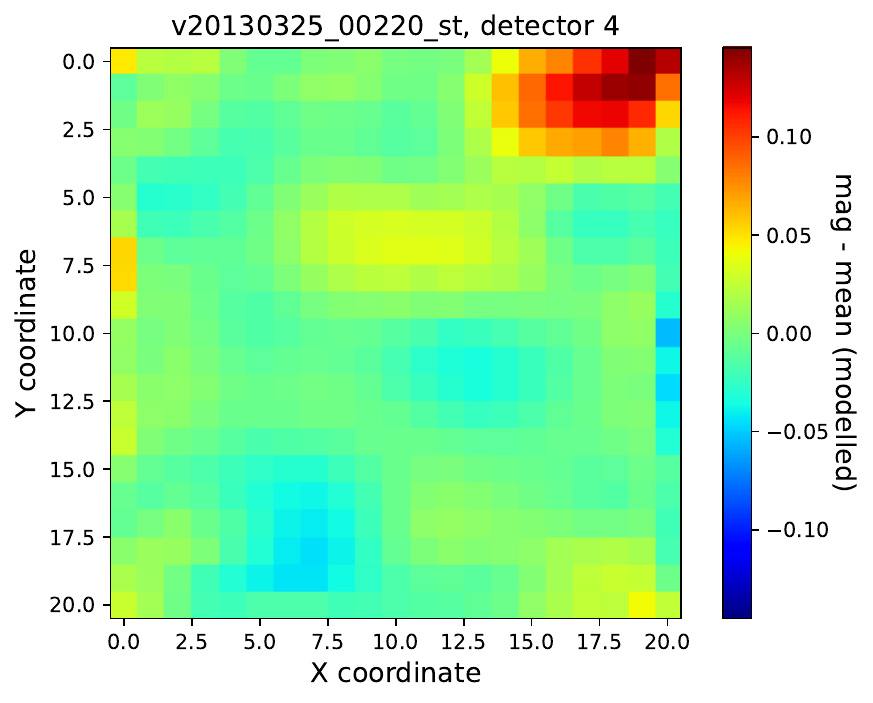}
    \includegraphics[width=.45\textwidth,keepaspectratio]{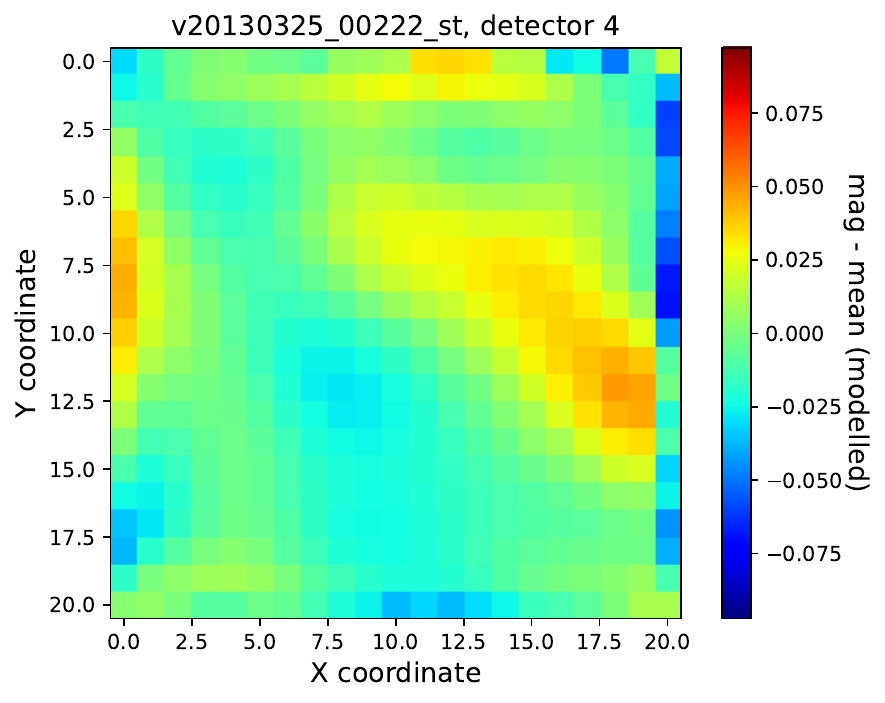}\\
    \includegraphics[width=.45\textwidth,keepaspectratio]{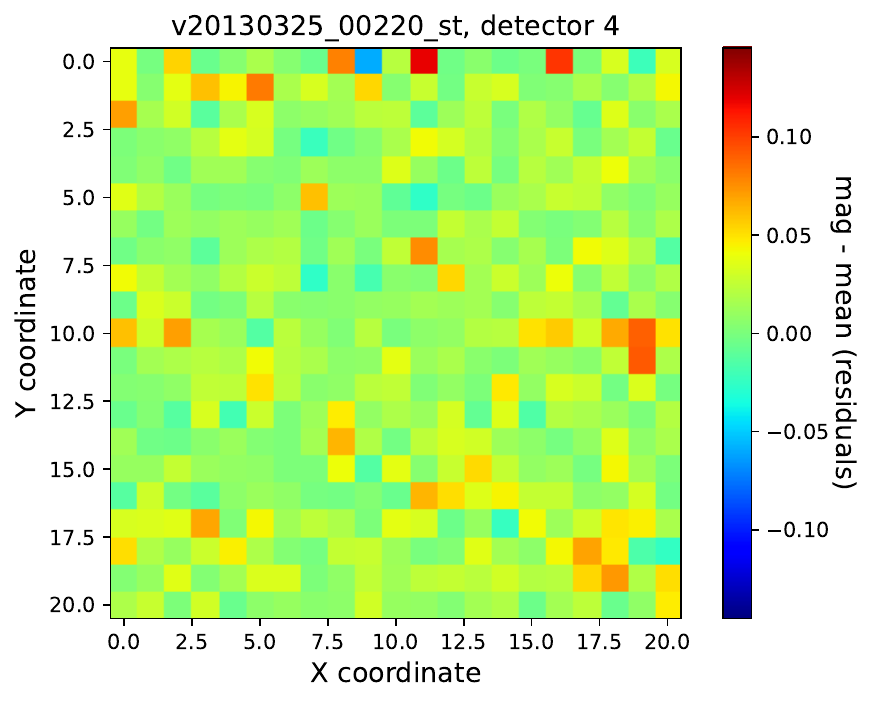}
    \includegraphics[width=.45\textwidth,keepaspectratio]{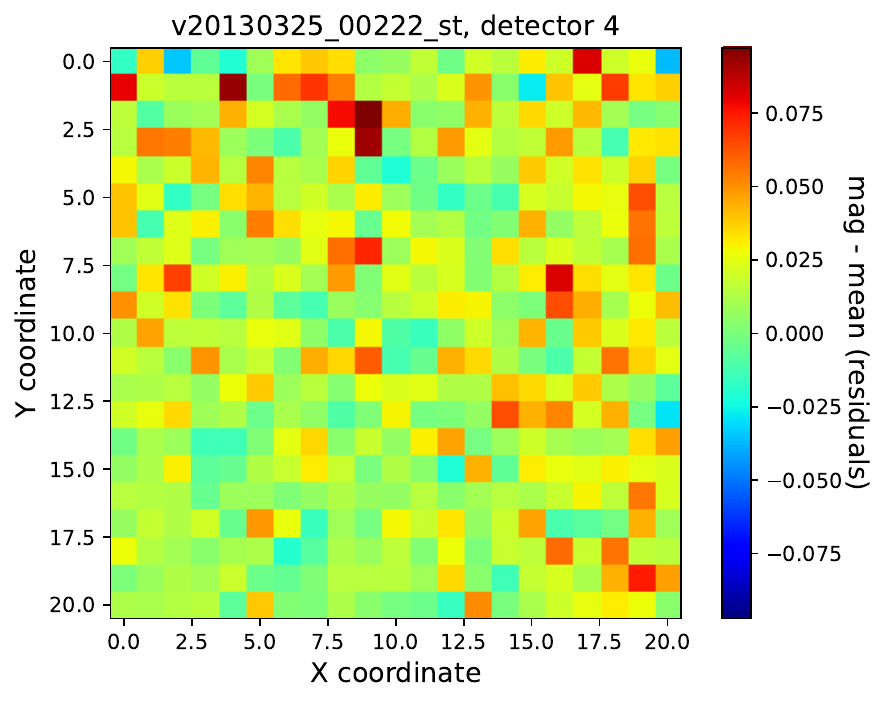}\\
    \caption{The left and right columns show results for detector 4 from two separate exposures, taken a few minutes apart and overlapping by half a detector. \textit{Upper panels}: Binned magnitude offsets of stars between their time-series averaged magnitude and those of these example exposures after the primary photometric calibration has been applied. The structure indicates there is additional systematic signal to be cleaned from the photometry. The difference between the left and right exposures indicates that the systematic signal varies on short timescales. \textit{Middle panels}: Our fitted model of the offsets shown in the upper panels. \textit{Lower panels}: Residuals remaining after subtraction of the fitted model signal from the input signal.}
    \label{hfad_plots}
  \end{center}
\end{figure*}

During inspection of the outcome of the primary calibration described in Section \ref{primarycal}, it became clear there remained coherent time-varying structures in maps of the residuals, see Figure \ref{hfad_plots}. Removing this structure with further processing was desirable, since doing so would reduce scatter in the light curves.

Visual inspection of a few hundred maps of the residuals to the primary calibration indicated that the patterns, spatial scales, and amplitudes of the coherent structures varied significantly over short timescales and among observations of the same field. In addition, the density of available high quality reference sources varied significantly between fields. Given these considerations we again decided to fit Chebyshev polynomials of varying degrees to the offset between source magnitudes in a given observation from their average magnitudes across all observations, in a similar manner to the astrometric calibration method detailed above (see Section \ref{sec:astrocalib}). Essentially, we were building a time-dependent illumination map, $I(xy,t)$ to reuse the notation of the primary calibration. However here it had a lower effective resolution as we had far fewer reference stars in a single observation, and the effective resolution varied between observations to account for the varying numbers and spatial distributions of reference stars.

We generated a reference star candidate pool from stars that met all of criteria i and iii. In addition they had to have two or more detections that met criteria v, vii, and viii, and that had \mbox{$>100$} reference sources for all coefficients of the initial photometric calibration and that were not at the edge of the illumination map. The primary calibration component benefitted from an abundance of reference stars, hence the selection criteria for the secondary component was less strict.

For each reference star candidate we adopted the median magnitude of detections meeting the detection-specific criteria listed above as the `true' magnitude. The median is more robust to outliers than other measures, and outliers were still present in our data to some degree despite the data quality requirements described above. 

For a given observation, we selected stars from the above pool by additionally imposing detection-specific criterion iv. This criterion tends to remove fainter detections, so incorporating it when computing the median magnitudes would have biased them bright. The various selection criteria listed above left us with a list of high quality detections of reliable reference sources, with robustly measured average magnitudes that were anchored to the VISTA photometric system.

The fit itself was performed using 5-fold cross-validation to robustly measure the residuals to the Chebyshev polynomial fits through a pre-determined sequence of 3, 5, 7, 10, 13, 17, 21, and 25 degrees. We stopped the sequence when the mean cross-validated standardised residual either: increased, in which case we used the previous value; or decreased by less than 1\% or we reached the end of the sequence, in which case we used the current value. The purpose of this procedure was to fit for what are fairly complex residual maps but to also avoid over-fitting our data. Once we had determined the optimal number of Chebyshev polynomial degrees to achieve this balance we removed reference sources with cross-validated residuals greater than $3\sigma$ and refitted using all remaining reference sources. Application of the calibration was simply a matter of evaluating the fitted polynomials for every detection in each observation and subtracting the obtained offset. An example structure model is shown in Figure \ref{hfad_plots}.

Finally, to calibrate our uncertainties we adopted a similar method to that of the astrometric uncertainty calibration (see Section \ref{sec:astrocalib}). For each detector in each observation we rescaled the \textsc{DoPhot} instrumental magnitude uncertainties by multiplying by a scaling factor and adding in quadrature a calibration uncertainty, as per Equation \ref{eqn:errorscale}. These values were acquired by minimising the function below over $N$ equal width magnitude bins:
\begin{equation}\label{eqn:magerrscale}
    \sum_{n=1}^{N}{\Biggl(\ln{\Biggl(
    k\cdot{}{\rm MAD}\Biggl(   \frac{m - \bar{m}}{\sigma{}_{\rm VISTA}}   \Biggr)\Biggr)
    \Biggr)^2}}
\end{equation}
\noindent where $k$ is the 1.4826 approximate scaling factor to be applied to the median absolute deviation (MAD) to obtain a reasonable estimate of the standard deviation that is robust against outliers; and 
$m-\bar{m}$ are the separations between the single epoch VISTA magnitudes and the median magnitudes across all selected observations. Rescaling the photometric uncertainties in this way served to bring the standardised residuals closer to unit Gaussian.

\subsection{Photometry Statistics}\label{photstats}

For individual detections we provide a photometric quality flag, to indicate where a detection has failed one or more criteria. These are supplied as integer values corresponding to 6 bit flags represented as an integer, and are as follows:
\begin{description}
    \item 000001 - 1: no secondary photometric calibration applied,
    \item 000010 - 2: $<5$ secondary photometric reference stars per coefficient,
    \item 000100 - 4: in an edge bin of the illumination map,
    \item 001000 - 8: $<100$ reference stars for the primary calibration $I(xy)$ coefficient,
    \item 010000 - 16: $<100$ reference stars for the primary calibration $F(c,t)$ coefficient,
    \item 100000 - 32: $<100$ reference stars for the primary calibration $ZP$ coefficient.
\end{description}
The photometric error flag supplied for a detection is the combination of its bit flags, i.e. the sum of the integer values listed above.

We provide some basic statistics for each bandpass to characterise the photometric time series of all sources. Statistics were computed using only high quality observations. The requirements for a detection to be included in the photometric sequence for the purposes of computing statistics were:
\begin{enumerate}
    \item zero photometric quality flag, i.e. it failed none of the criteria listed above,
    \item not flagged as ambiguous (see Section \ref{sec:ambmatch}), and
    \item not an astrometric outlier above the $5\sigma$ level.
\end{enumerate}
All three of these criteria had to be met. All detections, regardless of these quality criteria are included in the time series datasets, should one wish to recompute statistics with a different set of criteria. The number of detections that contributed to the statistics in each bandpass is provided in the catalogue.

Note that the set of observations contributing to the photometric statistics differed from that used for fitting the mean astrometry. Firstly, for astrometric purposes the photometric quality flags were ignored. Secondly, the $5\sigma{}$ astrometric outlier cut was applied using the cross-validated residuals in the case of mean astrometry fitting, and using residuals to the output mean astrometry in the case of photometric statistics computation.

The photometric statistics provided are the mean magnitude and the standard deviation of the magnitudes in each bandpass. In addition, for the $K_s$ bandpass only, we provide some additional statistics describing the photometric time series. These are: The modified Julian day of the first and last epochs; the skewness (corrected for statistical bias); a selection of percentiles -- 0, 1, 2, 4, 5, 8, 16, 25, 32, 50, 68, 75, 84, 92, 95, 96, 98, 99, and 100; the median absolute deviation from the median; the median photometric uncertainty; the Stetson I, J and K indices (\citealt{stetsonI, stetsonJK}); and the von Neumann ratio $\eta$ (\citealt{eta1, eta2}).
Note that when computing the Stetson indices we considered observations taken within 1 hour to be contemporaneous, for the purpose of identifying observation pairs. The number of observation pairs that contributed to the Stetson I and J indices for each source is also provided. For the Stetson J index we included unpaired observations with half weight.

\section{The catalogues}\label{catalogues}

The outcome of running the pipeline processes described in Section \ref{data} was a raw catalogue of 1\,390\,256\,078 tentative sources. The density of these on the sky is shown in the top panel of Figure \ref{skyplot_density}. Of these, 1\,025\,855\,108 had five-parameter astrometric solutions, and 364\,400\,970 had two-parameter astrometric solutions (average positions). Of the rows with five- and two-parameter solutions, $23\%$ and $12\%$ were flagged as probable duplicates respectively -- $20\%$ of the raw catalogue in total.

\subsection{Rejection of unreliable sources}
\label{sec:junk_rejection}

\begin{figure*}
  \begin{center}
    \includegraphics[width=\textwidth,keepaspectratio]{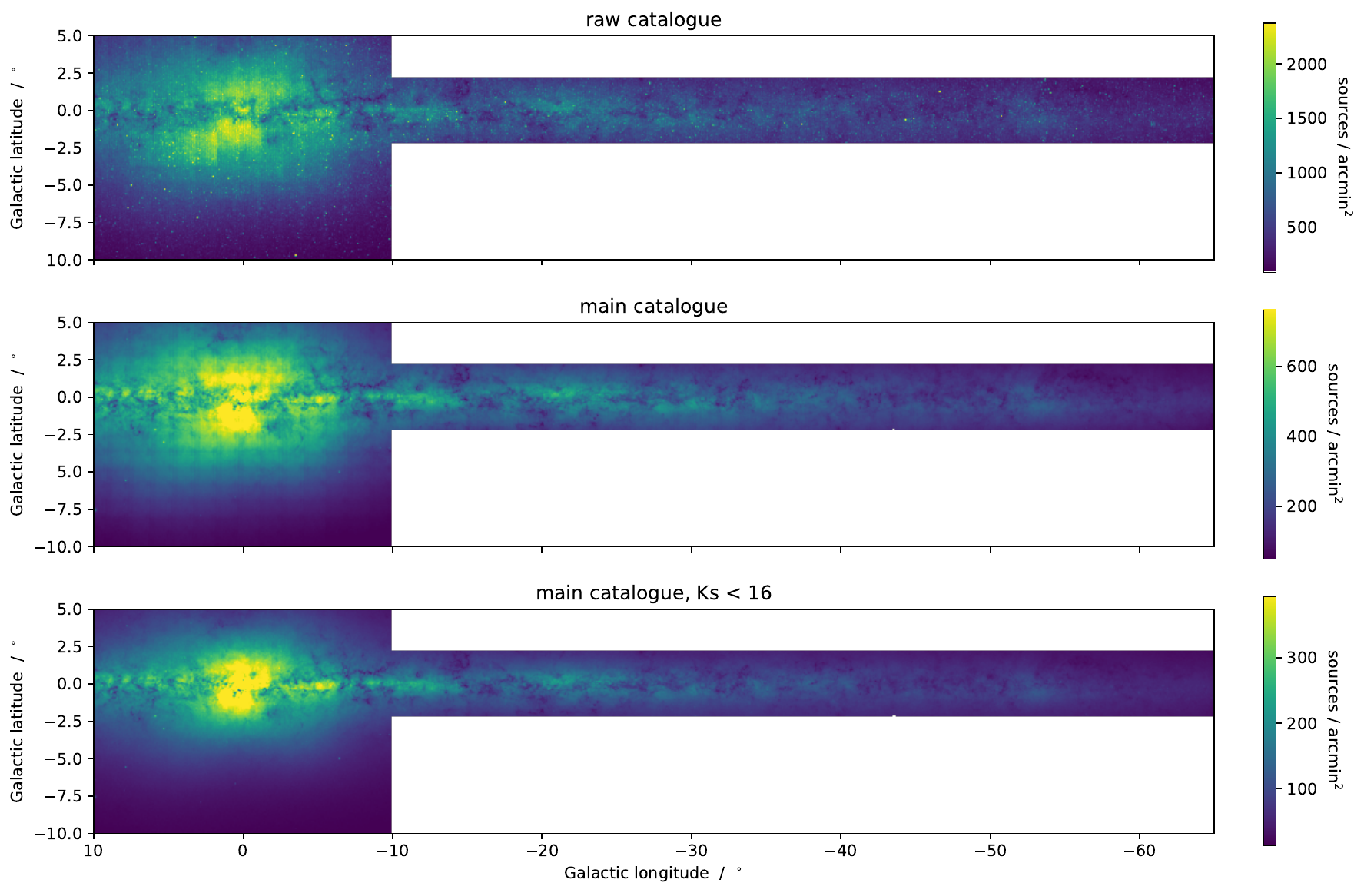}
    \caption{
    \textit{Upper}: Density map of sources in the raw catalogue. Visible are apparent cluster-like high density regions around bright sources caused by false detections in their wings. An additional lower level increased density due to \textsc{DoPhot} erroneously fitting sky noise as sources is also visible, particularly along the boundary of the high observing cadence region ($1.6\lesssim{}l^{\circ}\lesssim{}7.5$, $-3.6\lesssim{}b^{\circ}\lesssim{}-1.5$).
    \textit{Middle}: Density map of sources meeting the selection criteria used to define the main catalogue. The majority of the non-astrophysical inhomogeneity seen in the upper panel has been removed. What remains is largely due to varying detector sensitivity.
    \textit{Lower}: Density map of sources meeting the selection criteria used to define the main catalogue that also have mean $K_s$ band magnitude below 16, where completeness is universally high. This map is essentially free from survey-related density fluctuation.}
    \label{skyplot_density}
  \end{center}
\end{figure*}

The upper panel of Figure \ref{skyplot_density} shows the source density of the raw catalogue and exhibits many survey-related, or other non-astrophysical features which should ideally be removed.
Ground-based NIR observations tend to be relatively noisy compared to optical observations, and the configuration parameters we used for \textsc{DoPhot} caused it to occasionally report detection of sources that were simply image noise. In addition, the wings of bright stars also gave rise to many erroneous detections (see Figure \ref{imcore_vs_dophot}). Requiring multiple matched detections helped enormously with rejecting the erroneous ones, but with hundreds or thousands of coincident observations, false detections matched together by chance fairly frequently. This was particularly common where the false detections cluster together, which happened most frequently in the wings of bright sources even though the VISTA/VIRCAM diffraction pattern is not constant for a given orientation. These were the main sources of contaminants in the raw catalogue.

In addition to rejecting probable duplicates (sources within $0.339$\arcsec{} of another that share most of their detections), we employed two methods of identifying and rejecting probable contaminants from the raw catalogue. The first was a requirement that sources be detected in more than $20\%$ of the observations that cover them. It follows that the more observations there are, the more likely it is that erroneous detections will occur within the matching radius. This was particularly evident in the eight high-cadence tiles (covering $1.6\lesssim{}l^{\circ}\lesssim{}7.5$, $-3.6\lesssim{}b^{\circ}\lesssim{}-1.5$), where there were a few times more observations than was typical for the other bulge fields. Due to overlapping VIRCAM pawprints, some small regions have a little over 2000 observations. A requirement of a $>20\%$ detection fraction, rather than a flat number of detections, acted to reduce and homogenise the contamination rate across the survey. 
The other major discriminator of contaminants was their astrometric goodness-of-fit. The residuals to the astrometric solutions of well behaved sources are adequately characterised by their uncertainties, while this is typically not the case for the contaminants. Non-single stars have an additional component of motion that the 5-parameter astrometric model does not account for, but this is highly unlikely to be significant in our data. The unit weight error (UWE) is defined as:
\begin{equation}
    {\rm UWE} = \sqrt{\frac{\chi{}^2}{N - M}}
\end{equation}
\noindent where $\chi{}^2$ is the chi-squared statistic of the astrometric fit, the sum of the residuals squared divided by their uncertainties squared; $N$ is the number of measurements, in our case one per dimension per observation; and $M$ is the number of parameters solved for, in our case $M = 5$ (we computed goodness-of-fit statistics only for five astrometric parameter solutions). Sources whose residual scatter is well characterised by their uncertainties are expected to have ${\rm UWE} \approx{} 1$. In addition to the above selections we employed a ${\rm UWE} < 1.8$ cut, which rejected a further $3\%$ of the remaining sources.

\begin{figure}
  \begin{center}
    \includegraphics[width=.48\textwidth]{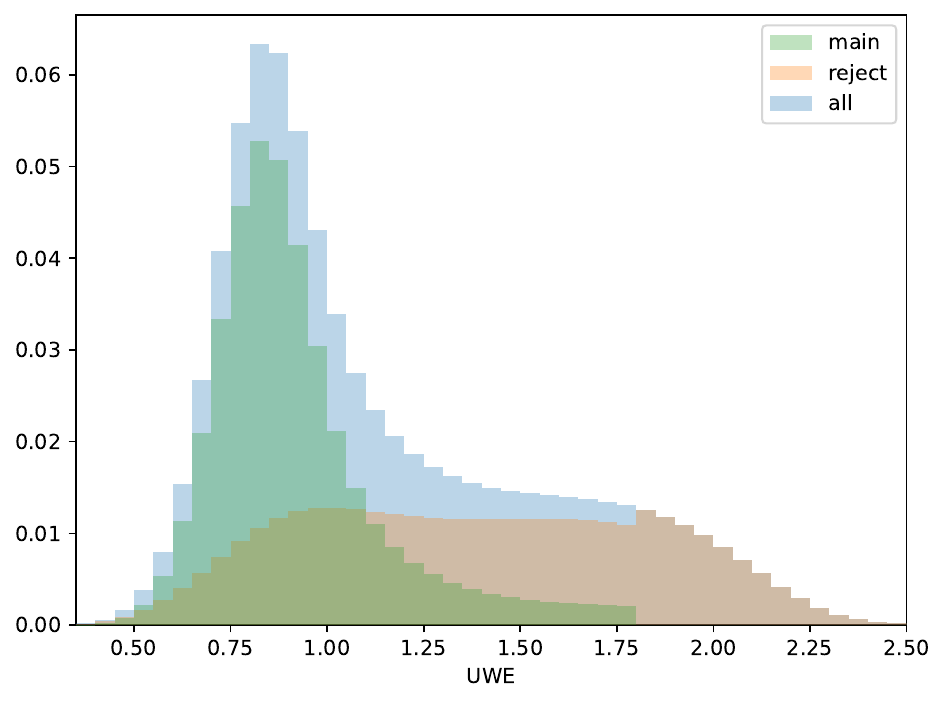}
    \caption{Distribution of the VIRAC2 ${\rm UWE}$ parameter for the raw catalogue (blue shaded), the main subset (green shaded), and the reject subset (orange shaded).}
    \label{fig:UWE}
  \end{center}
\end{figure}

Figure \ref{fig:UWE} shows the distribution of UWE for the raw catalogue and our (eventual, see below) main and reject subsets. The jump in the reject distribution at ${\rm UWE}=1.8$ occurs due to the aforementioned rejection of stars at ${\rm UWE}>1.8$.
The peak of the distribution occurs at ${\rm UWE}\approx{}0.85$, where it should ideally occur at ${\rm UWE}\approx{}1.0$. We speculate that this might be due to an overestimation of the VISTA per-epoch positional uncertainties (see Section \ref{sec:astrocalib}). It is well established the Gaia uncertainties tend to be underestimated (see discussion in Section \ref{sec:gdr3_validation}). These come into play through Equation~\ref{eqn:escale_ast}, as $\sigma{}^2_{\rm Gaia}$, whereby an underestimate would cause $\sigma{}^2_{\rm VISTA}$ to be overestimated. We are ultimately not concerned by the potential overestimation since the uncertainties on the mean astrometric parameters appear to be fairly reliable (see Section \ref{sec:gdr3_validation}).

Since the minimum number of observations covering any region of the survey was $\approx{}50$, and a five astrometric parameter solution is a de-facto $10$ epoch requirement, the $20\%$ detection fraction requirement is a de-facto five astrometric parameter solution requirement. In addition, we did not compute any goodness-of-fit statistics for two parameter astrometric solutions, meaning the UWE discriminator was unavailable for these sources. For these reasons, we elected to explicitly impose the five astrometric parameter solution requirement.

On application of the 5 astrometric parameter, non-duplicate, greater than $20\%$ detection fraction, and ${\rm UWE}<1.8$ selection we were left with a catalogue containing 545\,346\,537 sources, that was largely free from contamination. This is the main component of the VIRAC2 astrometric catalogue. The density map of this selection is shown in the middle panel of Figure \ref{skyplot_density}. We can now begin to pick out by eye some of the larger Galactic globular clusters that were previously hidden among the false detections -- NGCs 6440, 6544, 6553, 6626, 6656, and 6441 are all visible in the bulge on close inspection. The principal remaining source of non-astrophysical inhomogeneity is visible as a grid-like pattern across the survey and is due to varying detector sensitivity. If one wishes to remove this, and obtain a selection with a largely uniform high completeness and relatively low contamination then a further selection of ${\rm K_{s}} < 16~{\rm mag}$ will achieve this. Using artificial source injection, \citet{sanders22} determined that this selection is at least $90\%$ complete everywhere except the few square degrees around the Galactic centre (see their appendix C). This selection is demonstrated in the density map shown in the lower panel of Figure \ref{skyplot_density}, and is seen to be essentially free of non-astrophysical inhomogeneity.

We note that the $20\%$ detection fraction is the main cause of rejection of genuine sources. Stars at the faint end of the survey naturally tend not to be reliably detected, causing them to fail this selection where they might not necessarily fail the others. Transient stars also can suffer from rejection as a result of this criterion. In a survey of the $\Delta{}K_s > 4$~mag stars by \citet{lucas24}, which did not apply this criterion, approximately $10\%$ do not meet the detection fraction threshold. These tended to be novae and other short-timescale transients. The principal aim of this work was the production of a highly reliable astrometric catalogue, and our selection criteria were set to achieve this. However, we recognise that much useful data remains among the sources that we have rejected, and hence we publish the VIRAC2 reject catalogue alongside the main VIRAC2 catalogue.

\subsection{Astrometric precision}

\begin{figure*}
  \begin{center}
    \includegraphics[width=.45\textwidth,keepaspectratio]{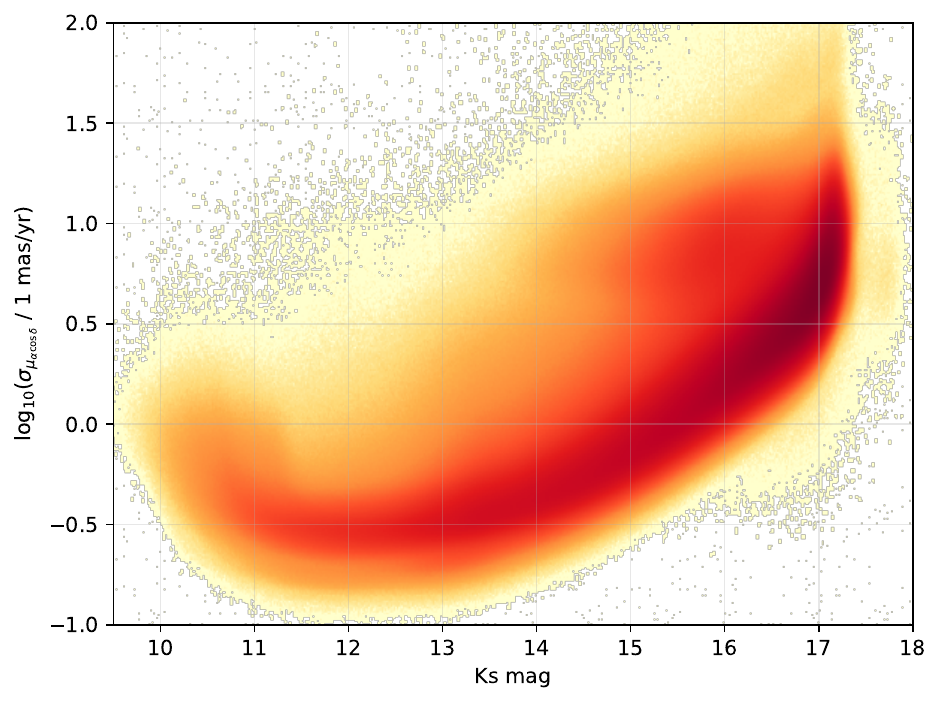}
    \includegraphics[width=.45\textwidth,keepaspectratio]{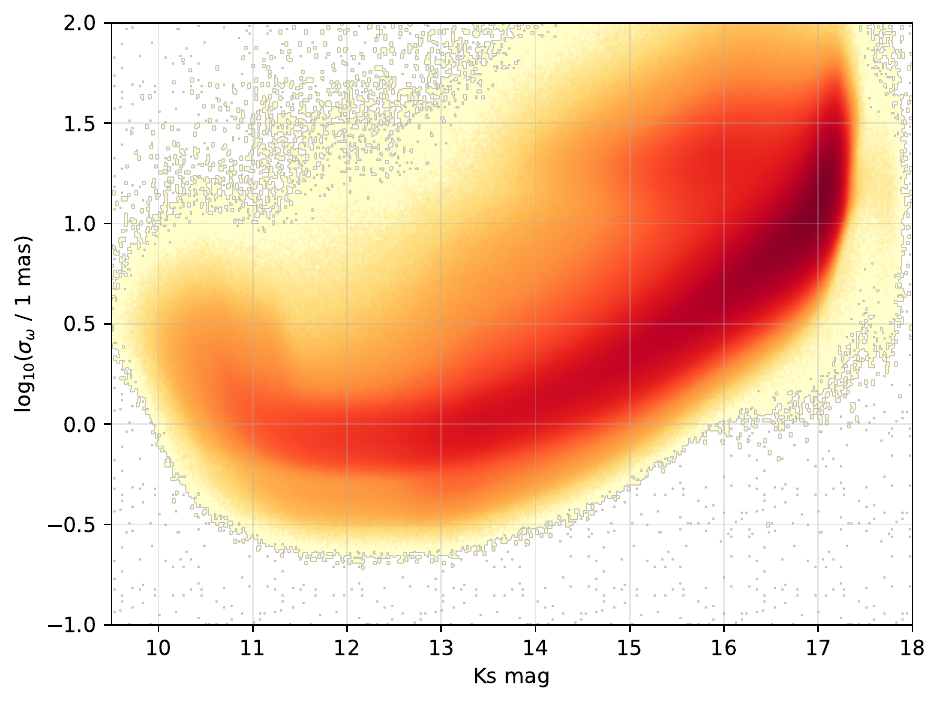}
    \caption{
    \textit{Left}: Uncertainty in $\mu{}_{\alpha{}\cos{}\delta{}}$ versus $K_s$ band mean magnitude for the main catalogue.
    \textit{Right}: Parallax uncertainty versus $K_s$ band mean magnitude for the main catalogue. For both plots the colour axis is logarithmic.}
    \label{mag_ast_errors}
  \end{center}
\end{figure*}

Figure \ref{mag_ast_errors} shows the distribution of the uncertainties on proper motion in right ascension and parallax as a function of $K_s$ band mean magnitude for the main catalogue. Peak astrometric performance is in the $11<K_s~{\rm mag}<14$ range, where proper motion uncertainties are typically better than $0.5~{\rm mas~yr}^{-1}$ per dimension (median $0.36$ and $0.38$ ${\rm mas~yr^{-1}}$ for $\mu{}_{\alpha\cos\delta}$ and $\mu_{\delta}$, respectively) and parallax uncertainties are typically around $1~{\rm mas}$ (median $1.02$ mas). Saturation impacts performance for stars brighter than this. Performance at $K_s = 16$ is typically around $1.5~{\rm mas~yr}^{-1}$ per dimension for proper motion and $5~{\rm mas}$ for parallax.
The faint cloud visible above the main locus in the $15<K_s~{\rm mag}<16.5$ range ($\sigma_{\mu_{\alpha\cos\delta}} > 5~{\rm mas~yr}^{-1}$, $\sigma_{\omega} > 10~{\rm mas}$) is primarily comprised of false sources that still pass our ${\rm UWE}<1.8$ cut. Application of a more strict selection (e.g. ${\rm UWE} < 1.4$) will generally remove them if a more robust sample is required, at the cost of some genuine sources.

\subsubsection{Validation with Gaia DR3}\label{sec:gdr3_validation}

We performed a 0.25\arcsec{} nearest neighbour crossmatch of the VIRAC2 main catalogue to Gaia DR3, finding that 233\,235\,552 VIRAC2 sources ($43\%$) have Gaia counterparts, of which 176\,954\,558 ($32\%$ of VIRAC2 sources) have 5 or 6 parameter mean astrometry in the Gaia catalogue.

\begin{figure}
  \begin{center}
    \includegraphics[width=.45\textwidth,keepaspectratio]{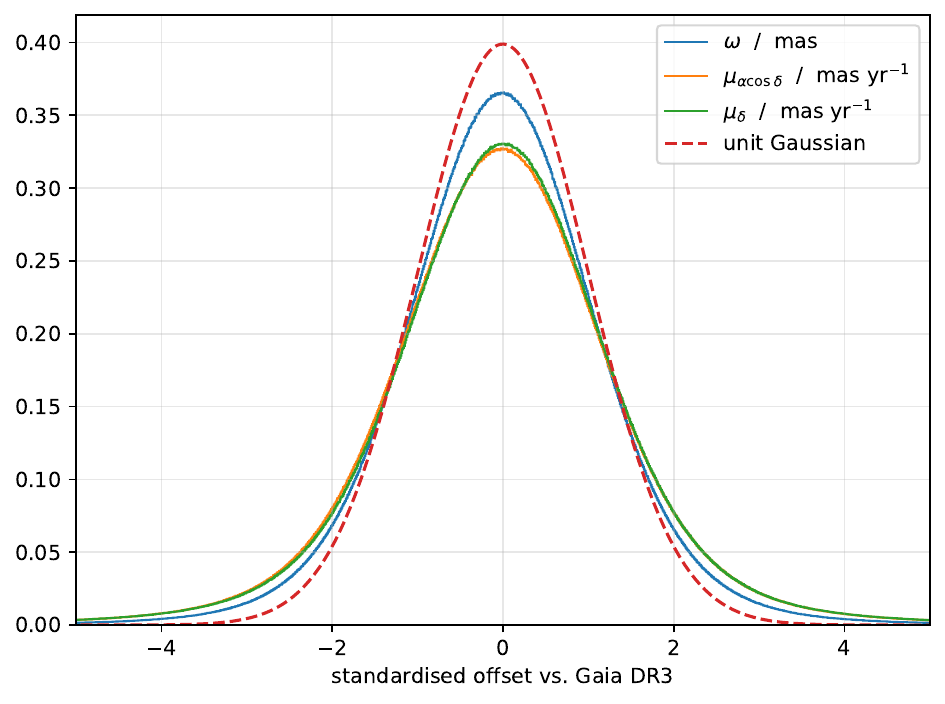}
    \caption{Standardised offsets between VIRAC2 and Gaia DR3 mean parallax and proper motion components. A unit Gaussian probability distribution function is shown for comparison. Gaussian distributions with $\sigma{} = 1.1$ and $\sigma{} = 1.2$ match closely the parallax and proper motion histograms, respectively. This implies that the uncertainties on the offsets are underestimated by approximately $10\%$ for parallax, and $20\%$ for proper motion, but caution should be used when interpreting this figure. See text for a more thorough analysis.}
    \label{gaia_comparison}
  \end{center}
\end{figure}

This sample will be broadly similar to the selection of reference stars used for the initial astrometric calibration (see Section \ref{sec:astrocalib}). For these, we should expect the VIRAC2 mean astrometry to match fairly closely that of Gaia DR3. Figure \ref{gaia_comparison} shows the 1D histograms of standardised offsets (i.e. offset divided by its error) in parallax and the two components of proper motion for this sample. If the uncertainties on the offsets are truly $1\sigma{}$ Gaussian, then we should find that the histograms match the probability distribution function of a unit Gaussian. In fact we find that the parallax distribution resembles a Gaussian distribution with $\sigma{}=1.1$, and the distributions of the two components of proper motion resemble a Gaussian distribution with $\sigma{}=1.2$. This modest difference implies that the uncertainty on the offsets is underestimated by approximately $10$ and $20$ per cent in parallax and proper motion, respectively.

The simplest explanation for an underestimate in the uncertainty on the offsets is that the uncertainties on the parameters themselves are underestimated. Through experimentation we found that in order for each distribution to resemble a unit Gaussian we had to either inflate Gaia uncertainties of all three parameters by a factor of approximately 2.25; or inflate VIRAC2 proper motion uncertainties by a factor of approximately 1.25 and parallax uncertainties by a factor of approximately 1.1. In practice, scaling factors should be applied to the per-epoch astrometric uncertainties and propagated through to the mean astrometry parameters. There is evidence presented in Section \ref{sec:junk_rejection} that potentially supports the conclusion that the VISTA per-epoch uncertainties are mild overestimates, namely that the unit weight error distribution peaks slightly below $1.0$. Further testing might enable us to apportion blame, though these data are not yet public for Gaia. \citet{elbadry21} and \citet{sanders23} find that Gaia DR3 parallax uncertainties are underestimated by up to a factor of 2, particularly for red sources (which are the dominant component of our sample) and those with ${\rm RUWE}>1.4$ (see their section 5). Refinement of the sample selection to have both Gaia DR3 ${\rm RUWE}<1.4$ and VIRAC2 ${\rm UWE}<1.4$ yields only a modest improvement, to around $\sigma=1.05$ and $\sigma=1.15$ for parallax and proper motion, respectively. This indicates that underestimated uncertainties, if present, are a more general problem not restricted only to the subset of poorly measured sources.

A common practice in least squares fitting is to scale the covariance matrix such that the reduced chi-squared equals unity. This is typically necessary when it is known that the uncertainties on the data points are unreliable, as was the case for VIRAC2$\beta{}$ (see Appendix \ref{app:virac2b}). For the main pipeline run we considered the centroid errors to be significantly more reliable, and hence the rescaling to be unnecessary. The peak of the distribution of the unit weight error being reasonably near unity indicates that there was indeed limited need. However, should users wish to apply this correction then they should simply multiply the individual parameter uncertainties by the unit weight error. We note that as the posterior probability density functions are relatively simple in this case, rescaling the parameter errors in this way approximates parameter uncertainties produced by Markov chain Monte Carlo methods that incorporate an error scaling factor.

Another explanation could be mismatches between VIRAC2 and Gaia DR3 sources. Crossmatches between catalogues with large bandwidth and resolution differences are never perfect. In principle we could significantly reduce the 0.25\arcsec{} matching radius used. Typically VIRAC2 positional uncertainties are at the level of a few milliarcsec, and Gaia DR3 is similar for the faintest sources and much better for brighter ones. However, in doing so we would bias the selection towards only those with agreement in mean astrometry, thereby invalidating this comparison.

We must also consider that this sample is only broadly similar to that used for the initial astrometric calibration. We did not impose any goodness-of-fit statistic requirements on the Gaia sources this time, nor was every VISTA detection that was used for astrometric calibration also used for mean astrometric parameter fitting. Technically, we were comparing mean astrometric parameters at epoch 2014.0 for VIRAC2 and epoch 2016.0 for Gaia DR3, though this should have a negligible impact in practice.

It is important to note that a modest underestimate in the errors on the offsets is unlikely to have only a single cause. Additionally, there are reasons to suspect that the offsets might be biased towards larger values. Given this, and that the apparent underestimates are relatively minor, we consider that the VIRAC2 catalogue mean astrometry and their uncertainties are valid.

\subsubsection{External validation with HST}

\citet{luna23} compared Hubble Space Telescope (HST) proper motions from three crowded fields in the Galactic bulge to those of VIRAC2. They measured the multiplicative factor by which VIRAC2 proper motion uncertainties must be inflated to account for the scatter observed in their offsets from the HST measurements as a function of $J$ band magnitude.
They found that a multiplicative factor of between $1.0$ and $1.5$, the larger value being necessary at brighter $J$ band magnitudes and in the densest field tested, but a value near unity was broadly correct. Ultimately they concluded that no inflation factor was necessary, potentially further indication that it is the Gaia uncertainties that are underestimates.

\subsubsection{Known failure modes}\label{failuremodes}

In the special case of large amplitude variable sources in crowded fields, the astrometric solution could become unreliable due to systematic changes in the location of the centroid found by \textsc{DoPhot}. This could occur if stars adjacent to the variable star were no longer detected when the latter became much brighter. This issue was noted by \citet{lucas24}, in the context of highly variable giant stars in the Nuclear Disc of the Milky Way. Some additional failure modes that were identified during nearby object searches are detailed in Appendix \ref{sec:app_failures}.

Population studies can be performed with relative confidence, but we caution the reader that attempts to select outliers (e.g. high proper motion or parallax sources) will tend to also preferentially select the erroneous examples. Care must be exercised in cleaning such samples.

\subsection{Time series data}

Photometric time series are supplied alongside their aggregated stats. Astrometric time series are also supplied, as both calibrated equatorial and array coordinates. The schema of the time series table is provided in Table \ref{tab:timeseries_schema}. 

We have found that erroneous photometric outliers are typically evident through inspection of the \textit{chi} and \textit{ast\_res\_chisq} statistics, and whether or not the detection is flagged as an ambiguous match. Where \textit{chi} was output directly from \textsc{dophot}, \textit{ast\_res\_chisq} is the $\chi^2$ of the astrometric residual (with 2 degrees of freedom), and the ambiguous match flag was set to `$1$' if that detection also appears in the time series of another source (e.g. through blending).

\begin{figure*}
  \begin{center}
    \includegraphics[width=\textwidth,keepaspectratio]{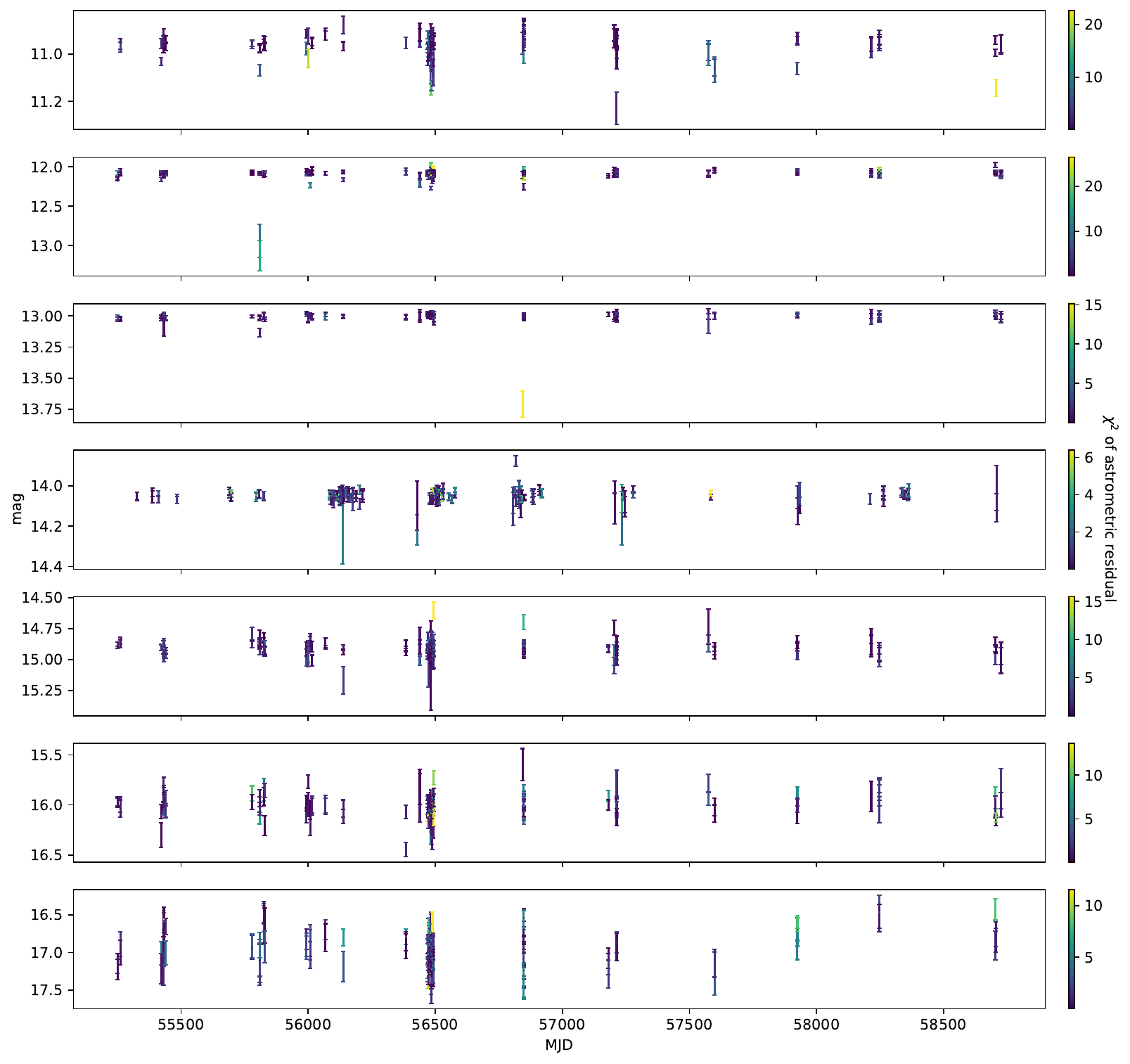}
    \caption{Example $K_s$ band light curves at each integer magnitudes 11 through 17 in the $K_s$ bandpass. Examples were selected from sources with $>100$ $K_s$ band epochs, within $0.1$~mag of the target magnitude, and with average $K_s$ mag uncertainty within 5~mmag of what is typical for that magnitude. The colour axis shows the $\chi^2$ (with 2 degrees of freedom) of the astrometric residual for each data point. It can be seen that outliers in magnitude space are often also outliers in the astrometry, making this a useful method for cleaning the photometric time series.}
    \label{fig:example_lcs}
  \end{center}
\end{figure*}

A set of example $K_s$ band photometric times series for a range of brightnesses is presented in Figure \ref{fig:example_lcs}. The sources shown were selected from a pool of those with typical photometric uncertainties for their average magnitude. From this plot it is evident that \textit{ast\_res\_chisq} can be a useful statistic for rejection of erroneous data points.

\subsection{Colour-magnitude diagram improvement}

\begin{figure}
  \begin{center}
    \includegraphics[width=0.45\textwidth,keepaspectratio]{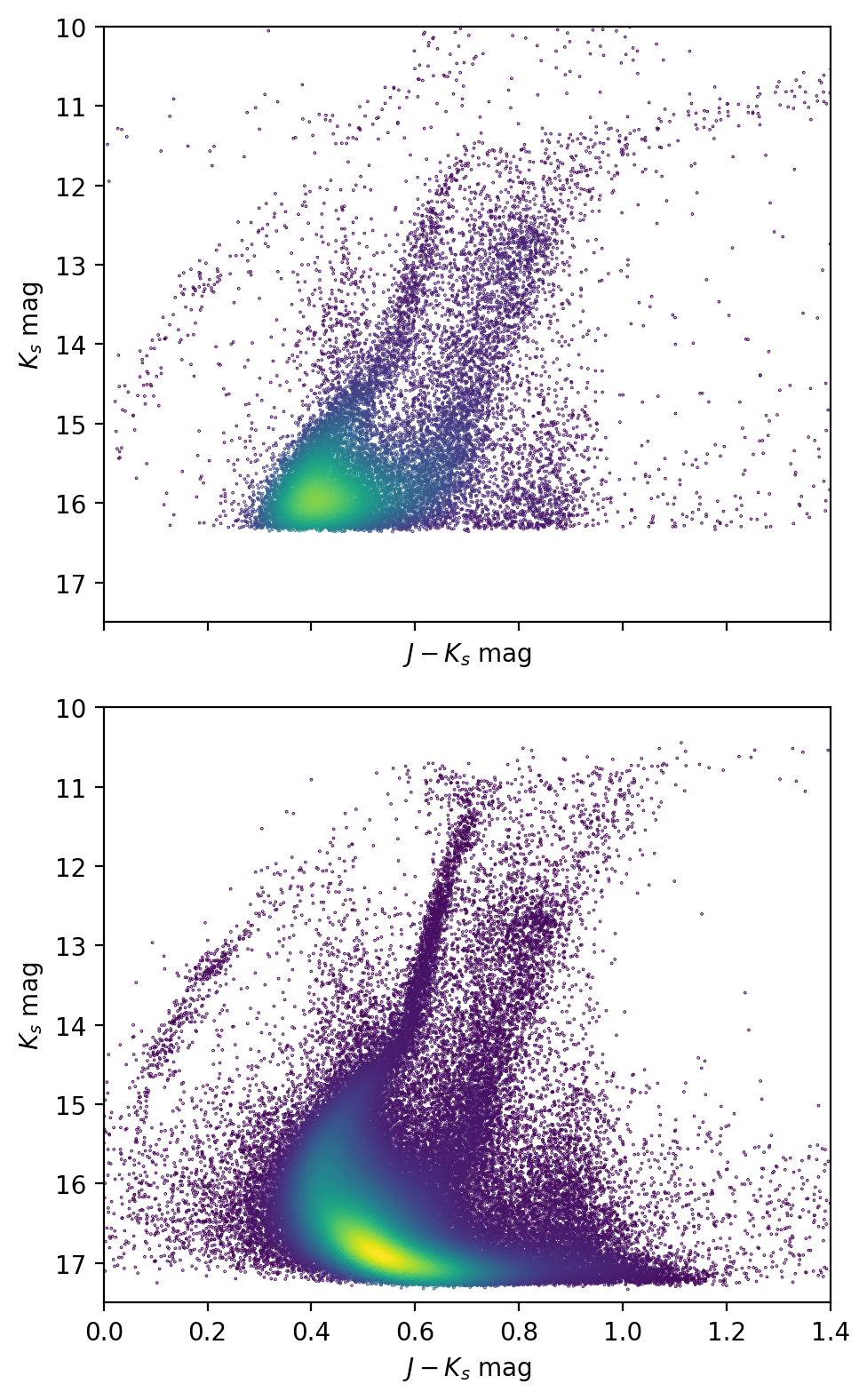}
    \caption{The $J-K_s$ vs. $K_s$ colour-magnitude diagram for the Galactic globular cluster NGC 6656 from VIRAC v1 (upper panel) and VIRAC2 (lower panel). The colour axis represents the density of points within the diagram (blue are isolated, yellow are highly grouped).}
    \label{fig:M22_cmds}
  \end{center}
\end{figure}

To demonstrate the improvements in the photometry and depth of VIRAC2 versus VIRAC v1, we provide color-magnitude diagrams from both catalogues for the Galactic globular cluster Messier 22 (NGC 6656), see Figure \ref{fig:M22_cmds}. All selected stars are within 0.2 degrees of the cluster centre. For VIRAC2 we included only stars in the main table (i.e. we did not include the \textit{reject} table), and for VIRAC v1 we only included stars flagged as reliable. The VIRAC v1 selection comprises 17\,075 sources, while the VIRAC2 selection comprises 100\,657 sources. The extra $\approx{}1$~mag of depth is evident, as is a general tightening up of the various stellar sequences. We note that while VIRAC v1 tends to be better for saturated stars, these are often flagged as `unreliable' and hence were not included in Figure \ref{fig:M22_cmds}.

We direct those interested in CMDs in these regions to \citet{AG18}, whose catalogue is available from the VISTA Science Archive. The \citet{AG18} catalogue required detections in three bandpasses, where VIRAC2 relies on astrometric goodness of fit and detection fraction to discern reliable sources from false. This means that sources which are non-detections in one or two bands, and hence are missing from the \citet{AG18} catalogue, may be present in VIRAC2.

\subsection{Catalogue description and access}

The complete catalogues are available from the ESO archive at \url{https://archive.eso.org}. The table identifiers are:
\begin{enumerate}
    \item VVVX\_VIRAC\_V2\_SOURCES,
    \item VVVX\_VIRAC\_V2\_REJECTED\_SOURCES,
    \item VVVX\_VIRAC\_V2\_LC,
    \item VVVX\_VIRAC\_V2\_REJECTED\_LC, and
    \item VVVX\_VIRAC\_V2\_OBS.
\end{enumerate}
Where tables (i) and (ii) are the aggregate source data (e.g. positions, proper motions, mean photometry), tables (iii) and (iv) are the time series data, and table (v) contains observation related information (e.g. seeing, airmass). Tables (i) and (iii) are the main selection, and tables (ii) and (iv) are the reject selection.

Section \ref{sec:junk_rejection} describes the makeup of the main and reject selections. Schemata and example rows for the source and time series tables are given in Appendix \ref{app:schema}.

The catalogues can be efficiently queried using the Table Access Protocol (TAP) via the ESO programmatic access tool\footnote{\href{http://archive.eso.org/programmatic/\#TAP}{archive.eso.org/programmatic}}, or e.g. TOPCAT\footnote{\href{https://www.star.bris.ac.uk/~mbt/topcat}{star.bris.ac.uk/~mbt/topcat}} \citep{topcat}. More information, including example queries, can be found in the document accompanying the ESO release.


\section{Example applications}\label{sec:eg_applications}

\subsection{New stars and ultracool dwarfs in the solar neighbourhood}
\label{sec:100pc}

VIRAC2 parallaxes and proper motions can be used to discover new nearby stars and brown dwarfs in a similar manner to VIRAC v1 \citep{LCS18}. This complements the Gaia Catalogue of Nearby Stars \citep[GCNS,][]{smart21, reyle21}
and the earlier Gaia-based ultracool dwarf search of \citet{reyle18} 
by providing slightly better sensitivity to L and T dwarfs, which are optically very faint. While most VIRAC2-selected nearby sources are already catalogued in GCNS or earlier studies \citep[see e.g.][]{mejias22}, we see below that the majority of VIRAC2 ultracool dwarf candidates with visually confirmed high proper motions are new discoveries.

We searched for new ultracool dwarfs and any other nearby sources that might have been missed by Gaia using the following initial selection in the VIRAC2 main source table:

\begin{enumerate}
    \item $\varpi > 10$~mas
    \item $\varpi/\sigma_{\varpi} \ge 5$
    \item $\itl{ks\_n\_det} > 0.5 \times \itl{ks\_n\_obs}$
\end{enumerate}

\noindent where $\varpi$ is the parallax and {\it ks\_n\_det} and {\it ks\_n\_obs} are the VIRAC2 parameters corresponding to the number of $K_s$ detections of a source and the approximate number of $K_s$ observations of the source's position, respectively.

Applying these four selection criteria returned 1\,576\,427 sources from the database, the great majority of which clearly had incorrect parallaxes. For example, the GCNS contains only 331\,312 sources at distances $d<100$~pc across the whole sky in the Gaia 3rd Data Release, so VIRAC2 should find far fewer in a survey of $\approx{}1.4$ per cent of the sky. 
Distinguishing bona fide nearby VIRAC2 sources from false positives is a difficult task, perhaps best suited to machine learning methods at this scale \citep{smart21}. Here we focus mainly on ultracool dwarfs (UCDs), rather than attempting to recover all nearby stars, but colour blind searches for very high proper motion stars and stars within 50~pc of the sun were also attempted. To note, the candidate nearby stars selected above are widely distributed across the survey area but there is some degree of clustering in the fields with the largest number of observations, e.g. the Galactic centre tile and a group of eight contiguous VVV bulge tiles at $1.6\lesssim{}l^{\circ}\lesssim{}7.5$, $-3.6\lesssim{}b^{\circ}\lesssim{}-1.5$. These fields all have a high source density.

\subsubsection{High proper motion search}
\label{sec:HPM}

From the initial selection above, we selected sources with
proper motion, $\mu > 500$~\masyr{}, where $\mu$ is 
computed from the sum in quadrature of the two components
of the proper motion. This yielded 20\,889 candidates, most of which have a significant fraction of ambiguous detections in the time series. In fact, the spatial distribution of the 20\,889 candidates shows a very strong clustering in the area of the eight contiguous high cadence tiles mentioned above, almost all of which are sources for which a large fraction of the detections are listed as ambiguous. Specifically, $f_{amb} > 0.9$, where $f_{amb}$ is the ratio of VIRAC2 parameters $\itl{ks\_n\_amb/ks\_n\_det}$ and
{\it ks\_n\_amb} is the number of detections with an ambiguous match to more than one VIRAC2 source, i.e. it is the fraction of ambiguous detections.

We applied the following three cuts to reduce the 20\,889 candidates to a small number for visible inspection:

\begin{enumerate}
    \item $f_{amb} < 0.4$
    \item $\Delta t >0.5$~yr
    \item $\itl{ks\_n\_det > 20}$
\end{enumerate}

\noindent where $\Delta t$ is the time span defined
by the VIRAC2 parameters {\it ks\_first\_epoch} and {\it ks\_last\_epoch}, the first and last modified Julian dates of the $K_s$ detections in the photometric time series set (see Section \ref{photstats}).
Bona fide high proper motion stars should pass these cuts since they would typically be expected to have only a small proportion of ambiguous detections, e.g. due to blending as the rapidly moving source passes close to another source in the field. They would also be expected to be detected over a time baseline of several years and to have a large number of detections. These cuts yielded a list of 21 candidates, 18 of which were confirmed as genuine high proper motion stars by visual inspection of a pair of images taken several years apart. The three false candidates were due in two cases to blended pairs of stars. In the third case, all the detections in the VIRAC2 time series were in 2010, save for a single detection in 2018 that appears likely to be noise, from inspection of the image.

The distribution of $f_{amb}$ for candidates satisfying all the cuts except the one on $f_{amb}$ is bimodal, with a very large number of false positives found near $f_{amb}=1$ and genuine high proper motion stars making up a smaller peak near $f_{amb}=0$. The cut at $f_{amb}=0.4$ was chosen to exclude the larger peak of false positive sources. Since only two genuine sources were found with $0.2 < f_{amb} < 0.4$, we expect that relatively few sources
with larger values of this parameter have been missed. However, this search did miss the T5 brown dwarf VVV~J165507.13-421755.8 \citep{schapera22}, for which $f_{amb}=0.78$, $\mu \approx 705$~\masyr{} and $\varpi \approx 66$~mas ($d \approx 15$~pc). Its motion was detected via a machine learning analysis \citep{caselden20} of the unWISE coadds \citep{meisner18} before then turning to VVV and VIRAC2 for clearer images and astrometry. Due to its large proper motion the VIRAC2 time baseline is
short, $\Delta t=3.9$~yr, because the source is blended with more than one star over the course of the VVV time series. This highlights the value of combining independent data sets for more complete searches.

The 18 genuine high proper motion sources include one new discovery, VVV~J181453.17-265453.6, and 17 stars that are already in the GCNS and earlier lists of higher proper motion sources \citep{terzan80,lepine05,lepine08,beamin13,ivanov13,luhman14a,luhman14b,schneider16,LCS18,reyle18,kluter18,gentile-fusillo21}. 
Data for the new source, hereafter VVV~J1814-2654, are given in Table \ref{tab:earlyT}. 

\begin{table}
    \centering
    \begin{tabular}{r|l}
        VIRAC2 source ID & 13333546009625 \\
        Name & VVV~J1814-2654 \\
        RA (hms) & 18 14 53.17 \\
        Dec (dms) & -26 54 53.6 \\
        RA ($^\circ$) & 273.721554 \\
        Dec ($^\circ$) & -26.914877 \\
        $l$ ($^\circ$) & 4.965352 \\ 
        $b$ ($^\circ$) & -4.610297 \\
        \pmra~(\masyr) & -235.1 $\pm$ 2.1 \\
        \pmdec~(\masyr )& -446.7 $\pm$ 2.2 \\
        $\varpi$~(mas) & 36.4 $\pm$ 4.2 \\
        $d$ (pc) & 27.1$^{+3.6}_{-2.9}$ \\
        $Z$ & $>$19.5 \\
        $Y$ & 18.49 $\pm$ 0.20 \\
        $J$ & 17.28 $\pm$ 0.12 \\
        $H$ & 16.35 $\pm$ 0.09 \\
        $K_s$ & 15.80 $\pm$ 0.10 \\
        $M_{K_s}$ & 13.60$^{+0.26}_{-0.28}$ \\
        $Z-J$ & $>$2.2 \\
        $Y-J$ & 1.21 $\pm$ 0.23\\
        $J-H$ & 0.93 $\pm$ 0.15\\
        $H-K_s$ & 0.55 $\pm$ 0.13\\
        $J-K_s$ & 1.48 $\pm$ 0.16\\
        $z-J$ & 2.54 $\pm$ 0.28
    \end{tabular}
    \caption{Data for the brown dwarf VVV~J1814-2654, discovered by the high proper motion search. The VIRAC2 J2000 coordinates are at epoch 2014.0. The absolute magnitude, $M_{K_s}$, is the median value and its error bars represent the 68\% confidence interval, after sampling over Gaussian distributions in $\varpi$ and $K_s$ with the quoted parameters to produce a probability distribution. The quoted $K_s$ magnitude is the median value, given as the $ks\_p50$ parameter in Table \ref{tab:sources_schema}.}
    \label{tab:earlyT}
\end{table}

Nearby stars have negligible interstellar reddening so the non-detection in the VISTA $Z$ passband, along with red $Y-J$, $J-H$ and $H-K_s$ colours, indicate that this is an ultracool dwarf with a spectral type in the range L0 to T2, see e.g. \citet{kirkpatrick21}. The absolute $K_s$ magnitude, $M_{K_s}$ suggests a spectral type between T1 and T3, according to the data in \citet{dupuy12}. VVV~J1814-2654 is therefore an early T dwarf candidate, though there is sufficient scatter in the $M_{K_s}$ vs. spectral type relation that a late L dwarf type is also quite possible. A cross-match to the Dark Energy Camera Plane Survey (DECaPS, \citealt{saydjari23, schlafly18}, see section \ref{sec:50_pc_search}) provides an optical $z$ band detection and we note
that the colour $z-J=2.54 \pm 0.28$ is relatively blue for a T dwarf (see section \ref{sec:Ldwfs}). However, the uncertainty is large and the data are consistent with a source near the L/T transition. We can therefore be confident that VVV~J1814-2654 is a brown dwarf.
The system is projected in the direction of the inner Galactic bulge, within 5$^\circ$ of the Galactic centre in both longitude and latitude (see Table \ref{tab:earlyT}). Only two confirmed brown dwarfs: VVV~BD001 and LTT~7251B \citep{beamin13, LCS18} have previously been found in the inner bulge, where the high stellar density (especially in the infrared) makes searches for high proper motion stars difficult. Given the current community effort to achieve a complete census of nearby stars and brown dwarfs \citep[see e.g.][]{smart21, kirkpatrick24}, the discovery of VVV~J1814-2654 at $d\approx27$~pc, is a helpful addition. 
Such systems also make good targets for adaptive optics imaging to search for companions, given the abundance of suitably bright reference stars.

A high proper motion source was detected at the same location in VIRAC v1 \citep{LCS18} but the proper motion was smaller and quite different (\pmra = -83.6 \masyr, \pmdec = -171.1 \masyr) and no parallax value was reported, owing to the 5$\sigma$ parallax threshold adopted in that work. Here we have visually
confirmed the motion, which was not done in \citet{LCS18} for this source.
The erroneous motion in VIRAC v1 was probably due to the effect of blending
with an adjacent star of similar brightness and the use of aperture photometry. There is no detection in the Galactic surveys by the {\it Spitzer} Space Telescope \citep{werner04}, see further discussion in section \ref{sec:50_pc_search}.

\subsubsection{T dwarf search}
\label{sec:Ts}


T dwarfs with spectral types later than T0 typically have fainter $M_{K_s}$ values and bluer ($J-K_s$) and ($J-H$) colours than brown dwarfs of earlier types. To search for T dwarfs, we began with the $\sim$1.6 million candidate nearby VIRAC2 sources with 5$\sigma$ parallax detections selected in section \ref{sec:100pc}. 

We then applied the following initial cuts: 

\begin{enumerate}

\item $M_{K_s} > 12$
\item $Y-J>0.8$
\item $J-{K_s}<0.9$
\item $Z-J>2$ or no $Z$ detection.
\item $\mu > 30$~\masyr{}
\item $\Delta t >3$~yr
\item $f_{amb}$ < 0.4
\item $\itl{ks\_n\_det > 20}$
\end{enumerate}

\begin{table}
    \centering
    \begin{tabular}{r|l|l}
        VIRAC2 source ID & 13415483005492 & 16122933004133 \\
        Name & VVV~J1820-2742 & VVV~J1253-6339 \\
        RA (hms) & 18 20 46.14 & 12 53 38.22 \\
        Dec (dms) & -27 42 39.0 & -63 39 47.0 \\
        RA ($^\circ$) & 275.192250 & 193.409232 \\
        Dec ($^\circ$) &  -27.710837 & -63.663048 \\
        $l$ ($^\circ$) & 4.8711048 & 303.175843 \\ 
        $b$ ($^\circ$) & -6.137462 & -0.792342 \\
        \pmra~(\masyr) & 10.0 $\pm$ 1.9 & -190.7 $\pm$ 2.0 \\
        \pmdec~(\masyr) & -175.9 $\pm$ 1.9 & -255.9 $\pm$ 2.1 \\
        $\varpi$~(mas) & 36.8 $\pm$ 4.3 & 66.6 $\pm$ 6.7 \\
        $d$ (pc) & 27.1$^{+3.6}_{-2.9}$ & 15.0$^{+1.7}_{-1.4}$ \\
        $Z$ & $>$19.5 & 19.77 $\pm$ 0.19 \\
        $Y$ & 17.81 $\pm$ 0.07 & 16.89 $\pm$ 0.01 \\
        $J$ & 16.77 $\pm$ 0.05 & 15.88 $\pm$ 0.01 \\
        $H$ & 16.32 $\pm$ 0.05 & 16.10 $\pm$ 0.01\\
        $K_s$ & 16.18 $\pm$ 0.10 & 16.29 $\pm$ 0.11 \\
        $M_{K_s}$ & 14.01$^{+0.26}_{-0.29}$ & 15.40$^{+0.24}_{-0.25}$\\
        $Z-J$ & $>$2.7 & 3.89 $\pm$ 0.19  \\
        $Y-J$ & 1.04 $\pm$ 0.09 & 1.01 $\pm$ 0.01  \\
        $J-H$ & 0.45 $\pm$ 0.07 & -0.22 $\pm$ 0.01   \\
        $H-K_s$ & 0.14 $\pm$ 0.11 & -0.19 $\pm$ 0.11   \\
        $J-K_s$ & 0.59 $\pm$ 0.11 & -0.41 $\pm$ 0.01    \\
        $K_s - \mathrm{[3.6]}$& - & 1.63 $\pm$ 0.15 \\  
        $\mathrm{[3.6] - [4.5]}$ & - & 0.87 $\pm$ 0.16 
    \end{tabular}
    \caption{Data for the T dwarfs VVV~1820-2742 and VVV~1253-6339, discovered by the T dwarf search. The VIRAC2 J2000 coordinates are at epoch 2014.0 and the $M_{K_s}$ values are 
    computed as in Table \ref{tab:earlyT}.}
    \label{tab:mid-lateT}
\end{table}

The absolute magnitude and colour cuts were defined mainly from inspection of the data in \citet{dupuy12} and \citet{kirkpatrick21} respectively. The L/T transition actually occurs at $M_{K_s} \approx 13$ rather than $M_{K_s}=12$ but we adopted a 1 mag brighter threshold in order to include equal mass binary systems and allow for the significant scatter that exists in the spectral type vs. absolute magnitude relation. The $Z-J$ constraint was also relaxed since the colour vs. type relation is not well defined in the literature for the VISTA $Z$ filter but we can be confident that T dwarfs will comfortably pass the $Z-J>2$ threshold in the Vega system, \citep[see e.g.][for a very similar $Z$ filter]{hewett06}. The proper motion cut helps to remove false positives amongst the 1.6 million candidate nearby VIRAC2 sources that are actually distant stars. After applying these cuts, there were 750 candidates in the database.\footnote{N.B. We do not include a cut on ($J$-$H$) or ($H$-$K_s$), partly for simplicity but also to remove the requirement for an $H$ magnitude in the VIRAC2 source table. Sources can lack flux measurements in the source table in one or more of the $Z$, $Y$, $J$, $H$ passbands, despite having detections in the time series in the missing band(s), if all the detections in that passband are marked as ambiguous matches, i.e. potentially matching to more than one source. An example of this is the second T dwarf listed in Table \ref{tab:mid-lateT}, for which no $H$ band magnitude is listed in the source table but there are 3 detections in the time series, all flagged as ambiguous matches.}

Two additional parameters can be used to identify the best candidates for visual inspection: (i) the unit weight error (${\rm UWE}$) of the 5 parameter astrometric fit and (ii) $V_{\rm tan}$, the tangential velocity of the source in the plane of the sky, which we computed simply as $V_{\rm tan} = 4.74 \mu/\varpi$. The {\rm UWE} of valid solutions should typically be near unity, but the distribution for the 750 initial candidates was found to rise steeply at ${\rm UWE} > 1$. Further investigation using a cross match to GCNS (see \ref{sec:50_pc_search}) found that
bona fide nearby stars typically have $0.7<UWE<1.1$ in VIRAC2, though there is a tail extending to higher values. For example, VVV~J165507.13-421755.8, the T5 dwarf that was missed in section \ref{sec:HPM}, has ${\rm UWE}=1.48$. Compared to bona fide nearby sources, candidates lacking a GCNS match have a distribution with a larger mode in ${\rm UWE}$, corresponding to unreliable astrometric solutions in most cases.
 
The $V_{\rm tan}$ parameter can be used to complement the cut on proper motion (item (v) above) by removing a large number of sources with significant proper motions but over-estimated parallaxes, these typically having unusually small values of  $V_{\rm tan}$ (a few km~s$^{-1}$). Some genuine nearby stars will have unusually small tangential velocities and another small proportion will have large UWE values. We therefore used two complementary selections to capture such cases.

\begin{description}
\item Selection 1: ${\rm UWE}<1.2$ 
\item
\item Selection 2: ${\rm UWE}<1.5$ and $V_{\rm tan}>10$~km s$^{-1}$
\end{description}

\begin{figure}
  \begin{center}
    \includegraphics[width=.48\textwidth,keepaspectratio]{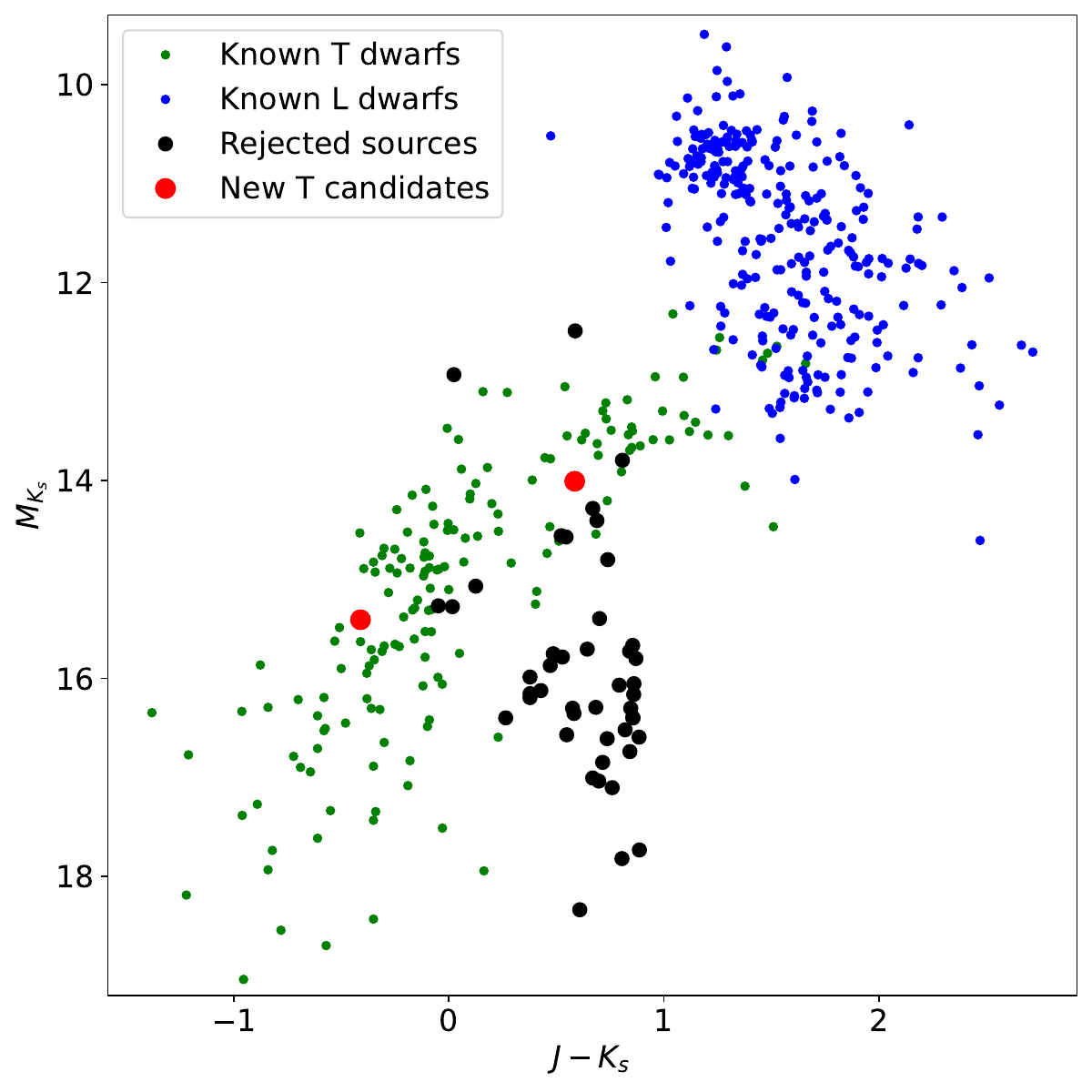}
    \caption{Colour vs. Absolute Magnitude diagram for the  sources identified in the T dwarf search. Known L dwarfs (blue points) and T dwarfs (green points) from the UltracoolSheet are over-plotted, with unresolved binaries excluded. The two new discoveries with visually confirmed proper motion (large red points) lie in the same region as known T dwarfs (green points). Candidates that failed the visual inspection (black points) mostly lie below this region.}
    \label{fig:CMD_Ts}
  \end{center}
\end{figure}

Selection 1 provided 36 candidates and Selection 2 provided 13 candidates. After visually inspecting pairs of cut-out images taken several years apart, only two candidates
showed clear proper motions and these turned out to be 
the only two located in the ``sweet spot'' where both selections are satisfied. The two mid-late T dwarf candidates are VVV~J182046.14-274239.0, hereafter VVV~1820-2742, and VVV~J125338.22-633947.0, 
hereafter VVV~1253-6339. In Figure \ref{fig:CMD_Ts} we plot the two in a colour vs. absolute magnitude diagram (large red circles). For comparison, we overplot L dwarfs (blue points) and T dwarfs (green points) from the UltracoolSheet\footnote{A regularly updated compilation of data for known brown dwarfs, \url{https://doi.org/10.5281/zenodo.10573247}.},
along with the candidates rejected by visual inspection (black points). Data for the two T dwarfs are given in Table \ref{tab:mid-lateT}. For VVV~1253-6339 we include colours that draw on {\it Spitzer} mid-infrared photometry from the Deep Galactic Legacy Infrared Mid-Plane Survey Extraordinaire survey \citep[Deep GLIMPSE;][]{whitney11}, having propagated the VIRAC2 coordinates to Epoch 2012.5 and matching with a 1\arcsec~ cross-match radius. Care must be taken when matching VVV and {\it Spitzer} detections because the spatial resolution of the latter is poorer ($\sim$2\arcsec), sometimes causing flux from more than one VVV source to be included in the beam. Inspection of the images indicates no such problem in this case though.
VVV~1820-2742 was not detected by any of {\it Spitzer} Galactic surveys.

VVV~1820-2742 is at a distance, $d\approx{}27$~pc, very similar to the early T dwarf candidate VVV~1814-2654 and it has a similar projected location within the Galactic bulge. However, the different proper motions indicate that a common origin is unlikely. The $J-H$, $H-K_s$ and $J-K_s$ colours resemble those of normal stars. Such colours suggest a spectral type of T2 to T3, based on inspection of the data provided by \citet{kirkpatrick21}. The $M_{K_s}$ value indicates a spectral type of T3 to T4.5, based on the data in the 2MASS passband provided by \citet{dupuy12}, consistent with the colours\footnote{The VISTA/VIRCAM filters are not identical to the 2MASS or MKO filters, which complicates the comparison. In $K_s$, VIRCAM closely resembles 2MASS $K_s$; in $H$, the VIRCAM, MKO and 2MASS filters are all quite similar; in $J$, VIRCAM resembles MKO rather than the unusually broad 2MASS $J$ filter. By comparing VIRAC2 results with either MKO or 2MASS, as appropriate, the effects of different filter systems should be smaller than the intrinsic scatter at each spectral type.}.

VVV~1253-6339 is at a distance, $d\approx 15$~pc, very similar to 
the white dwarf VVV~J141159.32-592045.7, the nearest star to be discovered
by VIRAC v1 \citep{LCS18}. The negative $J-H$, $H-K_s$ and $J-K_s$ colours and fainter $M_{K_s}$ value indicate a later spectral type than VVV~1820-2742. Comparison with the colour data in \citet{kirkpatrick21} indicates a type later than T4 and the $M_{K_s}$ value suggests a T6 or T7 type \citep{dupuy12}. The $\mathrm{[3.6] - [4.5]}$ colour is consistent with a type between T4 and T6.5, based on comparison with data in the UltracoolSheet, which were drawn mainly from \citet{patten06, kirkpatrick11, kirkpatrick19, kirkpatrick21, mace13, meisner20a, meisner20b} and \citet{griffith12}. Taken together, these constraints imply a spectral type of $\sim$T6.

While these two systems are technically only candidates, their
T dwarf nature is almost certain, given their colours, high proper motions and parallaxes. The only plausible alternative, in the case of VVV~1820-2742, would be an L-type sub dwarf, see e.g. \citet{zhang17}. However, this would require the parallax to be greatly over-estimated. Such objects are in any case much rarer than T dwarfs. For VVV~1253-6339, the red $\mathrm{[3.6] - [4.5]}$ colour appears to rule out the L sub-dwarf interpretation.

A limitation of this search is that it was difficult to confirm or reject some candidates with $\mu < 40$~\masyr{} through visual inspection of the images, since the candidates are mostly quite faint and total motion over the course of the survey is small. We only report sources with obvious motion here and we also imposed a hard lower limit at $\mu = 30$~\masyr{} so it is possible that some slow-moving T dwarfs were missed.

\subsubsection{$d<50$~pc search}
\label{sec:50_pc_search}

We performed a parallax-based search for new sources within $50$~pc without any colour-based criteria. From the initial selection of $\sim$1.6 million candidate nearby ($\varpi>10$~mas) VIRAC2 sources with 5$\sigma$ parallax detections described in Section \ref{sec:100pc}, we applied the following additional cuts:

\begin{enumerate}
    \item $\varpi > 20$~mas
    \item $f_{amb} < 0.4$
    \item $\Delta t >6$~yr
    \item $\itl{ks\_n\_det > 20}$
    \item $\mu > 30$~\masyr{}
    \item ${\rm UWE} <1.2$
    \item $V_{\rm tan} > 5$~km s$^{-1}$
\end{enumerate}

Cuts (ii) to (vii) in this list are based on the same parameters that we used to select T dwarfs in section \ref{sec:Ts}. The choice of a 6~yr minimum time baseline and a maximum ${\rm UWE}$ limit of 1.2 were informed by a cross-match against GCNS \citep{smart21} that also helped to identify previously known nearby systems. The cross-match was performed using a set of 1.34 million VIRAC2 sources, a subset of the initial 1.6 million candidate nearby sources with these additional criteria: $f_{amb} < 0.4$, $\Delta t >0.5$~yr and $\itl{ks\_n\_det > 20}$.
A small cross-match radius of 0.5\arcsec was used to minimise the number of chance alignments, having first propagated all GCNS source coordinates back from the Gaia reference epoch at 2016.0 to the VIRAC2 reference epoch at 2014.0. This cross-match yielded 2178 matches. All of the matched sources have $\Delta t>4$~yr and $\sim$99 per cent have $\Delta t > 6$~yr.

\begin{figure}
  \begin{center}
    \includegraphics[width=.48\textwidth]{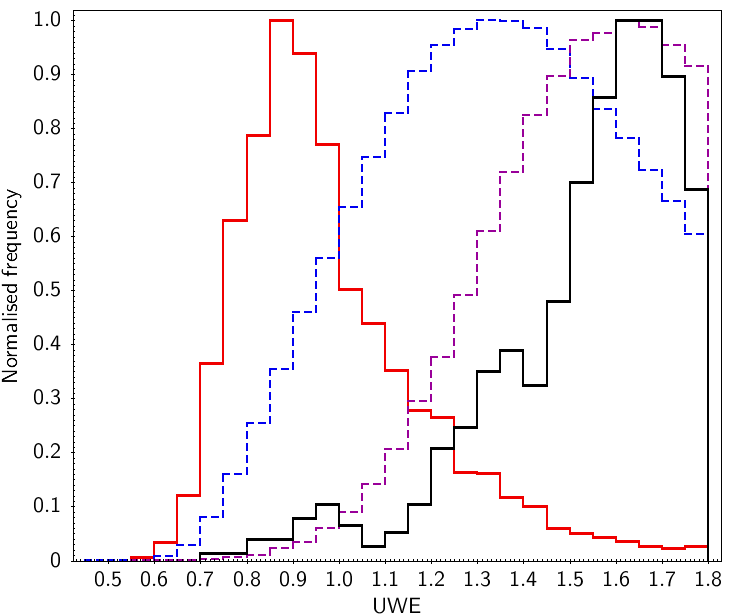}
    \caption{Normalised distributions of the VIRAC2 ${\rm UWE}$ parameter for VIRAC2 candidate nearby sources ($\varpi>10$~mas) having a counterpart in GCNS (red line) or not (other lines). The selections in the latter case are: (i) all sources (dashed blue line); (ii) $\mu>30$~\masyr{} (dashed purple line); (iii) $\mu>200$~\masyr{} (black line). Sources with no GCNS match typically have incorrect parallaxes and larger ${\rm UWE}$ values than
    GCNS matches, though the $\mu>200$~\masyr{} subset shows a small peak at ${\rm UWE} \approx 1$, suggesting scope for new discoveries.}
    \label{fig:UWE_UCD}
  \end{center}
\end{figure}

The distribution of ${\rm UWE}$ for matched sources is shown by the solid red line in Figure \ref{fig:UWE_UCD}. These bona fide nearby stars ($d<100$~pc) typically have proper motions $\mu > 30$~\masyr.
The other lines in Figure \ref{fig:UWE_UCD} show the distributions of ${\rm UWE}$ for (i) unmatched VIRAC2 sources (dashed blue line), (ii) unmatched VIRAC2 sources with $\mu > 30$~\masyr{} (dashed magenta line) and (iii) unmatched VIRAC2 sources with $\mu > 200$~\masyr{} (solid black line). We see that bona fide nearby stars have a peak at ${\rm UWE=0.9}$, slightly above the typical value of $0.85$ for
the VIRAC2 catalogue as a whole (see Figure \ref{fig:UWE}). The larger typical values of ${\rm UWE}$ for unmatched sources show that most of them have unreliable astrometric solutions. Our cut at ${\rm UWE}=1.2$ should include most bona fide nearby stars, whilst excluding the bulk of the unreliable solutions. 
Erroneous astrometric solutions with large proper motions are frequently due to mis-matches in the time series, i.e. a fit to two different stars. Encouragingly, the solid black line in Figure \ref{fig:UWE_UCD} shows a small secondary peak just below unity. We can presume this feature corresponds to genuine new high proper motion stars that are not in GCNS.

Having justified criteria (ii) to (vi), cut (vii) on $V_{\rm tan}$ was necessary to reduce the number of VIRAC2 candidates to a manageable level for individual inspection. Without it, cuts (i) to (vi) leave 10~526 candidates with $\varpi>20$~mas. A 5~km s$^{-1}$ cut was imposed in order to retain slow moving stars such as the nearby white dwarf VVV~J141159.32-592045.70 ($V_{\rm tan}$=7 km/s) which was discovered with VIRAC v1 \citep{LCS18}. After applying this cut, we had 628 candidates. This selection had 150 matches in GCNS,
using a 0.5\arcsec matching radius at epoch 2014.0 as above, which left 478 new candidate nearby sources. This is still a large number for visual inspection so we decided to increase the parallax significance threshold from $\varpi / \sigma_{\varpi} > 5$ to $\varpi / \sigma_{\varpi} > 7$. This left 89 candidates for assessment, after excluding the three T dwarf candidates already listed in Tables \ref{tab:earlyT} and \ref{tab:mid-lateT}.

In a plot of $\sigma_{\varpi}$ vs. $\varpi$ (not shown), these 89 candidates split into two groups: (i) a smaller group of mainly composed of brighter sources ($13.5 < K_s < 16$) with smaller parallaxes and errors ($20 < \varpi/\textrm{mas} < 50$, $1 < \sigma_{\varpi}/\textrm{mas} < 6$) that we might expect to be accurate solutions; (ii) a larger group of fainter sources ($K_s > 16$) mostly having larger parallaxes and errors ($\varpi > 60$~mas, $\sigma_{\varpi} > 7$~mas) in which we would expect inaccurate solutions to predominate. We visually inspected pairs of cut-out images
taken several years apart to confirm the high proper motions, using
the \textsc{SAOImage DS9} software to mark the positions expected
at the date of observation. Marking the expected positions was necessary because there were cases where there was a discernible proper motion that differed from the VIRAC2 values,
owing to the effect of blending (or blending and stellar variability) on the source matching and the astrometric solution. After this check, only 12 nearby
candidates with visually confirmed high proper motions remained. An
additional check on the astrometric solutions was made by plotting
RA vs. time and Dec vs. time, to confirm the expected linear trends.

The 12 candidate nearby sources are listed in Table \ref{tab:50pc}.  The final column contains an indicative UCD spectral type based on the $M_{K_s}$ vs. spectral type relation in \citet{dupuy12}, provided that the colours and absolute magnitude are consistent. The best-fitting sub-type is shown in brackets, to indicate that there is significant uncertainty. A simple question mark indicates that colours are not consistent with a UCD
interpretation, whereas ``L(?)" type indicates indicates a UCD for which $M_{K_s}$ is fainter than the usual range for L0-T2 types.

We caution that verification of high proper motion does not guarantee that the parallaxes are correct. E.g. the 150 sources with $\varpi>20~$mas and a match in GCNS include four cases where the VIRAC2 parallax is larger than the Gaia parallax by a factor between two and three, corresponding to discrepancies at the 3.3$\sigma$ to 4.5$\sigma$ level.
We tried inspecting plots of the annual parallactic motion in RA and Dec, after subtracting the proper motion. In a few cases with high parallax significance it was possible to clearly confirm the motion by phase-folding the time series on a one year period and then binning the detections by phase to reduce the large scatter on individual points. More often though, plots of parallax fits with below $\sim$10$\sigma$ significance (not shown) appear plausible rather than compelling, due to large error bars and the fact that the observing season for a given source often has a phase coverage of only about half a year. By way of cautionary examples, it initially appeared that we might have 14 bona fide new nearby sources where the proper motion was visually confirmed in the images. However, two of these systems had unusual properties and they were ruled out on further investigation of all available information, including the plots of position vs. time. Details of these failure modes are given in Appendix \ref{sec:app_failures}, in case they are relevant to further work with VIRAC2. 

In Table \ref{tab:50pc} we see that all sources are in the range \mbox{$10.8<M_{K_s}<13.9$} (neglecting uncertainties in parallax and photometry). In this range we would expect to find L dwarfs, T dwarfs and white dwarfs. 
Nine of the 12 sources were also found in our focussed search for red UCDs, discussed in section \ref{sec:Ldwfs}. To avoid repetition, we discuss only the other three sources here: source nos. 1, 8 and 9.


To provide more information, we cross-matched the sources in Table \ref{tab:50pc} with the second data release of DECaPS, \citealt{saydjari23}), the deepest optical survey available for the southern Galactic plane. A positional match within 0.5\arcsec~cross-match radius was required after propagating the VIRAC2 coordinates to the mean observation date of the DECaPS catalogue entry. 
All 12 sources except source no. 9 had a catalogue match.

We also cross-matched the sources in Table \ref{tab:50pc} with the {\it Spitzer} GLIMPSE I, GLIMPSE II, GLIMPSE 3-D, Deep GLIMPSE, GLIMPSE Proper and ApoGLIMPSE mid-infrared Galactic surveys \citep{benjamin03, churchwell09, whitney11, benjamin15, benjamin16}. A 1\arcsec~cross-match radius was used, after first propagating the VIRAC2 coordinates to suitable dates when the VVV area was observed in each survey. Six of the 12 sources in Table \ref{tab:50pc} have no match, including source nos. 8 and 9, 
and a further two sources with matches (listed in section \ref{sec:Ldwfs}) have unreliable {\it Spitzer} photometry because inspection of the images indicates that two VVV sources are blended into a single 
{\it Spitzer} source. However, four of the 12 sources in Table \ref{tab:50pc} have unblended matches in the $I1$ (3.6~$\mu$m) and $I2$ (4.5~$\mu$m) filters: source nos. 1, 3, 4 and 6. 

Source no. 1 (VVV J121047.63-622729.2, hereafter VVV~1210-6227) has $M_{K_s}=13.84$, corresponding to a T3 type, but its $J$-$H$ and $H$-$K_s$ colours are unusually red even for an L dwarf and far too red for a T3 type. The {\it Spitzer} matches give us the colours $K_s - [3.6] = 1.63 \pm 0.13$ and $[3.6] - [4.5] = 0.31 \pm 0.22$, which are most consistent with a late L type. DECaPS has catalogued detections in $i$, $z$ and $y$ filters and the colour $z-J=2.41$ is also more consistent with an L type than a T type, from comparison with typical UCD colours given in \citet{carnero19} for Dark Energy Camera and VISTA photometry. Inspection of the images shows no sign of the source in $i$ however: the entry was based on a single detection, whereas most bona fide entries have two or more detections.
One possibility is that VVV~1210-6227 is a young late-type L dwarf with very low gravity, given that these sources typically have redder near infrared colours than normal field L dwarfs \citep{faherty16} and their $M_{K_s}$ values appear to extend into the T dwarf range \citep{liu16}, similar to young planets such as HR8799b, HR8799d and 2MASS~1207-3932b \citep{marois08, chauvin04, ducourant08}. Alternatively, if the parallax is over-estimated by $\sim$2.5$\sigma$ then it could simply be a late L dwarf with unusually red colours. The \textsc{BANYAN$\Sigma$} software tool \citep{gagne18} finds a 91 per cent probability that that the system is a field object, rather than member of a known young moving group. However, a number of very low gravity L dwarfs are known that are not members of known young moving groups \citep{faherty16, liu16} so this check is not conclusive.

Source no. 8 (VVV~J173302.42-245559.9) is an interesting case. The $M_{K_s}$ value is in the range for L dwarfs and most of the VIRAC2 colours are red, consistent with this. However, the $Z$-$J$ colour (and to a lesser extent the $Y$-$J$ colour) is unusually blue for such sources. The DECaPS data show that the optical colours are also relatively blue: $r-i = 1.40$, $i-z = 0.79$.
Some possible explanations for this transition from bluer optical colours to redder near infrared colours are: (i) a blend of a UCD and a more distant, optically brighter star that does not contribute significantly to the astrometric solution due to being fainter in the infrared; (ii) a
white dwarf $+$ L dwarf binary pair.

Source no. 9 (VVV~J174337.03-215018.1) has $M_{K_s}=13.47$, in the range for early T dwarfs. However,  the blue $Z$-$J$ and $Y$-$J$ colours are not consistent with a brown dwarf: even very metal poor UCDs have $Y-J>0.5$ \citep{zhang18},
suggesting that it may be a white dwarf. Unfortunately, source no. 9 is just outside the DECaPS footprint. The only available optical detections are a single passband datum from Gaia DR3 ($G = 20.36$), without a parallax, and a $y$ band detection in the second data release of the Panoramic Survey Telescope and Rapid Response System \citep[Pan-STARRS1 DR2;][]{chambers16, flewelling20}. The latter detection adds little to the VIRAC2 photometry but the colour $G-J=3.71$ and absolute magnitude $M_G = 18.06$ are not in the range occupied by typical UCDs \citep{reyle18}. The absolute magnitude is also very faint for a white dwarf, see e.g. \citet{golovin24}. This source therefore appears to be an unusual system, provided that the parallax measurement is correct.

\begin{landscape}
\begin{table}
\footnotesize
    \centering
    \caption{Additional new sources within 50~pc of the Sun. Coordinates are for Epoch 2014.0, Equinox J2000.0. Approximate distances and $M_{K_s}$ values are based on the reciprocal of parallax. Errors on fluxes are available in the VIRAC2 source table. Nominal spectral types, based mainly on $M_{K_s}$, are indicated in the ``SpT'' column, with the best-fit sub-type in brackets, see main text.}
    \label{tab:50pc}
    \begin{tabular}{rllllrrcllllllllll}

    \# & Source ID & VVV Designation & RA & Dec & $\mu_{\alpha*}$ & $\mu_{\delta}$ & $\varpi$ & $d$ & $K_s$ & $M_{K_s}$ &
    $Z$-$J$ & $Y$-$J$ & $J$-$H$ & $H$-$K_s$ & $J$-$K_s$ & SpT\\

    & & & ($^{\circ}$) & ($^{\circ}$) & (\masyr) & (\masyr) & (mas) & (pc) & & & & & & & &\\ \hline

1 & 16064415000927 & VVV J121047.63-622729.2 & 182.69846 & -62.4581 & -152.1 $\pm$ 1.6 & 31.1 $\pm$ 1.7 & 35.0 $\pm$ 4.2 & 28 & 16.13 & 13.84 &  &  & 1.32 & 1.15 & 2.47 & L(?) \\
2 & 16028292006473 & VVV J133647.47-614648.4 & 204.19778 & -61.78012 & 123.4 $\pm$ 0.4 & 54.0 $\pm$ 0.4 & 31.5 $\pm$ 1.3 & 31 & 14.58 & 12.06 & 2.50 & 1.19 & 0.84 & 0.51 & 1.36 & L(6)\\ 
3 & 15983628001410 & VVV J140116.84-605554.0 & 210.32018 & -60.93168 & -130.4 $\pm$ 0.7 & -132.9 $\pm$ 0.7 & 30.2 $\pm$ 2.1 & 33 & 15.15 & 12.55 & 2.99 & 1.47 & 1.25 & 0.82 & 2.07 & L(8)\\ 
4 & 16081224003667 & VVV J140705.59-624656.0 & 211.77328 & -62.78223 & -153.8 $\pm$ 0.7 & -123.9 $\pm$ 0.9 & 30.9 $\pm$ 2.2 & 32 & 15.39 & 12.84 & 2.43 & 1.05 & 0.96 & 0.56 & 1.51 & L(9)\\ 
5 & 15214987007021 & VVV J163823.65-484211.5 & 249.59856 & -48.70319 & 0.6 $\pm$ 0.5 & -87.5 $\pm$ 0.5 & 28.3 $\pm$ 1.4 & 35 & 14.17 & 11.43 & 2.52 & 1.21 & 0.83 & 0.54 & 1.37 & L(3)\\ 
6 & 14394269006924 & VVV J172522.93-382122.1 & 261.34555 & -38.35615 & -34.0 $\pm$ 0.9 & -106.7 $\pm$ 0.8 & 28.0 $\pm$ 2.9 & 35 & 15.05 & 12.29 & 1.98 & 0.87 & 0.74 & 0.42 & 1.17 & L(7)\\ 
7 & 13218726001114 & VVV J172821.46-254333.2 & 262.08942 & -25.72589 & 105.8 $\pm$ 2.1 & -18.3 $\pm$ 2.0 & 32.2 $\pm$ 4.0 & 31 & 15.63 & 13.15 & 2.77 & 1.14 & 0.84 & 0.60 & 1.44 & T(1)\\ 
8 & 13140915004830 & VVV J173302.42-245559.9 & 263.2601 & -24.93331 & -6.0 $\pm$ 1.1 & -30.8 $\pm$ 1.1 & 20.4 $\pm$ 2.8 & 49 & 14.37 & 10.94 & 1.46 & 0.78 & 0.94 & 0.48 & 1.42 & ?\\ 
9 & 12829649008054 & VVV J174337.03-215018.1 & 265.90427 & -21.83836 & -71.1 $\pm$ 2.2 & -32.4 $\pm$ 2.5 & 34.2 $\pm$ 4.4 & 29 & 15.77 & 13.47 & 0.96 & 0.38 & 0.56 & 0.32 & 0.88 & ?\\ 
10 & 14222340003203 & VVV J180133.75-362315.0 & 270.39062 & -36.38749 & 108.5 $\pm$ 0.7 & -47.0 $\pm$ 0.8 & 21.2 $\pm$ 1.4 & 47 & 14.19 & 10.82 & 2.49 & 1.18 & 0.76 & 0.59 & 1.34 & L(1)\\ 
11 & 14124045003421 & VVV J180450.03-351920.2 & 271.20844 & -35.32227 & -8.7 $\pm$ 0.7 & -145.6 $\pm$ 0.8 & 20.5 $\pm$ 1.6 & 48 & 15.04 & 11.61 &  & 1.53 & 1.15 & 0.85 & 2.00 & L(4)\\ 
12 & 13673522002486 & VVV J181736.80-302618.9 & 274.40332 & -30.43858 & -259.4 $\pm$ 0.4 & -133.9 $\pm$ 0.4 & 22.0 $\pm$ 1.0 & 45 & 14.33 & 11.04 & 2.30 & 1.03 & 0.59 & 0.48 & 1.07 & L(2)

    \end{tabular}
\end{table}

\begin{table}
\footnotesize
    \centering
    \caption{Red UCD candidates from VIRAC2. Coordinates are for epoch 2014.0, Equinox J2000.0. Approximate distances and $M_{K_s}$ values are based on the reciprocal of parallax. Errors on fluxes are available in the VIRAC2 source table. Most of these are new discoveries, save for some candidates found in VIRAC v1, see main text. Nominal spectral types are indicated in the ``SpT'' column, as in Table \ref{tab:50pc}.}
    \label{tab:redUCDs}
    \begin{tabular}{rllllrrclllllllllll}

    \# & Source ID & VVV Designation & RA & Dec & $\mu_{\alpha*}$ & $\mu_{\delta}$ & $\varpi$ & $d$ & $K_s$ & $M_{K_s}$ &
    $Z$-$J$ & $Y$-$J$ & $J$-$H$ & $H$-$K_s$ & $J$-$K_s$ & SpT \\
    & & & ($^{\circ}$) & ($^{\circ}$) & (\masyr) & (\masyr) & (mas) & (pc) & & & & & & & & \\ \hline
1 & 16052453002106 & VVV J122708.14-621420.1 & 186.7839 & -62.2389 & -36.3 $\pm$ 0.7 & -11.4 $\pm$ 0.7 & 14.5 $\pm$ 2.0 & 69 & 14.60 & 10.41 & 2.11 & 0.89 & 0.66 & 0.60 & 1.26 & L(0)\\
2 & 16023286011236 & VVV J123646.95-613855.4 & 189.1956 & -61.64873 & -34.7 $\pm$ 0.5 & -11.2 $\pm$ 0.5 & 13.3 $\pm$ 1.8 & 75 & 14.82 & 10.43 & 2.32 & 1.00 & 0.66 & 0.59 & 1.25 & L(0)\\
3 & 16054900003583 & VVV J125158.50-621637.9 & 192.99374 & -62.27719 & 51.0 $\pm$ 1.1 & -73.0 $\pm$ 1.1 & 21.6 $\pm$ 3.4 & 46 & 15.54 & 12.22 & 2.34 & 1.23 & 0.85 & 0.65 & 1.50 & L(6)\\
4 & 16050114000271 & VVV J130944.97-620825.3 & 197.43736 & -62.14036 & -318.3 $\pm$ 0.9 & 46.1 $\pm$ 0.9 & 15.7 $\pm$ 3.0 & 64 & 15.61 & 11.59 & 2.42 & 1.26 & 0.75 & 0.49 & 1.24 & L(4)\\
5 & 16127533006131 & VVV J131358.89-634426.4 & 198.49536 & -63.74066 & -50.0 $\pm$ 0.8 & -47.1 $\pm$ 0.8 & 22.0 $\pm$ 2.3 & 46 & 14.80 & 11.51 & 1.93 & 0.86 & 0.55 & 0.47 & 1.03 & L(4)\\
6 & 15960732002707 & VVV J132919.47-603025.9 & 202.33112 & -60.50718 & -69.8 $\pm$ 0.9 & 80.8 $\pm$ 1.0 & 19.7 $\pm$ 2.6 & 51 & 15.54 & 12.01 & 1.97 & 1.00 & 0.67 & 0.33 & 1.00 & L(5)\\
7 & 16028292006473 & VVV J133647.47-614648.4 & 204.19778 & -61.78012 & 123.4 $\pm$ 0.4 & 54.0 $\pm$ 0.4 & 31.5 $\pm$ 1.3 & 32 & 14.57 & 12.06 & 2.50 & 1.19 & 0.84 & 0.51 & 1.36 & L(6)\\
8 & 15983628001410 & VVV J140116.84-605554.0 & 210.32018 & -60.93168 & -130.4 $\pm$ 0.7 & -132.9 $\pm$ 0.7 & 30.2 $\pm$ 2.1 & 33 & 15.15 & 12.55 & 2.99 & 1.47 & 1.25 & 0.82 & 2.07 & L(8)\\
9 & 16081224003667 & VVV J140705.59-624656.0 & 211.77328 & -62.78223 & -153.8 $\pm$ 0.7 & -123.9 $\pm$ 0.9 & 30.9 $\pm$ 2.2 & 32 & 15.38 & 12.84 & 2.43 & 1.05 & 0.96 & 0.56 & 1.51 & L(9)\\
10 & 15895686003323 & VVV J142226.89-591831.0 & 215.61203 & -59.30861 & -54.4 $\pm$ 0.7 & -40.0 $\pm$ 0.7 & 16.9 $\pm$ 2.4 & 59 & 15.49 & 11.63 & 2.28 & 1.03 & 0.71 & 0.49 & 1.20 & L(4)\\
11 & 16028385004618 & VVV J143132.70-614527.2 & 217.88625 & -61.75756 & -23.8 $\pm$ 0.5 & -29.7 $\pm$ 0.5 & 18.9 $\pm$ 1.9 & 53 & 14.92 & 11.30 & 2.38 & 1.12 & 0.80 & 0.53 & 1.34 & L(3)\\
12 & 15786418001139 & VVV J154206.53-572529.8 & 235.52722 & -57.42495 & -26.3 $\pm$ 1.1 & -28.6 $\pm$ 1.0 & 14.7 $\pm$ 2.6 & 68 & 15.50 & 11.33 & 2.47 & 1.07 & 0.69 & 0.45 & 1.14 & L(3)\\
13 & 15625547005863 & VVV J155405.53-544536.3 & 238.52303 & -54.76008 & -36.0 $\pm$ 0.7 & -67.4 $\pm$ 0.6 & 14.1 $\pm$ 2.3 & 71 & 15.11 & 10.86 & 2.51 & 1.25 & 0.89 & 0.64 & 1.53 & L(1)\\
14 & 15694314006505 & VVV J155946.54-555039.7 & 239.94391 & -55.84435 & -102.5 $\pm$ 0.8 & -101.4 $\pm$ 0.8 & 13.2 $\pm$ 2.0 & 76 & 15.37 & 10.97 & 2.48 & 1.20 & 0.68 & 0.50 & 1.18 & L(2)\\
15 & 15298621003341 & VVV J161328.24-495552.9 & 243.36765 & -49.93135 & -45.2 $\pm$ 0.8 & -3.9 $\pm$ 0.8 & 16.3 $\pm$ 2.7 & 61 & 15.30 & 11.36 & 2.31 & 1.14 & 0.70 & 0.59 & 1.29 & L(3)\\
16 & 15363256005368 & VVV J161802.53-504842.3 & 244.51053 & -50.81176 & -61.2 $\pm$ 0.6 & -152.2 $\pm$ 0.6 & 17.8 $\pm$ 1.7 & 56 & 15.11 & 11.37 & 2.52 & 1.32 & 0.92 & 0.63 & 1.56 & L(3)\\
17 & 15312381006935 & VVV J162816.50-500635.8 & 247.06874 & -50.10994 & 26.5 $\pm$ 1.5 & -17.2 $\pm$ 1.2 & 20.0 $\pm$ 3.6 & 50 & 15.66 & 12.16 & 2.75 & 1.36 & 0.90 & 0.49 & 1.39 & L(6) \\
18 & 15214987007021 & VVV J163823.65-484211.5 & 249.59856 & -48.70319 & 0.6 $\pm$ 0.5 & -87.5 $\pm$ 0.5 & 28.3 $\pm$ 1.4 & 35 & 14.17 & 11.38 & 2.52 & 1.21 & 0.83 & 0.54 & 1.37 & L(3) \\
19 & 15168695001226 & VVV J163953.34-480440.6 & 249.97226 & -48.07795 & 36.7 $\pm$ 0.6 & -30.1 $\pm$ 0.6 & 19.3 $\pm$ 1.8 & 52 & 14.09 & 10.52 & 2.56 & 1.28 & 0.86 & 0.54 & 1.40 & L(0) \\
20 & 13910884004131 & VVV J170513.58-325803.3 & 256.3066 & -32.96757 & -130.4 $\pm$ 1.4 & -176.1 $\pm$ 1.2 & 23.0 $\pm$ 3.8 & 43 & 15.56 & 12.37 & 2.53 & 1.31 & 0.74 & 0.58 & 1.32 & L(7) \\
21 & 14758019000425 & VVV J170533.90-424519.7 & 256.39124 & -42.75546 & 35.2 $\pm$ 1.8 & 18.0 $\pm$ 1.7 & 30.2 $\pm$ 5.0 & 33 & 16.22 & 13.62 & 2.49 & 1.40 & 0.88 & 0.65 & 1.53 & L(?)\\
22 & 14680939001576 & VVV J170743.15-414702.9 & 256.92978 & -41.78413 & -2.9 $\pm$ 0.7 & -92.1 $\pm$ 0.7 & 15.6 $\pm$ 2.3 & 64 & 15.21 & 11.18 & 1.95 & 0.84 & 0.61 & 0.42 & 1.03 & L(2)\\
23 & 14308244000412 & VVV J172204.74-372553.2 & 260.51974 & -37.43145 & 68.5 $\pm$ 1.0 & 17.8 $\pm$ 1.0 & 23.9 $\pm$ 3.5 & 42 & 15.97 & 12.87 & 2.42 & 1.06 & 1.14 & 0.67 & 1.81 & L(9)\\
24 & 14394269006924 & VVV J172522.93-382122.1 & 261.34555 & -38.35615 & -34.0 $\pm$ 0.9 & -106.7 $\pm$ 0.8 & 28.0 $\pm$ 2.9 & 36 & 15.05 & 12.29 & 1.98 & 0.87 & 0.74 & 0.42 & 1.17 & L(7)\\
25 & 13218726001114 & VVV J172821.46-254333.2 & 262.08942 & -25.72589 & 105.8 $\pm$ 2.1 & -18.3 $\pm$ 2.0 & 32.2 $\pm$ 4.0 & 31 & 15.61 & 13.15 & 2.77 & 1.14 & 0.84 & 0.60 & 1.44 & T(1)

    \end{tabular}
\end{table}
\end{landscape}

\begin{landscape}
\renewcommand{\thetable}{\arabic{table} - continued}
\addtocounter{table}{-1}
\begin{table}
\footnotesize
    \centering
    \caption{}
    \begin{tabular}{rllllrrclllllllllll}

    \# & Source ID & VVV Designation & RA & Dec & $\mu_{\alpha*}$ & $\mu_{\delta}$ & $\varpi$ & $d$ & $K_s$ & $M_{K_s}$ &
    $Z$-$J$ & $Y$-$J$ & $J$-$H$ & $H$-$K_s$ & $J$-$K_s$ & SpT \\
    & & & ($^{\circ}$) & ($^{\circ}$) & (\masyr) & (\masyr) & (mas) & (pc) & & & & & & & & \\ \hline

26 & 12956633004119 & VVV J174619.15-230825.6 & 266.5798 & -23.14046 & 114.1 $\pm$ 2.5 & -12.8 $\pm$ 2.4 & 20.9 $\pm$ 3.9 & 48 & 15.68 & 12.28 & 1.76 & 0.98 & 0.68 & 0.53 & 1.21 & L(6)\\
27 & 14521323002380 & VVV J175228.62-395300.3 & 268.11926 & -39.88342 & 15.7 $\pm$ 0.7 & -272.9 $\pm$ 0.7 & 11.5 $\pm$ 1.4 & 87 & 15.24 & 10.54 & 2.24 & 1.03 & 0.64 & 0.47 & 1.11 & L(0)\\
28 & 14377969003025 & VVV J175454.11-381356.3 & 268.72543 & -38.23231 & 22.6 $\pm$ 2.3 & -101.6 $\pm$ 2.4 & 24.7 $\pm$ 4.2 & 40 & 16.21 & 13.18 & 2.14 & 0.89 & 0.62 & 0.38 & 1.00 & L(?)\\
29 & 14545907003369 & VVV J175519.51-401056.0 & 268.8313 & -40.18223 & -16.0 $\pm$ 0.8 & -90.2 $\pm$ 0.8 & 10.1 $\pm$ 1.5 & 99 & 15.60 & 10.63 & 2.27 & 1.19 & 0.78 & 0.58 & 1.36 & L(1)\\
30 & 12952563001044 & VVV J175534.06-230515.8 & 268.89194 & -23.08773 & -14.0 $\pm$ 1.8 & -43.5 $\pm$ 1.6 & 23.7 $\pm$ 4.4 & 42 & 15.02 & 11.90 & 1.99 & 0.81 & 0.74 & 0.56 & 1.29 & L(5)\\
31 & 14042109008844 & VVV J175906.92-342531.3 & 269.77884 & -34.42537 & 35.8 $\pm$ 0.9 & -37.3 $\pm$ 0.9 & 12.1 $\pm$ 1.9 & 82 & 15.43 & 10.86 & 2.04 & 0.90 & 0.69 & 0.39 & 1.08 & L(1)\\
32 & 13014014002076 & VVV J175920.68-234045.3 & 269.83618 & -23.67924 & 45.9 $\pm$ 2.1 & 5.1 $\pm$ 1.9 & 31.7 $\pm$ 5.4 & 32 & 16.14 & 13.65 & 2.34 & 1.12 & 0.73 & 0.51 & 1.24 & T(2) \\
33 & 13710336009213 & VVV J175955.42-305014.8 & 269.98093 & -30.83745 & 56.4 $\pm$ 0.8 & -115.7 $\pm$ 0.9 & 12.9 $\pm$ 1.8 & 78 & 15.01 & 10.57 & 2.24 & 1.02 & 0.74 & 0.41 & 1.14 & L(1)\\
34 & 14660608000029 & VVV J175959.38-413153.5 & 269.99744 & -41.53152 & -13.4 $\pm$ 0.9 & -114.8 $\pm$ 0.9 & 12.2 $\pm$ 2.0 & 82 & 15.60 & 11.04 & 1.95 & 0.84 & 0.58 & 0.43 & 1.01 & L(2) \\
35 & 13820931005795 & VVV J180123.11-320007.0 & 270.34628 & -32.00194 & -12.3 $\pm$ 1.0 & -73.0 $\pm$ 1.0 & 12.8 $\pm$ 2.0 & 78 & 15.29 & 10.83 & 1.90 & 0.82 & 0.67 & 0.37 & 1.03 & L(1)\\
36 & 14222340003203 & VVV J180133.75-362315.0 & 270.39062 & -36.38749 & 108.5 $\pm$ 0.7 & -47.0 $\pm$ 0.8 & 21.2 $\pm$ 1.4 & 47 & 14.18 & 10.82 & 2.49 & 1.18 & 0.76 & 0.59 & 1.34 & L(1)\\
37 & 13685769012201 & VVV J180315.34-303321.8 & 270.8139 & -30.55605 & -24.3 $\pm$ 0.7 & -245.8 $\pm$ 0.8 & 17.3 $\pm$ 1.4 & 58 & 14.42 & 10.61 & 1.98 & 0.86 & 0.66 & 0.41 & 1.08 & L(1)\\
38 & 14124045003421 & VVV J180450.03-351920.2 & 271.20844 & -35.32227 & -8.7 $\pm$ 0.7 & -145.6 $\pm$ 0.8 & 20.5 $\pm$ 1.6 & 49 & 15.05 & 11.61 &  & 1.53 & 1.15 & 0.85 & 2.00 & L(4)\\
39 & 13566997000219 & VVV J180736.70-291700.1 & 271.90292 & -29.28336 & -59.6 $\pm$ 4.7 & -45.2 $\pm$ 3.5 & 50.7 $\pm$ 9.4 & 20 & 16.09 & 14.62 & 2.50 & 0.83 & 0.51 & 0.54 & 1.05 & L(?)\\
40 & 13710359003254 & VVV J180804.69-304948.6 & 272.01956 & -30.83018 & 4.7 $\pm$ 0.6 & -164.4 $\pm$ 0.6 & 19.9 $\pm$ 1.2 & 50 & 14.97 & 11.46 & 2.40 & 1.04 & 0.67 & 0.51 & 1.18 & L(3)\\
41 & 14496793000682 & VVV J180847.83-393713.4 & 272.1993 & -39.62039 & -38.8 $\pm$ 0.9 & 2.9 $\pm$ 0.9 & 11.3 $\pm$ 2.2 & 89 & 15.48 & 10.74 & 2.20 & 1.04 & 0.63 & 0.47 & 1.09 & L(1)\\
42 & 12686363006518 & VVV J180927.63-202725.3 & 272.36514 & -20.45704 & 55.7 $\pm$ 0.8 & -3.4 $\pm$ 0.7 & 10.4 $\pm$ 2.0 & 96 & 15.52 & 10.61 & 2.37 & 1.12 & 0.73 & 0.59 & 1.32 & L(1)\\
43 & 14119964002532 & VVV J180941.71-351601.9 & 272.42377 & -35.2672 & -104.3 $\pm$ 1.8 & -81.0 $\pm$ 2.0 & 18.1 $\pm$ 3.5 & 55 & 16.20 & 12.49 &  & 1.09 & 0.99 & 0.68 & 1.67 & L(7)\\
44 & 14255135000834 & VVV J181104.13-364709.7 & 272.7672 & -36.78602 & -6.2 $\pm$ 1.0 & -36.5 $\pm$ 0.9 & 14.4 $\pm$ 1.9 & 69 & 15.53 & 11.33 & 2.23 & 0.98 & 0.63 & 0.49 & 1.13 & L(3)\\
45 & 13165603009949 & VVV J181231.37-251351.7 & 273.1307 & -25.23102 & -19.7 $\pm$ 0.7 & -28.3 $\pm$ 0.7 & 10.9 $\pm$ 1.5 & 92 & 15.51 & 10.70 & 1.90 & 0.81 & 0.62 & 0.38 & 1.00 & L(1)\\
46 & 13931562003722 & VVV J181445.46-331032.0 & 273.68942 & -33.17556 & -2.6 $\pm$ 1.3 & -34.2 $\pm$ 1.4 & 15.3 $\pm$ 2.8 & 65 & 16.20 & 12.13 &  & 1.24 & 0.98 & 0.76 & 1.74 & L(6)\\
47 & 13333546009625 & VVV J181453.17-265453.6 & 273.72156 & -26.91488 & -235.1 $\pm$ 2.1 & -446.7 $\pm$ 2.2 & 36.4 $\pm$ 4.2 & 28 & 15.80 & 13.60 &  & 1.20 & 0.93 & 0.56 & 1.49 & T(2)\\
48 & 13030445003531 & VVV J181549.13-234845.1 & 273.95468 & -23.81254 & 76.3 $\pm$ 2.2 & -174.5 $\pm$ 2.3 & 29.7 $\pm$ 3.6 & 34 & 15.71 & 13.07 & 2.68 & 1.32 & 0.66 & 0.70 & 1.36 & L7\\
49 & 13304881005250 & VVV J181728.87-263606.4 & 274.37027 & -26.60177 & 11.4 $\pm$ 1.0 & -63.7 $\pm$ 1.1 & 10.8 $\pm$ 2.0 & 93 & 15.25 & 10.42 & 2.06 & 0.89 & 0.60 & 0.44 & 1.04 & L(0)\\
50 & 13673522002486 & VVV J181736.80-302618.9 & 274.40332 & -30.43858 & -259.4 $\pm$ 0.4 & -133.9 $\pm$ 0.4 & 22.0 $\pm$ 1.0 & 46 & 14.33 & 11.04 & 2.30 & 1.03 & 0.59 & 0.48 & 1.07 & L(2)\\
51 & 13976651001568 & VVV J182633.53-334138.3 & 276.6397 & -33.69398 & -48.0 $\pm$ 1.4 & -99.7 $\pm$ 1.6 & 22.6 $\pm$ 2.5 & 44 & 14.54 & 11.35 &  & 1.25 & 0.79 & 0.69 & 1.48 & L(3)\\
52 & 13382741000588 & VVV J182959.78-272446.4 & 277.49908 & -27.41289 & 16.7 $\pm$ 0.7 & -74.5 $\pm$ 0.7 & 11.8 $\pm$ 1.5 & 85 & 15.07 & 10.43 & 1.63 & 0.82 & 0.54 & 0.48 & 1.03 & L(0)\\
53 & 13435995001647 & VVV J183208.77-275804.3 & 278.03653 & -27.96786 & 62.1 $\pm$ 1.1 & -13.1 $\pm$ 1.2 & 12.6 $\pm$ 2.2 & 79 & 15.60 & 11.10 & 2.32 & 1.08 & 0.57 & 0.43 & 1.00 & L(2)\\
54 & 13145182001396 & VVV J183303.00-245943.5 & 278.2625 & -24.99543 & -104.1 $\pm$ 0.7 & 1.7 $\pm$ 0.7 & 16.1 $\pm$ 1.6 & 62 & 15.12 & 11.15 & 2.38 & 1.10 & 0.69 & 0.49 & 1.17 & L(2)\\
55 & 13493343002304 & VVV J183326.20-283243.9 & 278.35916 & -28.54553 & -39.7 $\pm$ 0.5 & -110.6 $\pm$ 0.5 & 15.7 $\pm$ 1.2 & 64 & 14.65 & 10.62 & 2.02 & 0.92 & 0.57 & 0.43 & 1.00 & L(1)\\
56 & 12993638002107 & VVV J183608.67-232842.2 & 279.03613 & -23.47839 & -57.3 $\pm$ 1.0 & -3.0 $\pm$ 0.9 & 13.0 $\pm$ 2.1 & 77 & 15.46 & 11.03 & 1.76 & 0.80 & 0.60 & 0.36 & 0.96 & L(2)

    \end{tabular}
\end{table}
\end{landscape}

\subsubsection{Red UCD search}
\label{sec:Ldwfs}

Finally, we searched for new red UCDs using cuts on absolute magnitude and colour that should include sources with spectral types in the L0 to T2 range, though some late M-type UCDs can be expected to scatter in. We began with the $\sim$1.6 million candidate nearby VIRAC2 sources ($\varpi>10$~mas, i.e. within $\sim$100~pc) with 5$\sigma$ parallax detections that were selected in section \ref{sec:100pc} and we then applied the following initial cuts: 

\begin{enumerate}
\item $M_{K_s} > 10.4$
\item $Y-J>0.8$
\item $J-{K_s}>0.9$
\end{enumerate}

This yielded 262\,903 sources, so it was clear that additional cuts are needed to remove the numerous reddened distant stars in the Galactic plane with incorrect parallax estimates. We performed three selections from this set using $f_{amb}$, ${\rm UWE}$, $\Delta t$, $\mu$ and $V_{\rm tan}$ to define quality cuts that are likely to include most genuine nearby sources. These selections (denoted 3a, 3b and 4) complement the colour-blind 50~pc search in section \ref{sec:50_pc_search} mainly by including sources at $d=50$ to 100~pc and sources with only 5$\sigma$ parallax significance, as opposed to 7$\sigma$ previously. 

Selections 3a and 3b first required $V_{\rm tan} > 10$~km~s$^{-1}$, which reduced the sample from 262\,903 sources to 619. A cross-match to GCNS with a 0.5\arcsec cross-match radius at epoch 2014.0 coordinates, found that 17 of these were known, leaving 602 potentially new candidates.

Selection 3a then applied these cuts:
\begin{enumerate}
\item $\Delta t >3$~yr
\item $f_{amb} < 0.1$
\item ${\rm UWE}<1.5$
\end{enumerate}

This yielded 54 candidates, of which 49 were visually confirmed as high proper motion sources. Of the remaining five, three were definite false positives and two cases were unclear due to low proper motion.

Selection 3b complemented 3a by relaxing the threshold on $f_{amb}$ whilst tightening the constraint on ${\rm UWE}$:

\begin{enumerate}
\item $\Delta t >3$~yr
\item $0.1 < f_{amb} < 0.4$
\item ${\rm UWE}<1.2$
\end{enumerate}

This yielded eight candidates, only one of which, VVV~J181549.13-234845.1, passed the visual proper motion check. This is in fact a known L dwarf, LTT~7251~B (= HD~167359~B), discovered by \citet{LCS18} using VIRAC v1 via a match of high proper motion sources to stars in the Tycho-{\it Gaia} Astrometric Solution (TGAS) catalogue \citep{michalik15}. LTT~7251~B is not detected by Gaia and hence it is not in the GCNS main table. It is also not in the GCNS table of missing objects since it has no parallax in the SIMBAD database at present. VIRAC v1 did not provide a parallax but VIRAC2 now does: $\varpi=29.7 \pm 3.6$~mas. This is consistent with the Gaia DR3 value measured for the G8-type primary, LTT~7251~A: $\varpi=26.54 \pm 0.02$~mas,  

Selection 4 complemented selections 3a and 3b by reducing the threshold value of $V_{\rm tan}$ from 10~km~s$^{-1}$ to 5~km~s$^{-1}$. To avoid including a very large number of false positives, this required us to impose tight constraints on all the other data quality parameters and apply a proper motion cut also:

\begin{enumerate}
\item $\Delta t >6$~yr
\item $f_{amb} < 0.1$
\item ${\rm UWE}<1.2$
\item $\mu>30$~\masyr{}
\item $5 < V_{\rm tan}/\mathrm{km s^{-1}} < 10$
\end{enumerate}

Selection 4 yielded 19 candidates, six of which were visually confirmed as (relatively) high proper motion sources. Of the remainder, 11 were clearly erroneous and two were unconfirmed due to the modest proper motions and blending.

\begin{figure}
  \begin{center}
    \includegraphics[width=.48\textwidth,keepaspectratio]{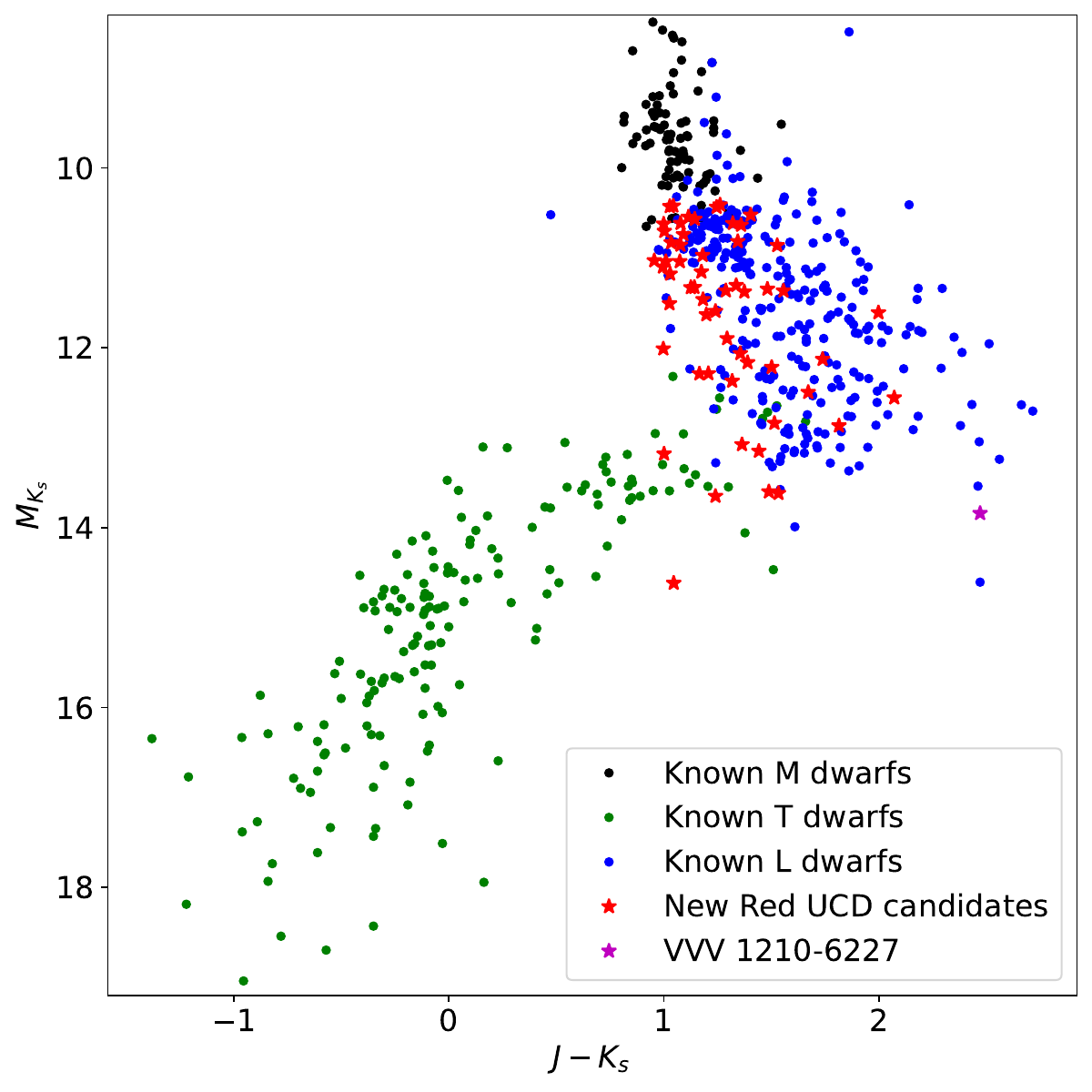}
    \caption{Colour vs. Absolute Magnitude diagram for the 56 red UCD candidates identified by our search (red stars). The low gravity late L dwarf candidate VVV~1210-6227 is also plotted (purple star). Known late M dwarfs (black points), L dwarfs (blue points) and T dwarfs (green points) from the UltracoolSheet are over-plotted, with unresolved binaries excluded. }
    \label{fig:CMD_redUCDs}
  \end{center}
\end{figure}

In Table \ref{tab:redUCDs} we list the 56 red UCD candidates from selections 3a, 3b and 4 that have visually confirmed high proper motions and are not in GCNS. A colour vs. absolute magnitude diagram for these sources is plotted in Figure \ref{fig:CMD_redUCDs} (red stars) and VVV~1210-6227, the low-gravity late L dwarf candidate from Table \ref{tab:50pc} is also marked (purple star). Known M dwarfs (black points), L (blue points) and T dwarfs (green points) from the UltracoolSheet are over-plotted for comparison.

Of the 56, nine were also found in the colour-blind 50~pc search (see Table \ref{tab:50pc}) and the selection also includes the high proper motion early T dwarf candidate VVV~J1814-2654 (see Table \ref{tab:earlyT}). We retain these sources in Table \ref{tab:redUCDs} to give the complete colour selection in one place. 53 of the 56 sources were included in VIRAC v1 but only 15 had parallaxes in that catalogue, in which 5$\sigma$ significance was required; only nine of these were listed as L dwarf
candidates since the VIRAC v1 aperture photometry is less reliable than VIRAC2 \textsc{Dophot} photometry. For the 15 with parallaxes, they are all consistent between VIRAC v1 and VIRAC2.
The uncertainties in VIRAC2 $\varpi$, $\mu_{\alpha*}$ and $\mu_{\delta}$ are smaller by 28 per cent, 43 per cent and 45 per cent respectively (median values). This reflects the fact that the longer time baseline of VIRAC2 improves proper motion precision more than parallax precision. A further three of the 56 sources were identified as UCDs using VIRAC v1 proper motion and colours alone \citep{LCS18}. Such an approach could be tried with VIRAC2 but it is beyond the scope of this work. After accounting for the VIRAC~v1 discoveries, including LTT~7251~B mentioned earlier, we are left with 
43 new red ultracool dwarf candidates. None of these systems are in the UltracoolSheet.

An indicative spectral type is given in the final column of Table \ref{tab:redUCDs} in a similar manner and format to Table \ref{tab:50pc}, based on the $M_{K_s}$ vs. type relation. A minor change was made for source no. 32 (discussed further below) where the $M_{K_s}$ vs. spectral type relation indicated a best-fit type of T3: the value was changed to T2 to be more consistent with the red colours, while remaining consistent with the absolute magnitude. ``L(?)" is given in cases where there is good reason to suspect the parallax is substantially over-estimated, detailed below. For LTT~7251~B (source no. 48), the known spectral type of L7 is given.

Two systems in Table \ref{tab:redUCDs} are binaries. VVV J163823.65-484211.5 (source no. 18) has a Gaia DR3 parallax, $\varpi=26.8 \pm 1.6$~mas, very similar to the VIRAC2 value, $28.3 \pm 1.4$~mas. It is only a new discovery in the partial sense that it was not in the GCNS list of nearby sources, despite having a parallax in Gaia eDR3. Our visual inspection showed that it is part of a binary pair: there is a brighter component, Gaia eDR3 source 5940919607037021568 (VVV~J163823.52-484211.7), that is included in VIRAC2 and GCNS, both of which indicate a proper motion and parallax similar to the secondary. The VIRAC2 colours and the absolute magnitudes indicate that the primary is also an L dwarf. The separation of 1.4\arcsec~ corresponds to $\sim$50~au at the 
36~pc distance of the primary (using the Gaia DR3 parallax); this is unusually wide for a brown dwarf pair \citep{burgasser07}.

 The second binary system is composed of VVV~J182633.53-334138.3 (source no. 51) and VVV~J182633.48-334138.4 (listed in GCNS). Again, the component listed in GCNS is the primary and VVV~J182633.53-334138.3 is the secondary (which has only a two-parameter astrometric solution in Gaia DR3). The absolute magnitudes suggest spectral types near L0 and L3 for the pair. They are blended in the VVV images with a separation of only 0.63\arcsec~ but despite this the components have consistent parallaxes and proper motions, with values in fair agreement with the Gaia DR3 values for the primary. 
The projected physical separation of the pair is $\sim$28~au. Again this is unusually wide for an L dwarf pair, but seeing limited surveys such as VVV have a strong bias towards these relatively wide pairs.

From inspection of the {\it ks\_eta} and {\it ks\_Stetson\_I} light curve indices given in the source
table for the 56 red UCDs, one source, VVV J180315.34-303321.8 (source no. 38) stands out from the rest with 
$ks\_eta = 0.43$, $ks\_Stetson\_I = 3.6$, compared to typical values near two and zero for non-variable sources, respectively. The light curve of this source, an early L dwarf candidate, shows a slow $\sim$0.2 mag rise in $K_s$ from 2010 to 2014 before stabilising and fading slightly in later years (not shown). A few other sources in the list also show hints of slow small amplitude variability, indicating that the VIRAC2 data may be useful to study long-term photometric variations in UCDs, whether due to magnetic activity or changes in weather patterns. To our knowledge, previous studies of variability in UCDs, e.g. \citet{vos22}, \citet{oliveros-gomez22} and references therein, have only examined changes on much shorter timescales. Examination of small amplitude variability will require careful statistical analysis so we defer the topic to a future work.

\begin{figure}
  \begin{center}
    \includegraphics[width=.48\textwidth,keepaspectratio]{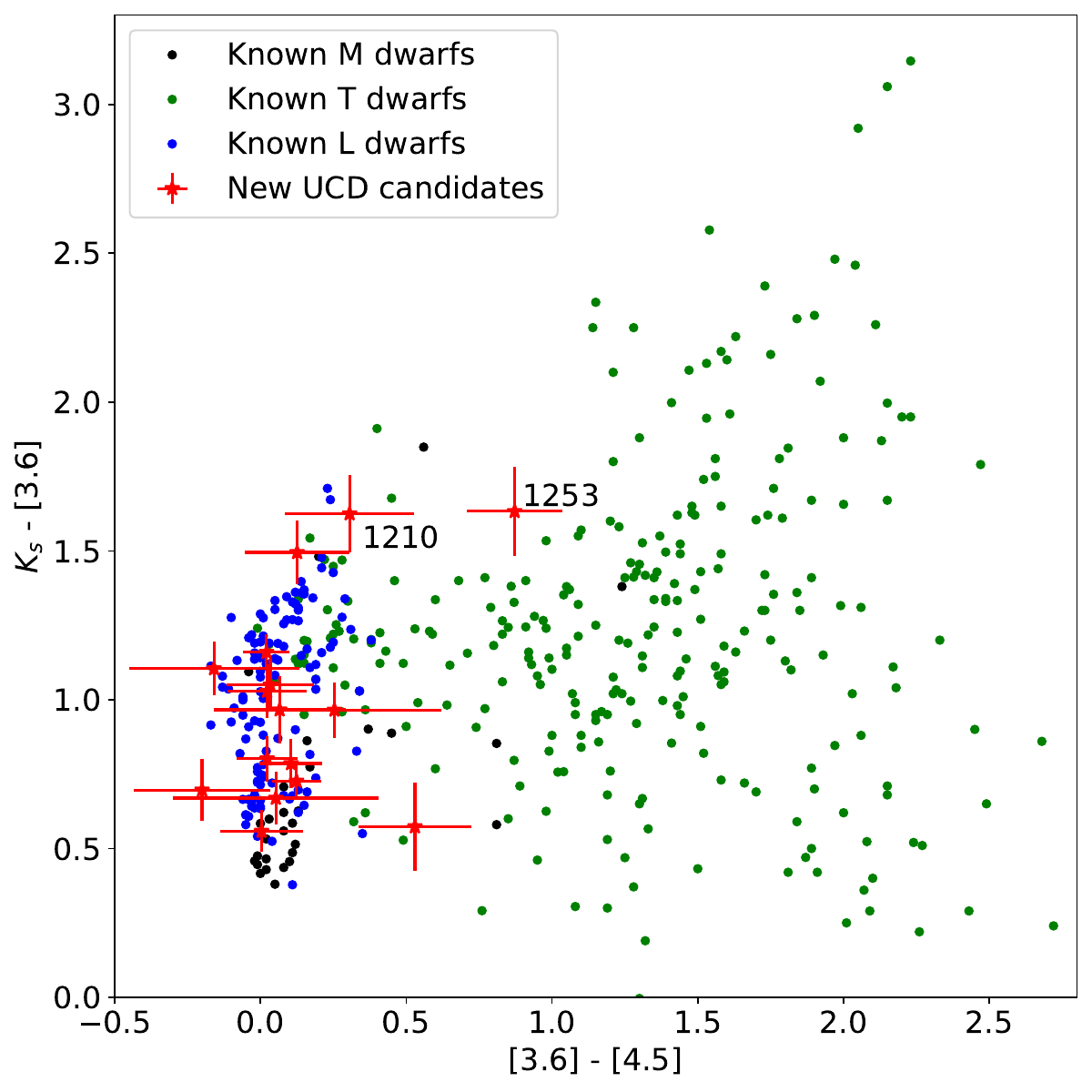}
    \caption{Two colour diagram incorporating {\it Spitzer} photometry. Red stars mark the new VIRAC2 UCD candidates with 3.6~$\mu$m and 4.5~$\mu$m detections in this work, including the low gravity late L dwarf candidate VVV~1210-6227 from Table \ref{tab:50pc} and the T dwarf candidate VVV~1253-6339 from Table \ref{tab:mid-lateT}. Known late M dwarfs (black points), L dwarfs (blue points) and T dwarfs (green points) from the UltracoolSheet are over-plotted, with unresolved binaries excluded.}
    \label{fig:TCD}
  \end{center}
\end{figure}

\begin{figure}
  \begin{center}
    \includegraphics[width=.48\textwidth,keepaspectratio]{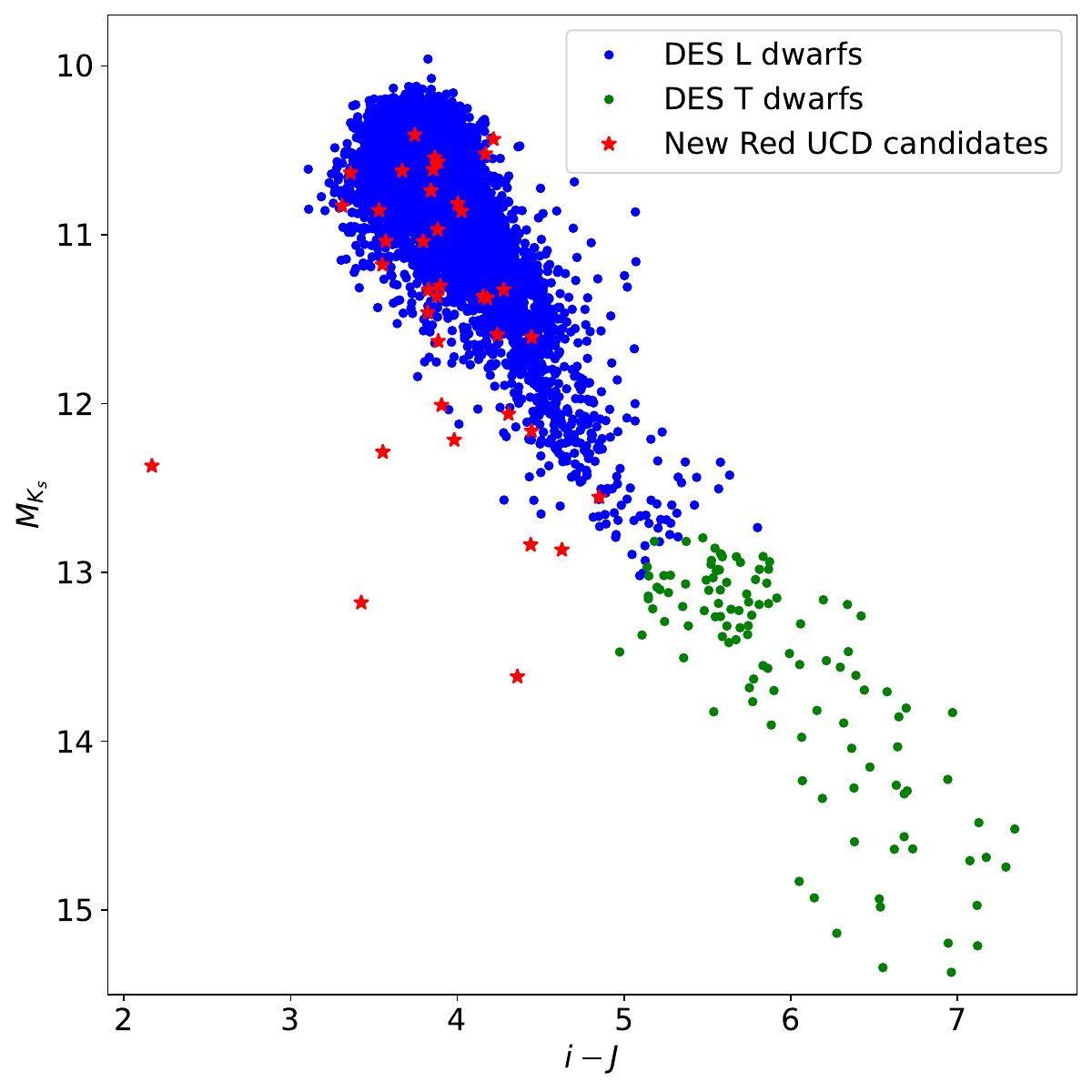}
    \includegraphics[width=.48\textwidth,keepaspectratio]{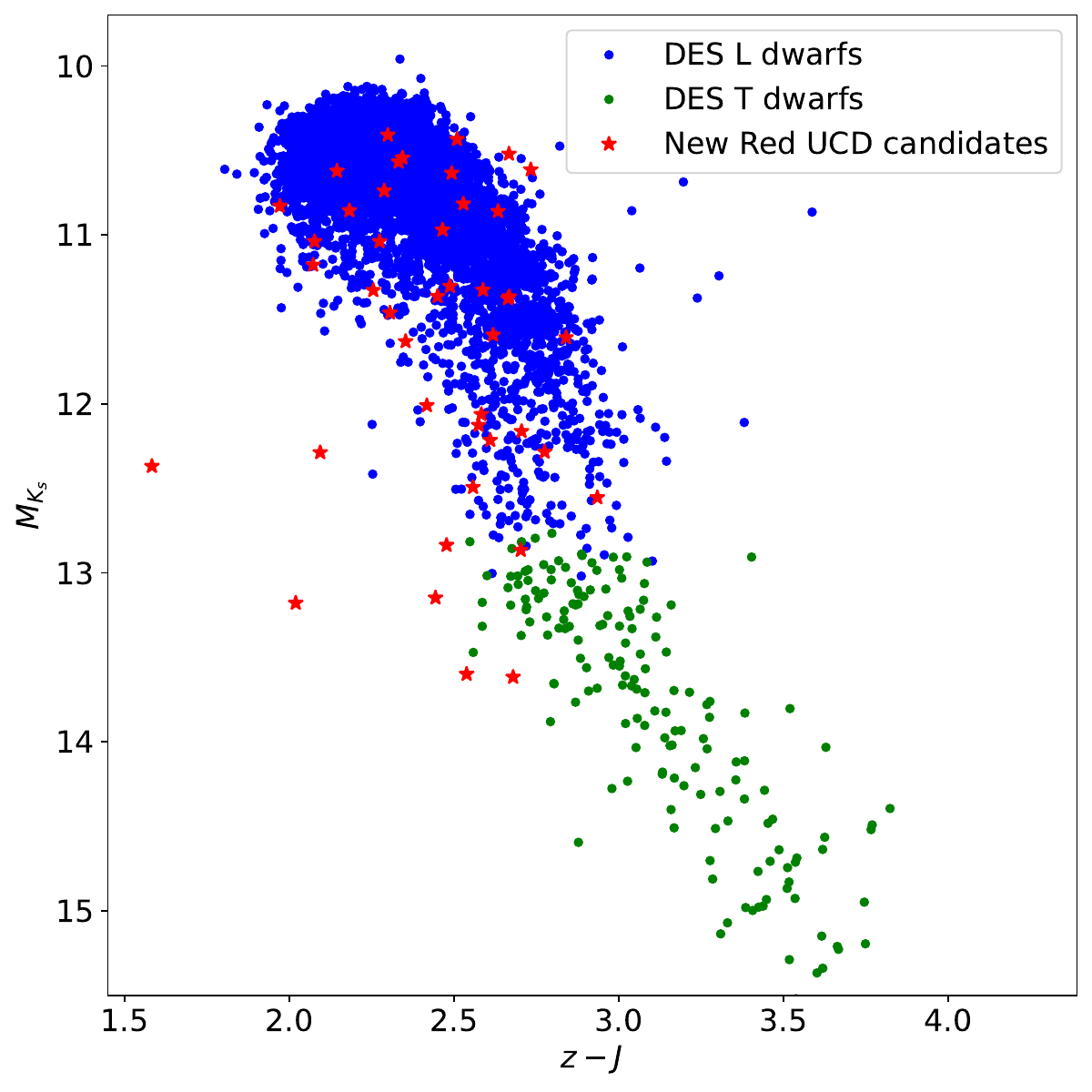}
    \caption{Colour vs. Absolute Magnitude diagrams incorporating DECaPS photometry. Red stars mark the new VIRAC2 UCD candidates with DECaPS counterparts. L dwarf candidates (blue points) and T dwarf candidates (green points) from DES \citep{carnero19} are also plotted. That study did not provide parallaxes so the $M_{K_s}$ values for the blue and green points are photometric estimates.}
    \label{fig:DECaPS}
  \end{center}
\end{figure}

We cross-matched the sources in Table \ref{tab:redUCDs} to the {\it Spitzer} Galactic survey archive tables and DECaPS in the same manner as for Table \ref{tab:50pc}. In the {\it Spitzer} surveys 21/56 candidates had a match and 18 of these were judged to be largely unaffected by blending of multiple VVV sources in the {\it Spitzer} beam, from inspection of the images. The three blended cases are source nos. 7, 18 and 21. A $[3.6]-[4.5]$ vs. $K_s - [3.6]$ two colour diagram is plotted in Figure \ref{fig:TCD} for the 14/18 sources with detections in both mid-infrared passbands. Within the uncertainties, these 14 sources have $K-[3.6]$ and $[3.6]-[4.5]$ colours that are consistent with those of L dwarfs, from comparison with the relevant data from the UltracoolSheet that is also plotted. These colour 
ranges overlap with those of late M dwarfs and early T dwarfs. The T dwarf candidate VVV~1253-6339 and the candidate low gravity late L dwarf VVV~1210-6227 are also plotted in Figure \ref{fig:TCD}. The red $[3.6]-[4.5]$ 
colour of  VVV~1253-6339 supports a T dwarf nature and the colours of VVV~1210-6227 support a late L type. In the latter case, there is insufficient {\it Spitzer} data available on low gravity late L dwarfs to judge how these colours are typically affected.

The DECaPS cross-match provided red optical photometry for 45 sources, see Figure \ref{fig:DECaPS}, though a few outlying data points (three in $i$ and 2 in $z$) were removed from the two panels after inspection of the images indicated unreliable blended detections or non-detections. We see that most of the red UCD candidates in this work have $i-J$ and $z-J$ colours and absolute magnitudes that follow the expected trends, as illustrated by previously published candidates from the Dark Energy Survey \citep[DES;][]{carnero19}. A few of the VIRAC2 candidates lie significantly below the expected trends, suggesting that a small number of the parallaxes may be over-estimated. We caution that the absolute magnitudes plotted for the DES sample are only photometric estimates based on likely spectral types inferred from
multi-filter photometry, not parallax measurements. Inspection of similar plots using data from the UltracoolSheet (which provides parallaxes and Pan-STARRS photometry in somewhat different $i$ and $z$ filters) suggests that the actual vertical spread in $M_{K_s}$ is likely to be larger than the blue and green data points suggest because the intrinsic spread of the UCD population is not fully captured.

Source no. 20 (VVV J170513.58-325803.3) is the most significant blue outlier in Figure \ref{fig:DECaPS}, with $i-J$ and $z-J$ colours inconsistent with a typical UCD. However the DECaPS image profiles show signs of possible blending or distortion and the VIRAC2 colours are typical of an L dwarf so we retain it as a candidate. Other outliers in Figure \ref{fig:DECaPS}, those with $M_{K_s}>13$, are discussed below.

While 49 of the 56 red UCD candidates have absolute magnitudes in the range of normal L dwarfs ($10.4<M_{K_s}<13$), seven sources ostensibly have lower luminosities ($13<M_{K_s}<15$), putting them in the T dwarf range. Unfortunately, none of these have reliable {\it Spitzer} photometry. We have mentioned two of the seven already: (i) VVV~1814-2654 ($M_{K_s}=13.6$) was discussed in section \ref{sec:HPM} as an early T dwarf candidate; (ii) LTT~7251~B (source no. 48 in Table \ref{tab:redUCDs}) was found in selection 3b. 
The VIRAC2 parallax of LTT~7251~B corresponds to $M_{K_s}=13.07$ but adopting the more precise Gaia parallax of the primary shifts the value to $M_{K_s}=12.83$, slightly over to the L dwarf side of the L/T boundary. LTT~7251~B was assigned a spectral type of L7 in \citet{LCS18}, where it was noted as a slightly under-luminous L dwarf, attributed to the mildly metal poor nature of the system \citep{casagrande11}. 

In these two cases the parallaxes have 8$\sigma{}$ significance, allowing us to have some confidence in the absolute magnitudes. Moreover, the high proper motion of VVV~J1814-2654, $\mu=505$~\masyr{}, translates to $V_{\rm tan}=66$~km~s$^{-1}$ at the nominal distance. If the source were a more distant L-type dwarf then its velocity would approach or exceed 100~km~s$^{-1}$, a rare occurrence for nearby members of the Thin Disc of the Milky Way.

A third system also has 8$\sigma{}$ parallax significance: source no. 25, VVV J172821.46-254333.2 (hereafter VVV~1728-2543) has
$M_{K_s}=13.15$, placing it near the early T/late L boundary. The VIRAC2 colours are all consistent with this interpretation.
VVV~1728-2543 has DECaPS photometry in the $i$, $z$ and $y$ filters. Inspection of the images shows no sign of a detection in $i$ (the catalogue entry was based on a single detection) but the counterparts in $z$ and $y$ were confirmed. Combining the DECaPS $z$ data and VIRAC2 $J$ data yields a colour $z-J=2.44 \pm 0.15$, which is slightly bluer than typical values for L/T transition objects but consistent within the error bars.

The parallaxes of the other four red sources with $M_{K_s}>13$ have only 5$\sigma$ to 6$\sigma$ significance. This means that a parallax over-estimate is more likely, especially given the bias imparted by our 5$\sigma$ parallax threshold: the error distribution will allow sources with over-estimated parallaxes to scatter into the sample, whilst sources with under-estimated parallaxes can scatter out.
We must therefore consider the possibility that they are in fact L dwarfs at slightly larger distances. These four sources are:

\begin{enumerate}
\item  VVV J175454.11-381356.3, hereafter VVV~1754-3813. With $M_{K_s}=13.18^{+0.34}_{-0.41}$, it lies at the L/T boundary, like VVV~1728-2543. At the nominal distance, $d \approx 40$~pc, the proper motion, $\mu = 104$~\masyr{}, corresponds to $V_{\rm tan} \approx 20$~km~s$^{-1}$. This source has a rather blue VIRAC2 $Z-J$ colour: $Z-J=2.14$, only just meeting the relaxed $Z-J=2$ threshold adopted in our earlier T dwarf search. Moreover, DECaPS detections
show that it is a clear outlier in Figure \ref{fig:DECaPS}, with 
$i-J=3.42$ and $z-J=2.02$. It therefore appears quite likely that this is an L-type or late M-type dwarf with over-estimated parallax, though an alternative possibility is that the blue colours are due to a metal poor nature \citep{zhang18}.

\item VVV J170533.90-424519.7, hereafter VVV~1705-4245. The absolute magnitude of $M_{K_s}=13.62^{+0.33}_{-0.40}$ suggests a spectral type in the T0--T4 range, though our red colour selection makes a T0--T2 type more likely. At the nominal distance, $d \approx 33$~pc, the proper motion, $\mu = 40$~\masyr{}, corresponds to $V_{\rm tan} \approx 6$~km~s$^{-1}$. DECaPS data show that it is a clear outlier in the upper panel of Figure \ref{fig:DECaPS} with $i-J=4.36$, though the colour $z-J=2.68$ is consistent with an L/T transition object. We suspect that this is a late L type dwarf with an over-estimated parallax.

\item VVV J175920.68-234045.3, hereafter VVV~1759-2340. The absolute magnitude of $M_{K_s}=13.65^{+0.34}_{-0.40}$ again suggests a spectral type in the T0--T4 range, though our red colour selection makes a T0--T2 type more likely. At the nominal distance, $d \approx 32$~pc, the proper motion, $\mu = 46$~\masyr{}, corresponds to $V_{\rm tan} \approx 7$~km~s$^{-1}$. The system is not detected in DECaPS.

\item VVV J180736.70-291700.1, hereafter VVV~1807-2917. The absolute magnitude of $M_{K_s}=14.62^{+0.37}_{-0.45}$ suggests a spectral type between T3.5 and T6.5. However, the low luminosity is at odds with the red colour selection, unless the system is a very low gravity L dwarf, as we suggested for VVV~J1210-6227 in section \ref{sec:50_pc_search}. The \textsc{BANYAN$\Sigma$} software tool finds a 99 per cent probability that this is a field object rather than a member of a young moving group, though as noted previously this is not a conclusive test of youth.
At the nominal distance, $d \approx 20$~pc, the proper motion, $\mu = 75$~\masyr{}, corresponds to $V_{\rm tan} \approx 7$~km~s$^{-1}$. The system lies outside the DECaPS footprint.
\end{enumerate}

All four of these sources are faint, $16.0<K_s<16.5$, hence the modest significance of the parallax solutions. Moreover, three of the four have small tangential velocities of 6 or 7~km~s$^{-1}$, if located at their nominal distances. These velocities are unremarkable individually but taken together they are smaller than is typical of L or T dwarfs in the local field \citep[e.g.][]{schmidt10, seifahrt10, burningham13}. This gives us reason to suspect that VVV~1705-4245, VVV~1759-2340 and VVV~1807-2917 are in fact L dwarfs with over-estimated parallaxes. In particular, VVV~1807-2917 has the largest parallax uncertainty in Table \ref{tab:redUCDs}, $\sigma_{\varpi}=9.4$~mas, approximately a factor of two larger than the other three candidates despite their similar brightness. This, coupled with the mismatch between colour and absolute magnitude, causes us to doubt the reliability of the solution even though the unit weight error, $\mathrm{UWE} = 1.06$, does not indicate a problem. 

While the 56 sources are not in GCNS, 13 have matches in {\it Gaia} DR3 within 0.5~arcsec, using VIRAC2 coordinates propagated to epoch 2016.0. Only two of these have five parameter solutions: (i) source no. 18 in Table \ref{tab:redUCDs} (VVV~J163823.65-484211.5) has {\it Gaia} parallax and proper motion in close agreement with VIRAC2; (ii) source no. 52 (VVV~J182959.78-272446.4) has a negative parallax in {\it Gaia} DR3, a much smaller proper motion and a relatively large reduced unit weight error ($\mathrm{ruwe} = 1.33$ in {\it Gaia}, compared to $\mathrm{UWE} = 1.03$ in VIRAC2). This disagreement may be due to the fact that towards the end of the VVV time series the source begins to be blended with a neighbouring star that is fainter in $K_s$ but brighter in $Z$ (and presumably brighter in the {\it Gaia} $G$ passband also). {\it Gaia} DR3 includes only one of the two sources and the coordinates are offset from the forward-propagated VIRAC2 coordinates by 0.39 arcsec, at an intermediate location. Therefore we speculate that the {\it Gaia} solution could be influenced by the neighbour and we retain source no. 52 as a candidate.

Pan-starrs1 DR2 and VPHAS$+$ DR2 \citep{drew14} provide red optical detections of a few additional sources outside the DECaPS footprint, at Galactic longitudes $l>5^{\circ}$. The 11 Pan-starrs1 matches (9 of which are not covered by DECaPS) all have red optical colours consistent with UCD status. VPHAS$+$ also has 11 matches, all of which have counterparts in DECaPS or Pan-starrs1.
Of these, 10/11 have very red $i-J$ colours: $3.28<i-J<4.39$ (Vega system), consistent with a UCD nature. The 11th, VVV~J131358.89-634426.4 appeared to be a bluer source with $i-J=2.59$
and $r-i=1.13$. However, inspection of the DECaPS images, taken several years after the VPHAS$+$ images, showed that this was due a blend of the UCD candidate and a bluer star that had begun to be resolved at the later epoch as the UCD moved away.

\subsection{Synergies with VIVACE}

Many types of variable star are prized for their well established relationships between period and luminosity. These relationships can be exploited to determine approximate distances to those stars. Combined with positions and proper motions, five dimensions of kinematic information are available, enabling the study of the dynamics of their various populations.

\citet{molnar22} developed an automated variable star classification pipeline for VIRAC2$\beta{}$ photometric time series data, thereby producing the VIrac VAriable Classiﬁcation Ensemble (VIVACE) catalogue of nearly 1.3 million variable stars. \citet{molnar22} discuss and demonstrate the science potential of the catalogue, which is outside the scope of this work.

VIVACE was produced using VIRAC2$\beta{}$ (see Appendix \ref{app:virac2b} for details), a preliminary version of the VIRAC2 catalogue based on a Gaia DR2 reference catalogue. The move to using Gaia DR3 for astrometric calibration meant a rerun of the main pipeline was necessary, and resulted in a low level of changes to the time series and an entirely new set of source IDs. It is therefore not quite as straightforward as it may seem to match between VIVACE and the VIRAC2 catalogue presented here. Variable stars in particular can be tricky astrometric targets in dense fields (for examples see Section \ref{failuremodes} and example 2 of Appendix \ref{sec:app_failures}).
To facilitate the exploitation of the combination of VIVACE with VIRAC2 astrometry, we felt it best to perform this matching ourselves using the additional information available in the (proprietary) VIRAC2$\beta{}$ catalogue.

To provide the cleanest possible VIRAC2 counterparts to VIVACE sources we started with a positional crossmatch to the main source table, requiring separations less than 100~mas, and that there be only one VIRAC2 match within this radius. At 100~mas there was a clear separation in the bimodal distribution of distances to all matches within 1~arcsec. This yielded potential matches for 1~338~664 VIVACE sources out of 1~364~732. We further required that the number of $K_s$ band detections in the VIRAC2 catalogue was no more than 5 per cent different to that of the VIRAC2$\beta{}$ catalogue, leaving 1~315~758 matches. The resultant mapping of VIVACE ID to VIRAC2 source ID is given in Table \ref{tab:vivace_xm} (full version available in on-line data).


\begin{table}
\centering
\caption{Crossmatches of VIVACE sources to VIRAC2 sources. The full table is available in the on-line data.}
\label{tab:vivace_xm}
\begin{tabular}{cc}
\hline
  \multicolumn{1}{c}{VIVACE ID} &
  \multicolumn{1}{c}{VIRAC2 source ID} \\
\hline
  0 & 14793959001490\\
  1 & 14797941003386\\
  2 & 14797942001475\\
  3 & 14793960001484\\
  4 & 14793960002043\\
  5 & 14793960001117\\
  6 & 14793960001577\\
  7 & 14789975003594\\
  8 & 14789975001112\\
  9 & 14793961001753\\
  \multicolumn{2}{c}{...}\\
\hline\end{tabular}
\end{table}

Of the $\approx{}$49~000 VIVACE sources without unambiguous counterparts in VIRAC2, 11.5\% have VIVACE variable type classification probability better than 0.9. This contrasts with 31.5\% in the complete VIVACE dataset.

\section{Summary}\label{summary}

We have undertaken a PSF fitting reduction of VVV and VVVX near-infrared images of $560~{\rm deg}^2$ of the southern Galactic plane and Galactic bulge. We described in detail our complete pipeline, from CASU reduced images through source detection, astrometric and photometric calibration, and time series production. Using various quality control criteria we further reduced the raw 1.4 billion row output down to a more reliable catalogue. The resultant VIRAC2 catalogue we present provides 5-parameter mean and time series astrometry, mean and time series photometry, and variability indices for 545\,346\,537 unique sources. Equivalent data are also provided for several hundred million additional \textit{reject} sources. This selection does contain some interesting real sources (e.g. transients), but is highly contaminated with duplicated sources and other erroneous data and so we must stress that it should only be used with a high degree of caution. All tables are available from the ESO archive at \url{https://archive.eso.org}.

Peak astrometric performance was achieved in the $11<K_s~{\rm mag}<14$ range, where proper motion uncertainties are typically $\approx{}0.37~{\rm mas~yr}^{-1}$ per dimension and parallax uncertainties are typically around $1~{\rm mas}$. At $K_s = 16$, where the catalogue is still typically 90\% complete, astrometric uncertainties are around $1.5~{\rm mas~yr}^{-1}$ per dimension for proper motion and $5~{\rm mas}$ for parallax.
VIRAC2 astrometric uncertainties were checked against Gaia DR3, and externally against HST measurements, and found to be valid.

We performed an initial search of the VIRAC2 catalogue for nearby sources with significant parallaxes, thereby demonstrating the use of the various included quality control parameters for selecting high quality candidates. This search led to the identification of a number of new  candidates in crowded Galactic star fields, including several projected in the inner Galactic bulge, where the census of nearby sources is much less complete than elsewhere. These discoveries include two new T dwarfs that are identified with high confidence and several other sources with redder colours that appear to lie close to L/T transition. In total, 26 new sources were discovered at likely distance $d<50$~pc, including a T dwarf at $d\sim15$~pc and two sources with relatively blue optical colours that are of uncertain nature. Further searches for nearby sources in VIRAC2 can be undertaken, including searches for fainter brown dwarfs and white dwarfs near the sensitivity limit that would rely more on proper motion than parallax.


VIRAC2 covers a region of the Milky Way in which Gaia, the Rubin observatory LSST and other optical surveys are essentially blind, and hence complements them superbly. It might also provide a useful early epoch for potential future Roman Space Telescope, JASMINE and GaiaNIR proper motion surveys, thereby retaining value for many years to come.

\section*{Acknowledgements}

We thank the anonymous referee for their careful reading and comments.

We gratefully acknowledge the use of data from the ESO Public Survey program IDs 179.B-2002 and 198.B-2004 taken with the VISTA telescope and data products from the Cambridge Astronomical Survey Unit (CASU) and the VISTA Science Archive (VSA) and the ESO Science Archive. VVV and VVVX data are published in the ESO Science Archive in the data collections identified by the following DOIs: https://doi.eso.org/10.18727/archive/67 and https://doi.eso.org/10.18727/archive/68.

This work has made use of the University of Hertfordshire's high-performance computing facility, we are grateful to its architects and maintainers.

This paper made use of the Whole Sky Database (wsdb) created and maintained by Sergey Koposov at the Institute of Astronomy, Cambridge with financial support from the Science \& Technology Facilities Council (STFC) and the European Research Council (ERC). It also made use of the Q3C software \citep{q3c} and the SIMBAD database \citep{wenger00}, operated at CDS, Strasbourg, France.
This work has benefited from The UltracoolSheet at http://bit.ly/UltracoolSheet, maintained by Will Best, Trent Dupuy, Michael Liu, Aniket Sanghi, Rob Siverd, and Zhoujian Zhang, and developed from compilations by \citet{dupuy12, dupuy13, deacon14, liu16, best18, best21, sanghi23} and \citet{schneider23}. We also used services and data provided by the Astro Data Lab \citep{fitzpatrick14,nikutta20}, which is part of the Community Science and Data Center (CSDC) Program of NSF NOIRLab. NOIRLab is operated by the Association of Universities for Research in Astronomy (AURA), Inc. under a cooperative agreement with the U.S. National Science Foundation.

L.C.S. acknowledges support from UKRI-STFC grants ST/X001628/1 and ST/X001571/1.

P.W.L. acknowledges support by STFC grant ST/Y000846/1.

S.E.K. acknowledges support from the Science \& Technology Facilities Council (STFC) grant ST/Y001001/1.

J.A.-G. acknowledges support from Fondecyt Regular 1201490 and  by ANID – Millennium Science Initiative Program – ICN12\_009 awarded to the Millennium Institute of Astrophysics MAS.

D.M. gratefully acknowledges support from the Center for Astrophysics and Associated Technologies CATA by the ANID BASAL projects ACE210002 and FB210003, by Fondecyt Project No. 1220724.

J.L.S. acknowledges support from the Royal Society (URF\textbackslash R1\textbackslash191555).

RK acknowledges partial support from ANID’s FONDECYT Regular grant \#1240249 and ANID’s Millennium Science Initiative through grants ICN12\_009 and AIM23-0001, awarded to the Millennium Institute of Astrophysics (MAS).

R.K.S. acknowledges support from CNPq/Brazil through projects 308298/2022-5 and 421034/2023-8.


\section*{Data Availability}

The VIRAC2 catalogue and the underlying image data are available from the ESO archive at \url{https://archive.eso.org}. 
The cross match of VIRAC2 to VIVACE is available from the journal online.



\bibliographystyle{mnras}
\bibliography{refs}

\begin{thebibliography}{}
\makeatletter
\relax
\def\mn@urlcharsother{\let\do\@makeother \do\$\do\&\do\#\do\^\do\_\do\%\do\~}
\def\mn@doi{\begingroup\mn@urlcharsother \@ifnextchar [ {\mn@doi@}
  {\mn@doi@[]}}
\def\mn@doi@[#1]#2{\def\@tempa{#1}\ifx\@tempa\@empty \href
  {http://dx.doi.org/#2} {doi:#2}\else \href {http://dx.doi.org/#2} {#1}\fi
  \endgroup}
\def\mn@eprint#1#2{\mn@eprint@#1:#2::\@nil}
\def\mn@eprint@arXiv#1{\href {http://arxiv.org/abs/#1} {{\tt arXiv:#1}}}
\def\mn@eprint@dblp#1{\href {http://dblp.uni-trier.de/rec/bibtex/#1.xml}
  {dblp:#1}}
\def\mn@eprint@#1:#2:#3:#4\@nil{\def\@tempa {#1}\def\@tempb {#2}\def\@tempc
  {#3}\ifx \@tempc \@empty \let \@tempc \@tempb \let \@tempb \@tempa \fi \ifx
  \@tempb \@empty \def\@tempb {arXiv}\fi \@ifundefined
  {mn@eprint@\@tempb}{\@tempb:\@tempc}{\expandafter \expandafter \csname
  mn@eprint@\@tempb\endcsname \expandafter{\@tempc}}}

\bibitem[\protect\citeauthoryear{{Alonso-Garc{\'\i}a}, {Mateo}, {Sen},
  {Banerjee}, {Catelan}, {Minniti}  \& {von Braun}}{{Alonso-Garc{\'\i}a}
  et~al.}{2012}]{dophot2}
{Alonso-Garc{\'\i}a} J.,  {Mateo} M.,  {Sen} B.,  {Banerjee} M.,  {Catelan} M.,
   {Minniti} D.,   {von Braun} K.,  2012, \mn@doi [\aj]
  {10.1088/0004-6256/143/3/70}, \href
  {https://ui.adsabs.harvard.edu/abs/2012AJ....143...70A} {143, 70}

\bibitem[\protect\citeauthoryear{{Alonso-Garc{\'\i}a}
  et~al.,}{{Alonso-Garc{\'\i}a} et~al.}{2018}]{AG18}
{Alonso-Garc{\'\i}a} J.,  et~al., 2018, \mn@doi [\aap]
  {10.1051/0004-6361/201833432}, \href
  {https://ui.adsabs.harvard.edu/abs/2018A&A...619A...4A} {619, A4}

\bibitem[\protect\citeauthoryear{{Alonso-Garc{\'\i}a}
  et~al.,}{{Alonso-Garc{\'\i}a} et~al.}{2021}]{agarcia21}
{Alonso-Garc{\'\i}a} J.,  et~al., 2021, \mn@doi [\aap]
  {10.1051/0004-6361/202140546}, \href
  {https://ui.adsabs.harvard.edu/abs/2021A&A...651A..47A} {651, A47}

\bibitem[\protect\citeauthoryear{{Beam{\'\i}n} et~al.,}{{Beam{\'\i}n}
  et~al.}{2013}]{beamin13}
{Beam{\'\i}n} J.~C.,  et~al., 2013, \mn@doi [\aap]
  {10.1051/0004-6361/201322190}, \href
  {https://ui.adsabs.harvard.edu/abs/2013A&A...557L...8B} {557, L8}

\bibitem[\protect\citeauthoryear{{Benjamin} et~al.,}{{Benjamin}
  et~al.}{2003}]{benjamin03}
{Benjamin} R.~A.,  et~al., 2003, \mn@doi [\pasp] {10.1086/376696}, \href
  {https://ui.adsabs.harvard.edu/abs/2003PASP..115..953B} {115, 953}

\bibitem[\protect\citeauthoryear{{Benjamin}, {Babler}, {Churchwell},
  {Clarkson}, {Kirkpatrick}, {Meade}  \& {Whitney}}{{Benjamin}
  et~al.}{2015}]{benjamin15}
{Benjamin} R.,  {Babler} B.,  {Churchwell} E.,  {Clarkson} W.,  {Kirkpatrick}
  D.,  {Meade} M.,   {Whitney} B.,  2015, {GLIMPSE Proper: Mid-Infrared
  Observations of Proper Motion and Variability Towards Galactic Center},
  Spitzer Proposal ID 12023

\bibitem[\protect\citeauthoryear{{Benjamin}, {Babler}, {D'Onghia}, {Clarkson},
  {Churchwell}, {Kirkpatrick}, {Zasowski}  \& {Majewski}}{{Benjamin}
  et~al.}{2016}]{benjamin16}
{Benjamin} R.,  {Babler} B.,  {D'Onghia} E.,  {Clarkson} W.,  {Churchwell} E.,
  {Kirkpatrick} D.,  {Zasowski} G.,   {Majewski} S.,  2016, {Three Dimensional
  Stellar Kinematics of the Galactic Bar and Disk: Where APOGEE Meets GLIMPSE},
  Spitzer Proposal ID 13117

\bibitem[\protect\citeauthoryear{{Best} et~al.,}{{Best} et~al.}{2018}]{best18}
{Best} W. M.~J.,  et~al., 2018, \mn@doi [\apjs] {10.3847/1538-4365/aa9982},
  \href {https://ui.adsabs.harvard.edu/abs/2018ApJS..234....1B} {234, 1}

\bibitem[\protect\citeauthoryear{{Best}, {Liu}, {Magnier}  \& {Dupuy}}{{Best}
  et~al.}{2021}]{best21}
{Best} W. M.~J.,  {Liu} M.~C.,  {Magnier} E.~A.,   {Dupuy} T.~J.,  2021,
  \mn@doi [\aj] {10.3847/1538-3881/abc893}, \href
  {https://ui.adsabs.harvard.edu/abs/2021AJ....161...42B} {161, 42}

\bibitem[\protect\citeauthoryear{{Burgasser}, {Reid}, {Siegler}, {Close},
  {Allen}, {Lowrance}  \& {Gizis}}{{Burgasser} et~al.}{2007}]{burgasser07}
{Burgasser} A.~J.,  {Reid} I.~N.,  {Siegler} N.,  {Close} L.,  {Allen} P.,
  {Lowrance} P.,   {Gizis} J.,  2007, in {Reipurth} B.,  {Jewitt} D.,   {Keil}
  K.,  eds, Protostars and Planets V. p.~427 (\mn@eprint {arXiv}
  {astro-ph/0602122}), \mn@doi{10.48550/arXiv.astro-ph/0602122}

\bibitem[\protect\citeauthoryear{{Burningham} et~al.,}{{Burningham}
  et~al.}{2011}]{burningham11}
{Burningham} B.,  et~al., 2011, \mn@doi [\mnras]
  {10.1111/j.1745-3933.2011.01062.x}, \href
  {https://ui.adsabs.harvard.edu/abs/2011MNRAS.414L..90B} {414, L90}

\bibitem[\protect\citeauthoryear{{Burningham} et~al.,}{{Burningham}
  et~al.}{2013}]{burningham13}
{Burningham} B.,  et~al., 2013, \mn@doi [\mnras] {10.1093/mnras/stt740}, \href
  {https://ui.adsabs.harvard.edu/abs/2013MNRAS.433..457B} {433, 457}

\bibitem[\protect\citeauthoryear{{Carnero Rosell} et~al.,}{{Carnero Rosell}
  et~al.}{2019}]{carnero19}
{Carnero Rosell} A.,  et~al., 2019, \mn@doi [\mnras] {10.1093/mnras/stz2398},
  \href {https://ui.adsabs.harvard.edu/abs/2019MNRAS.489.5301C} {489, 5301}

\bibitem[\protect\citeauthoryear{{Casagrande}, {Sch{\"o}nrich}, {Asplund},
  {Cassisi}, {Ram{\'\i}rez}, {Mel{\'e}ndez}, {Bensby}  \&
  {Feltzing}}{{Casagrande} et~al.}{2011}]{casagrande11}
{Casagrande} L.,  {Sch{\"o}nrich} R.,  {Asplund} M.,  {Cassisi} S.,
  {Ram{\'\i}rez} I.,  {Mel{\'e}ndez} J.,  {Bensby} T.,   {Feltzing} S.,  2011,
  \mn@doi [\aap] {10.1051/0004-6361/201016276}, \href
  {https://ui.adsabs.harvard.edu/abs/2011A&A...530A.138C} {530, A138}

\bibitem[\protect\citeauthoryear{{Caselden}, {Colin}, {Lack}, {Marocco},
  {Kirkpatrick}  \& {Meisner}}{{Caselden} et~al.}{2020}]{caselden20}
{Caselden} D.,  {Colin} G.,  {Lack} L.,  {Marocco} F.,  {Kirkpatrick} J.,
  {Meisner} A.,  2020, in American Astronomical Society Meeting Abstracts
  \#235. p. 274.18

\bibitem[\protect\citeauthoryear{{Chambers} et~al.,}{{Chambers}
  et~al.}{2016}]{chambers16}
{Chambers} K.~C.,  et~al., 2016, \mn@doi [arXiv e-prints]
  {10.48550/arXiv.1612.05560}, \href
  {https://ui.adsabs.harvard.edu/abs/2016arXiv161205560C} {p. arXiv:1612.05560}

\bibitem[\protect\citeauthoryear{{Chauvin}, {Lagrange}, {Dumas}, {Zuckerman},
  {Mouillet}, {Song}, {Beuzit}  \& {Lowrance}}{{Chauvin}
  et~al.}{2004}]{chauvin04}
{Chauvin} G.,  {Lagrange} A.~M.,  {Dumas} C.,  {Zuckerman} B.,  {Mouillet} D.,
  {Song} I.,  {Beuzit} J.~L.,   {Lowrance} P.,  2004, \mn@doi [\aap]
  {10.1051/0004-6361:200400056}, \href
  {https://ui.adsabs.harvard.edu/abs/2004A&A...425L..29C} {425, L29}

\bibitem[\protect\citeauthoryear{{Churchwell} et~al.,}{{Churchwell}
  et~al.}{2009}]{churchwell09}
{Churchwell} E.,  et~al., 2009, \mn@doi [\pasp] {10.1086/597811}, \href
  {https://ui.adsabs.harvard.edu/abs/2009PASP..121..213C} {121, 213}

\bibitem[\protect\citeauthoryear{{Clarke}, {Wegg}, {Gerhard}, {Smith}, {Lucas}
  \& {Wylie}}{{Clarke} et~al.}{2019}]{clarke19}
{Clarke} J.~P.,  {Wegg} C.,  {Gerhard} O.,  {Smith} L.~C.,  {Lucas} P.~W.,
  {Wylie} S.~M.,  2019, \mn@doi [\mnras] {10.1093/mnras/stz2382}, \href
  {https://ui.adsabs.harvard.edu/abs/2019MNRAS.489.3519C} {489, 3519}

\bibitem[\protect\citeauthoryear{{Contreras Ramos} et~al.,}{{Contreras Ramos}
  et~al.}{2017}]{CR17}
{Contreras Ramos} R.,  et~al., 2017, \mn@doi [\aap]
  {10.1051/0004-6361/201731462}, \href
  {http://adsabs.harvard.edu/abs/2017A%26A...608A.140C} {608, A140}

\bibitem[\protect\citeauthoryear{{Deacon} et~al.,}{{Deacon}
  et~al.}{2014}]{deacon14}
{Deacon} N.~R.,  et~al., 2014, \mn@doi [\apj] {10.1088/0004-637X/792/2/119},
  \href {https://ui.adsabs.harvard.edu/abs/2014ApJ...792..119D} {792, 119}

\bibitem[\protect\citeauthoryear{{Drew} et~al.,}{{Drew} et~al.}{2014}]{drew14}
{Drew} J.~E.,  et~al., 2014, \mn@doi [\mnras] {10.1093/mnras/stu394}, \href
  {https://ui.adsabs.harvard.edu/abs/2014MNRAS.440.2036D} {440, 2036}

\bibitem[\protect\citeauthoryear{{Ducourant}, {Teixeira}, {Chauvin}, {Daigne},
  {Le Campion}, {Song}  \& {Zuckerman}}{{Ducourant} et~al.}{2008}]{ducourant08}
{Ducourant} C.,  {Teixeira} R.,  {Chauvin} G.,  {Daigne} G.,  {Le Campion}
  J.~F.,  {Song} I.,   {Zuckerman} B.,  2008, \mn@doi [\aap]
  {10.1051/0004-6361:20078886}, \href
  {https://ui.adsabs.harvard.edu/abs/2008A&A...477L...1D} {477, L1}

\bibitem[\protect\citeauthoryear{{Dupuy} \& {Kraus}}{{Dupuy} \&
  {Kraus}}{2013}]{dupuy13}
{Dupuy} T.~J.,  {Kraus} A.~L.,  2013, \mn@doi [Science]
  {10.1126/science.1241917}, \href
  {https://ui.adsabs.harvard.edu/abs/2013Sci...341.1492D} {341, 1492}

\bibitem[\protect\citeauthoryear{{Dupuy} \& {Liu}}{{Dupuy} \&
  {Liu}}{2012}]{dupuy12}
{Dupuy} T.~J.,  {Liu} M.~C.,  2012, \mn@doi [\apjs]
  {10.1088/0067-0049/201/2/19}, \href
  {https://ui.adsabs.harvard.edu/abs/2012ApJS..201...19D} {201, 19}

\bibitem[\protect\citeauthoryear{{El-Badry}, {Rix}  \& {Heintz}}{{El-Badry}
  et~al.}{2021}]{elbadry21}
{El-Badry} K.,  {Rix} H.-W.,   {Heintz} T.~M.,  2021, \mn@doi [\mnras]
  {10.1093/mnras/stab323}, \href
  {https://ui.adsabs.harvard.edu/abs/2021MNRAS.506.2269E} {506, 2269}

\bibitem[\protect\citeauthoryear{{Faherty} et~al.,}{{Faherty}
  et~al.}{2016}]{faherty16}
{Faherty} J.~K.,  et~al., 2016, \mn@doi [\apjs] {10.3847/0067-0049/225/1/10},
  \href {https://ui.adsabs.harvard.edu/abs/2016ApJS..225...10F} {225, 10}

\bibitem[\protect\citeauthoryear{{Fitzpatrick} et~al.,}{{Fitzpatrick}
  et~al.}{2014}]{fitzpatrick14}
{Fitzpatrick} M.~J.,  et~al., 2014, in {Peck} A.~B.,  {Benn} C.~R.,   {Seaman}
  R.~L.,  eds,  Society of Photo-Optical Instrumentation Engineers (SPIE)
  Conference Series Vol. 9149, Observatory Operations: Strategies, Processes,
  and Systems V. p. 91491T, \mn@doi{10.1117/12.2057445}

\bibitem[\protect\citeauthoryear{{Flewelling} et~al.,}{{Flewelling}
  et~al.}{2020}]{flewelling20}
{Flewelling} H.~A.,  et~al., 2020, \mn@doi [\apjs] {10.3847/1538-4365/abb82d},
  \href {https://ui.adsabs.harvard.edu/abs/2020ApJS..251....7F} {251, 7}

\bibitem[\protect\citeauthoryear{{Folkes}, {Pinfield}, {Kendall}  \&
  {Jones}}{{Folkes} et~al.}{2007}]{folkes07}
{Folkes} S.~L.,  {Pinfield} D.~J.,  {Kendall} T.~R.,   {Jones} H.~R.~A.,  2007,
  \mn@doi [\mnras] {10.1111/j.1365-2966.2007.11789.x}, \href
  {https://ui.adsabs.harvard.edu/abs/2007MNRAS.378..901F} {378, 901}

\bibitem[\protect\citeauthoryear{{Folkes} et~al.,}{{Folkes}
  et~al.}{2012}]{folkes12}
{Folkes} S.~L.,  et~al., 2012, \mn@doi [\mnras]
  {10.1111/j.1365-2966.2012.21132.x}, \href
  {https://ui.adsabs.harvard.edu/abs/2012MNRAS.427.3280F} {427, 3280}

\bibitem[\protect\citeauthoryear{{Gagn{\'e}} et~al.,}{{Gagn{\'e}}
  et~al.}{2018}]{gagne18}
{Gagn{\'e}} J.,  et~al., 2018, \mn@doi [\apj] {10.3847/1538-4357/aaae09}, \href
  {https://ui.adsabs.harvard.edu/abs/2018ApJ...856...23G} {856, 23}

\bibitem[\protect\citeauthoryear{{Gentile Fusillo} et~al.,}{{Gentile Fusillo}
  et~al.}{2021}]{gentile-fusillo21}
{Gentile Fusillo} N.~P.,  et~al., 2021, \mn@doi [\mnras]
  {10.1093/mnras/stab2672}, \href
  {https://ui.adsabs.harvard.edu/abs/2021MNRAS.508.3877G} {508, 3877}

\bibitem[\protect\citeauthoryear{{Golovin}, {Reffert}, {Vani}, {Bastian},
  {Jordan}  \& {Just}}{{Golovin} et~al.}{2024}]{golovin24}
{Golovin} A.,  {Reffert} S.,  {Vani} A.,  {Bastian} U.,  {Jordan} S.,   {Just}
  A.,  2024, \mn@doi [\aap] {10.1051/0004-6361/202347767}, \href
  {https://ui.adsabs.harvard.edu/abs/2024A&A...683A..33G} {683, A33}

\bibitem[\protect\citeauthoryear{{Gonz{\'a}lez-Fern{\'a}ndez}
  et~al.,}{{Gonz{\'a}lez-Fern{\'a}ndez} et~al.}{2018}]{GF18}
{Gonz{\'a}lez-Fern{\'a}ndez} C.,  et~al., 2018, \mn@doi [\mnras]
  {10.1093/mnras/stx3073}, \href
  {http://adsabs.harvard.edu/abs/2018MNRAS.474.5459G} {474, 5459}

\bibitem[\protect\citeauthoryear{{Griffith} et~al.,}{{Griffith}
  et~al.}{2012}]{griffith12}
{Griffith} R.~L.,  et~al., 2012, \mn@doi [\aj] {10.1088/0004-6256/144/5/148},
  \href {https://ui.adsabs.harvard.edu/abs/2012AJ....144..148G} {144, 148}

\bibitem[\protect\citeauthoryear{{Griggio}, {Libralato}, {Bellini}, {Bedin},
  {Anderson}, {Smith}  \& {Minniti}}{{Griggio} et~al.}{2024}]{griggio24}
{Griggio} M.,  {Libralato} M.,  {Bellini} A.,  {Bedin} L.~R.,  {Anderson} J.,
  {Smith} L.~C.,   {Minniti} D.,  2024, \mn@doi [\aap]
  {10.1051/0004-6361/202449560}, \href
  {https://ui.adsabs.harvard.edu/abs/2024A&A...687A..94G} {687, A94}

\bibitem[\protect\citeauthoryear{{Hewett}, {Warren}, {Leggett}  \&
  {Hodgkin}}{{Hewett} et~al.}{2006}]{hewett06}
{Hewett} P.~C.,  {Warren} S.~J.,  {Leggett} S.~K.,   {Hodgkin} S.~T.,  2006,
  \mn@doi [\mnras] {10.1111/j.1365-2966.2005.09969.x}, \href
  {https://ui.adsabs.harvard.edu/abs/2006MNRAS.367..454H} {367, 454}

\bibitem[\protect\citeauthoryear{{Husseiniova}, {McGill}, {Smith}  \&
  {Evans}}{{Husseiniova} et~al.}{2021}]{husseiniova21}
{Husseiniova} A.,  {McGill} P.,  {Smith} L.~C.,   {Evans} N.~W.,  2021, \mn@doi
  [\mnras] {10.1093/mnras/stab1882}, \href
  {https://ui.adsabs.harvard.edu/abs/2021MNRAS.506.2482H} {506, 2482}

\bibitem[\protect\citeauthoryear{{Irwin}}{{Irwin}}{1985}]{irwin85}
{Irwin} M.~J.,  1985, \mn@doi [\mnras] {10.1093/mnras/214.4.575}, \href
  {https://ui.adsabs.harvard.edu/abs/1985MNRAS.214..575I} {214, 575}

\bibitem[\protect\citeauthoryear{{Irwin} et~al.,}{{Irwin} et~al.}{2004}]{vdfs}
{Irwin} M.~J.,  et~al., 2004, {VISTA data flow system: pipeline processing for
  WFCAM and VISTA}.
pp 411--422, \mn@doi{10.1117/12.551449}

\bibitem[\protect\citeauthoryear{{Ivanov} et~al.,}{{Ivanov}
  et~al.}{2013}]{ivanov13}
{Ivanov} V.~D.,  et~al., 2013, \mn@doi [\aap] {10.1051/0004-6361/201321045},
  \href {https://ui.adsabs.harvard.edu/abs/2013A&A...560A..21I} {560, A21}

\bibitem[\protect\citeauthoryear{{Kaczmarek}, {McGill}, {Evans}, {Smith},
  {Wyrzykowski}, {Howil}  \& {Jab{\l}o{\'n}ska}}{{Kaczmarek}
  et~al.}{2022}]{kaczmarek22}
{Kaczmarek} Z.,  {McGill} P.,  {Evans} N.~W.,  {Smith} L.~C.,  {Wyrzykowski}
  {\L}.,  {Howil} K.,   {Jab{\l}o{\'n}ska} M.,  2022, \mn@doi [\mnras]
  {10.1093/mnras/stac1507}, \href
  {https://ui.adsabs.harvard.edu/abs/2022MNRAS.514.4845K} {514, 4845}

\bibitem[\protect\citeauthoryear{{Kaczmarek}, {McGill}, {Evans}, {Smith},
  {Golovich}, {Kerins}, {Specht}  \& {Dawson}}{{Kaczmarek}
  et~al.}{2024}]{Kaczmarek24}
{Kaczmarek} Z.,  {McGill} P.,  {Evans} N.~W.,  {Smith} L.~C.,  {Golovich} N.,
  {Kerins} E.,  {Specht} D.,   {Dawson} W.~A.,  2024, \mn@doi [\mnras]
  {10.1093/mnras/stae445}, \href
  {https://ui.adsabs.harvard.edu/abs/2024MNRAS.529.1308K} {529, 1308}

\bibitem[\protect\citeauthoryear{{Kirkpatrick} et~al.,}{{Kirkpatrick}
  et~al.}{2011}]{kirkpatrick11}
{Kirkpatrick} J.~D.,  et~al., 2011, \mn@doi [\apjs]
  {10.1088/0067-0049/197/2/19}, \href
  {https://ui.adsabs.harvard.edu/abs/2011ApJS..197...19K} {197, 19}

\bibitem[\protect\citeauthoryear{{Kirkpatrick} et~al.,}{{Kirkpatrick}
  et~al.}{2019}]{kirkpatrick19}
{Kirkpatrick} J.~D.,  et~al., 2019, \mn@doi [\apjs] {10.3847/1538-4365/aaf6af},
  \href {https://ui.adsabs.harvard.edu/abs/2019ApJS..240...19K} {240, 19}

\bibitem[\protect\citeauthoryear{{Kirkpatrick} et~al.,}{{Kirkpatrick}
  et~al.}{2021}]{kirkpatrick21}
{Kirkpatrick} J.~D.,  et~al., 2021, \mn@doi [\apjs] {10.3847/1538-4365/abd107},
  \href {https://ui.adsabs.harvard.edu/abs/2021ApJS..253....7K} {253, 7}

\bibitem[\protect\citeauthoryear{{Kirkpatrick} et~al.,}{{Kirkpatrick}
  et~al.}{2024}]{kirkpatrick24}
{Kirkpatrick} J.~D.,  et~al., 2024, \mn@doi [\apjs] {10.3847/1538-4365/ad24e2},
  \href {https://ui.adsabs.harvard.edu/abs/2024ApJS..271...55K} {271, 55}

\bibitem[\protect\citeauthoryear{{Kl{\"u}ter}, {Bastian}, {Demleitner}  \&
  {Wambsganss}}{{Kl{\"u}ter} et~al.}{2018}]{kluter18}
{Kl{\"u}ter} J.,  {Bastian} U.,  {Demleitner} M.,   {Wambsganss} J.,  2018,
  \mn@doi [\aap] {10.1051/0004-6361/201833978}, \href
  {https://ui.adsabs.harvard.edu/abs/2018A&A...620A.175K} {620, A175}

\bibitem[\protect\citeauthoryear{{Koposov} \& {Bartunov}}{{Koposov} \&
  {Bartunov}}{2006}]{q3c}
{Koposov} S.,  {Bartunov} O.,  2006, in {Gabriel} C.,  {Arviset} C.,  {Ponz}
  D.,   {Enrique} S.,  eds,  Astronomical Society of the Pacific Conference
  Series Vol. 351, Astronomical Data Analysis Software and Systems XV. p.~735

\bibitem[\protect\citeauthoryear{{Kurtev} et~al.,}{{Kurtev}
  et~al.}{2017}]{kurtev17}
{Kurtev} R.,  et~al., 2017, \mn@doi [\mnras] {10.1093/mnras/stw2357}, \href
  {https://ui.adsabs.harvard.edu/abs/2017MNRAS.464.1247K} {464, 1247}

\bibitem[\protect\citeauthoryear{{L{\'e}pine}}{{L{\'e}pine}}{2005}]{lepine05}
{L{\'e}pine} S.,  2005, \mn@doi [\aj] {10.1086/432161}, \href
  {https://ui.adsabs.harvard.edu/abs/2005AJ....130.1247L} {130, 1247}

\bibitem[\protect\citeauthoryear{{L{\'e}pine}}{{L{\'e}pine}}{2008}]{lepine08}
{L{\'e}pine} S.,  2008, \mn@doi [\aj] {10.1088/0004-6256/135/6/2177}, \href
  {https://ui.adsabs.harvard.edu/abs/2008AJ....135.2177L} {135, 2177}

\bibitem[\protect\citeauthoryear{{Lewis}, {Irwin}  \& {Bunclark}}{{Lewis}
  et~al.}{2010}]{lewis10}
{Lewis} J.~R.,  {Irwin} M.,   {Bunclark} P.,  2010, {Pipeline Processing for
  VISTA}.
p.~91

\bibitem[\protect\citeauthoryear{{Libralato} et~al.,}{{Libralato}
  et~al.}{2015}]{Libralato15}
{Libralato} M.,  et~al., 2015, \mn@doi [\mnras] {10.1093/mnras/stv674}, \href
  {http://adsabs.harvard.edu/abs/2015MNRAS.450.1664L} {450, 1664}

\bibitem[\protect\citeauthoryear{{Lindegren} et~al.,}{{Lindegren}
  et~al.}{2018}]{Lindegren18}
{Lindegren} L.,  et~al., 2018, \mn@doi [\aap] {10.1051/0004-6361/201832727},
  \href {http://adsabs.harvard.edu/abs/2018A%26A...616A...2L} {616, A2}

\bibitem[\protect\citeauthoryear{{Lindegren} et~al.,}{{Lindegren}
  et~al.}{2021}]{lindegren21}
{Lindegren} L.,  et~al., 2021, \mn@doi [\aap] {10.1051/0004-6361/202039709},
  \href {https://ui.adsabs.harvard.edu/abs/2021A&A...649A...2L} {649, A2}

\bibitem[\protect\citeauthoryear{{Liu}, {Dupuy}  \& {Allers}}{{Liu}
  et~al.}{2016}]{liu16}
{Liu} M.~C.,  {Dupuy} T.~J.,   {Allers} K.~N.,  2016, \mn@doi [\apj]
  {10.3847/1538-4357/833/1/96}, \href
  {https://ui.adsabs.harvard.edu/abs/2016ApJ...833...96L} {833, 96}

\bibitem[\protect\citeauthoryear{{Looper}, {Kirkpatrick}  \&
  {Burgasser}}{{Looper} et~al.}{2007}]{looper07}
{Looper} D.~L.,  {Kirkpatrick} J.~D.,   {Burgasser} A.~J.,  2007, \mn@doi [\aj]
  {10.1086/520645}, \href
  {https://ui.adsabs.harvard.edu/abs/2007AJ....134.1162L} {134, 1162}

\bibitem[\protect\citeauthoryear{{Lucas} et~al.,}{{Lucas}
  et~al.}{2010}]{lucas10}
{Lucas} P.~W.,  et~al., 2010, \mn@doi [\mnras]
  {10.1111/j.1745-3933.2010.00927.x}, \href
  {https://ui.adsabs.harvard.edu/abs/2010MNRAS.408L..56L} {408, L56}

\bibitem[\protect\citeauthoryear{{Lucas} et~al.,}{{Lucas}
  et~al.}{2024}]{lucas24}
{Lucas} P.~W.,  et~al., 2024, \mn@doi [\mnras] {10.1093/mnras/stad3929}, \href
  {https://ui.adsabs.harvard.edu/abs/2024MNRAS.528.1789L} {528, 1789}

\bibitem[\protect\citeauthoryear{{Luhman}}{{Luhman}}{2014}]{luhman14a}
{Luhman} K.~L.,  2014, \mn@doi [\apj] {10.1088/0004-637X/781/1/4}, \href
  {https://ui.adsabs.harvard.edu/abs/2014ApJ...781....4L} {781, 4}

\bibitem[\protect\citeauthoryear{{Luhman} \& {Sheppard}}{{Luhman} \&
  {Sheppard}}{2014}]{luhman14b}
{Luhman} K.~L.,  {Sheppard} S.~S.,  2014, \mn@doi [\apj]
  {10.1088/0004-637X/787/2/126}, \href
  {https://ui.adsabs.harvard.edu/abs/2014ApJ...787..126L} {787, 126}

\bibitem[\protect\citeauthoryear{{Luna}, {Marchetti}, {Rejkuba}  \&
  {Minniti}}{{Luna} et~al.}{2023}]{luna23}
{Luna} A.,  {Marchetti} T.,  {Rejkuba} M.,   {Minniti} D.,  2023, \mn@doi
  [\aap] {10.1051/0004-6361/202346257}, \href
  {https://ui.adsabs.harvard.edu/abs/2023A&A...677A.185L} {677, A185}

\bibitem[\protect\citeauthoryear{{Luna}, {Marchetti}, {Rejkuba}, {Leigh},
  {Alonso-Garc{\'\i}a}, {Valenzuela Navarro}, {Minniti}  \& {Smith}}{{Luna}
  et~al.}{2024}]{luna24}
{Luna} A.,  {Marchetti} T.,  {Rejkuba} M.,  {Leigh} N.~W.~C.,
  {Alonso-Garc{\'\i}a} J.,  {Valenzuela Navarro} A.,  {Minniti} D.,   {Smith}
  L.~C.,  2024, \mn@doi [\mnras] {10.1093/mnras/stae128}, \href
  {https://ui.adsabs.harvard.edu/abs/2024MNRAS.528.5495L} {528, 5495}

\bibitem[\protect\citeauthoryear{{Mace} et~al.,}{{Mace} et~al.}{2013}]{mace13}
{Mace} G.~N.,  et~al., 2013, \mn@doi [\apjs] {10.1088/0067-0049/205/1/6}, \href
  {https://ui.adsabs.harvard.edu/abs/2013ApJS..205....6M} {205, 6}

\bibitem[\protect\citeauthoryear{{Marois}, {Macintosh}, {Barman}, {Zuckerman},
  {Song}, {Patience}, {Lafreni{\`e}re}  \& {Doyon}}{{Marois}
  et~al.}{2008}]{marois08}
{Marois} C.,  {Macintosh} B.,  {Barman} T.,  {Zuckerman} B.,  {Song} I.,
  {Patience} J.,  {Lafreni{\`e}re} D.,   {Doyon} R.,  2008, \mn@doi [Science]
  {10.1126/science.1166585}, \href
  {https://ui.adsabs.harvard.edu/abs/2008Sci...322.1348M} {322, 1348}

\bibitem[\protect\citeauthoryear{{McGill}, {Smith}, {Evans}, {Belokurov}  \&
  {Lucas}}{{McGill} et~al.}{2019}]{mcgill19}
{McGill} P.,  {Smith} L.~C.,  {Evans} N.~W.,  {Belokurov} V.,   {Lucas} P.~W.,
  2019, \mn@doi [\mnras] {10.1093/mnrasl/slz073}, \href
  {https://ui.adsabs.harvard.edu/abs/2019MNRAS.487L...7M} {487, L7}

\bibitem[\protect\citeauthoryear{{Meisner}, {Lang}  \& {Schlegel}}{{Meisner}
  et~al.}{2018}]{meisner18}
{Meisner} A.~M.,  {Lang} D.,   {Schlegel} D.~J.,  2018, \mn@doi [\aj]
  {10.3847/1538-3881/aacbcd}, \href
  {https://ui.adsabs.harvard.edu/abs/2018AJ....156...69M} {156, 69}

\bibitem[\protect\citeauthoryear{{Meisner} et~al.,}{{Meisner}
  et~al.}{2020a}]{meisner20a}
{Meisner} A.~M.,  et~al., 2020a, \mn@doi [\apj] {10.3847/1538-4357/ab6215},
  \href {https://ui.adsabs.harvard.edu/abs/2020ApJ...889...74M} {889, 74}

\bibitem[\protect\citeauthoryear{{Meisner} et~al.,}{{Meisner}
  et~al.}{2020b}]{meisner20b}
{Meisner} A.~M.,  et~al., 2020b, \mn@doi [\apj] {10.3847/1538-4357/aba633},
  \href {https://ui.adsabs.harvard.edu/abs/2020ApJ...899..123M} {899, 123}

\bibitem[\protect\citeauthoryear{{Mej{\'\i}as}, {Minniti},
  {Alonso-Garc{\'\i}a}, {Beam{\'\i}n}, {Saito}  \& {Solano}}{{Mej{\'\i}as}
  et~al.}{2022}]{mejias22}
{Mej{\'\i}as} A.,  {Minniti} D.,  {Alonso-Garc{\'\i}a} J.,  {Beam{\'\i}n}
  J.~C.,  {Saito} R.~K.,   {Solano} E.,  2022, \mn@doi [\aap]
  {10.1051/0004-6361/202141759}, \href
  {https://ui.adsabs.harvard.edu/abs/2022A&A...660A.131M} {660, A131}

\bibitem[\protect\citeauthoryear{{Michalik}, {Lindegren}  \&
  {Hobbs}}{{Michalik} et~al.}{2015}]{michalik15}
{Michalik} D.,  {Lindegren} L.,   {Hobbs} D.,  2015, \mn@doi [\aap]
  {10.1051/0004-6361/201425310}, \href
  {https://ui.adsabs.harvard.edu/abs/2015A&A...574A.115M} {574, A115}

\bibitem[\protect\citeauthoryear{{Minniti} et~al.,}{{Minniti}
  et~al.}{2010}]{minniti10}
{Minniti} D.,  et~al., 2010, \mn@doi [\na] {10.1016/j.newast.2009.12.002},
  \href {https://ui.adsabs.harvard.edu/abs/2010NewA...15..433M} {15, 433}

\bibitem[\protect\citeauthoryear{{Minniti}, {Fern{\'a}ndez-Trincado},
  {G{\'o}mez}, {Smith}, {Lucas}  \& {Contreras Ramos}}{{Minniti}
  et~al.}{2021}]{minniti21}
{Minniti} D.,  {Fern{\'a}ndez-Trincado} J.~G.,  {G{\'o}mez} M.,  {Smith} L.~C.,
   {Lucas} P.~W.,   {Contreras Ramos} R.,  2021, \mn@doi [\aap]
  {10.1051/0004-6361/202141129}, \href
  {https://ui.adsabs.harvard.edu/abs/2021A&A...650L..11M} {650, L11}

\bibitem[\protect\citeauthoryear{{Minniti} et~al.,}{{Minniti}
  et~al.}{2024}]{minnitti24}
{Minniti} D.,  et~al., 2024, \mn@doi [\aap] {10.1051/0004-6361/202348100},
  \href {https://ui.adsabs.harvard.edu/abs/2024A&A...683A.150M} {683, A150}

\bibitem[\protect\citeauthoryear{{Molnar}, {Sanders}, {Smith}, {Belokurov},
  {Lucas}  \& {Minniti}}{{Molnar} et~al.}{2022}]{molnar22}
{Molnar} T.~A.,  {Sanders} J.~L.,  {Smith} L.~C.,  {Belokurov} V.,  {Lucas} P.,
    {Minniti} D.,  2022, \mn@doi [\mnras] {10.1093/mnras/stab3116}, \href
  {https://ui.adsabs.harvard.edu/abs/2022MNRAS.509.2566M} {509, 2566}

\bibitem[\protect\citeauthoryear{{Nieuwmunster} et~al.,}{{Nieuwmunster}
  et~al.}{2024}]{Nieuwmunster24}
{Nieuwmunster} N.,  et~al., 2024, \mn@doi [\aap] {10.1051/0004-6361/202349000},
  \href {https://ui.adsabs.harvard.edu/abs/2024A&A...685A..93N} {685, A93}

\bibitem[\protect\citeauthoryear{{Nikutta}, {Fitzpatrick}, {Scott}  \&
  {Weaver}}{{Nikutta} et~al.}{2020}]{nikutta20}
{Nikutta} R.,  {Fitzpatrick} M.,  {Scott} A.,   {Weaver} B.~A.,  2020, \mn@doi
  [Astronomy and Computing] {10.1016/j.ascom.2020.100411}, \href
  {https://ui.adsabs.harvard.edu/abs/2020A&C....3300411N} {33, 100411}

\bibitem[\protect\citeauthoryear{{Oliveros-Gomez}, {Manjavacas}, {Ashraf},
  {Bardalez-Gagliuffi}, {Vos}, {Faherty}, {Karalidi}  \&
  {Apai}}{{Oliveros-Gomez} et~al.}{2022}]{oliveros-gomez22}
{Oliveros-Gomez} N.,  {Manjavacas} E.,  {Ashraf} A.,  {Bardalez-Gagliuffi}
  D.~C.,  {Vos} J.~M.,  {Faherty} J.~K.,  {Karalidi} T.,   {Apai} D.,  2022,
  \mn@doi [\apj] {10.3847/1538-4357/ac96f2}, \href
  {https://ui.adsabs.harvard.edu/abs/2022ApJ...939...72O} {939, 72}

\bibitem[\protect\citeauthoryear{{Padmanabhan} et~al.,}{{Padmanabhan}
  et~al.}{2008}]{ubercal}
{Padmanabhan} N.,  et~al., 2008, \mn@doi [\apj] {10.1086/524677}, \href
  {https://ui.adsabs.harvard.edu/abs/2008ApJ...674.1217P} {674, 1217}

\bibitem[\protect\citeauthoryear{{Patten} et~al.,}{{Patten}
  et~al.}{2006}]{patten06}
{Patten} B.~M.,  et~al., 2006, \mn@doi [\apj] {10.1086/507264}, \href
  {https://ui.adsabs.harvard.edu/abs/2006ApJ...651..502P} {651, 502}

\bibitem[\protect\citeauthoryear{{Pe{\~n}a Ram{\'\i}rez}, {Smith},
  {Ram{\'\i}rez Alegr{\'\i}a}, {Chen{\'e}}, {Gonz{\'a}lez-Fern{\'a}ndez},
  {Lucas}  \& {Minniti}}{{Pe{\~n}a Ram{\'\i}rez} et~al.}{2022}]{pramirez22}
{Pe{\~n}a Ram{\'\i}rez} K.,  {Smith} L.~C.,  {Ram{\'\i}rez Alegr{\'\i}a} S.,
  {Chen{\'e}} A.~N.,  {Gonz{\'a}lez-Fern{\'a}ndez} C.,  {Lucas} P.~W.,
  {Minniti} D.,  2022, \mn@doi [\mnras] {10.1093/mnras/stac1296}, \href
  {https://ui.adsabs.harvard.edu/abs/2022MNRAS.513.5799P} {513, 5799}

\bibitem[\protect\citeauthoryear{{Phan-Bao} et~al.,}{{Phan-Bao}
  et~al.}{2008}]{phan-bao08}
{Phan-Bao} N.,  et~al., 2008, \mn@doi [\mnras]
  {10.1111/j.1365-2966.2007.12564.x}, \href
  {https://ui.adsabs.harvard.edu/abs/2008MNRAS.383..831P} {383, 831}

\bibitem[\protect\citeauthoryear{{Prusti} et~al.,}{{Prusti}
  et~al.}{2016}]{prusti16}
{Prusti} T.,  et~al., 2016, \mn@doi [\aap] {10.1051/0004-6361/201629272}, \href
  {https://ui.adsabs.harvard.edu/abs/2016A&A...595A...1G} {595, A1}

\bibitem[\protect\citeauthoryear{{Reyl{\'e}}}{{Reyl{\'e}}}{2018}]{reyle18}
{Reyl{\'e}} C.,  2018, \mn@doi [\aap] {10.1051/0004-6361/201834082}, \href
  {https://ui.adsabs.harvard.edu/abs/2018A&A...619L...8R} {619, L8}

\bibitem[\protect\citeauthoryear{{Reyl{\'e}}, {Jardine}, {Fouqu{\'e}},
  {Caballero}, {Smart}  \& {Sozzetti}}{{Reyl{\'e}} et~al.}{2021}]{reyle21}
{Reyl{\'e}} C.,  {Jardine} K.,  {Fouqu{\'e}} P.,  {Caballero} J.~A.,  {Smart}
  R.~L.,   {Sozzetti} A.,  2021, \mn@doi [\aap] {10.1051/0004-6361/202140985},
  \href {https://ui.adsabs.harvard.edu/abs/2021A&A...650A.201R} {650, A201}

\bibitem[\protect\citeauthoryear{{Saito} et~al.,}{{Saito} et~al.}{2024}]{vvvx}
{Saito} R.~K.,  et~al., 2024, \mn@doi [\aap] {10.1051/0004-6361/202450584},
  \href {https://ui.adsabs.harvard.edu/abs/2024A&A...689A.148S} {689, A148}

\bibitem[\protect\citeauthoryear{{Sanders}}{{Sanders}}{2023}]{sanders23}
{Sanders} J.~L.,  2023, \mn@doi [\mnras] {10.1093/mnras/stad1431}, \href
  {https://ui.adsabs.harvard.edu/abs/2023MNRAS.523.2369S} {523, 2369}

\bibitem[\protect\citeauthoryear{{Sanders}, {Smith}, {Evans}  \&
  {Lucas}}{{Sanders} et~al.}{2019}]{sanders19}
{Sanders} J.~L.,  {Smith} L.,  {Evans} N.~W.,   {Lucas} P.,  2019, \mn@doi
  [\mnras] {10.1093/mnras/stz1630}, \href
  {https://ui.adsabs.harvard.edu/abs/2019MNRAS.487.5188S} {487, 5188}

\bibitem[\protect\citeauthoryear{{Sanders}, {Smith},
  {Gonz{\'a}lez-Fern{\'a}ndez}, {Lucas}  \& {Minniti}}{{Sanders}
  et~al.}{2022a}]{sanders22}
{Sanders} J.~L.,  {Smith} L.,  {Gonz{\'a}lez-Fern{\'a}ndez} C.,  {Lucas} P.,
  {Minniti} D.,  2022a, \mn@doi [\mnras] {10.1093/mnras/stac1367}, \href
  {https://ui.adsabs.harvard.edu/abs/2022MNRAS.514.2407S} {514, 2407}

\bibitem[\protect\citeauthoryear{{Sanders}, {Matsunaga}, {Kawata}, {Smith},
  {Minniti}  \& {Lucas}}{{Sanders} et~al.}{2022b}]{sandersmira22}
{Sanders} J.~L.,  {Matsunaga} N.,  {Kawata} D.,  {Smith} L.~C.,  {Minniti} D.,
   {Lucas} P.~W.,  2022b, \mn@doi [\mnras] {10.1093/mnras/stac2274}, \href
  {https://ui.adsabs.harvard.edu/abs/2022MNRAS.517..257S} {517, 257}

\bibitem[\protect\citeauthoryear{{Sanders}, {Kawata}, {Matsunaga}, {Sormani},
  {Smith}, {Minniti}  \& {Gerhard}}{{Sanders} et~al.}{2024}]{sanders24}
{Sanders} J.~L.,  {Kawata} D.,  {Matsunaga} N.,  {Sormani} M.~C.,  {Smith}
  L.~C.,  {Minniti} D.,   {Gerhard} O.,  2024, \mn@doi [\mnras]
  {10.1093/mnras/stae711}, \href
  {https://ui.adsabs.harvard.edu/abs/2024MNRAS.530.2972S} {530, 2972}

\bibitem[\protect\citeauthoryear{{Sanghi} et~al.,}{{Sanghi}
  et~al.}{2023}]{sanghi23}
{Sanghi} A.,  et~al., 2023, \mn@doi [\apj] {10.3847/1538-4357/acff66}, \href
  {https://ui.adsabs.harvard.edu/abs/2023ApJ...959...63S} {959, 63}

\bibitem[\protect\citeauthoryear{{Saydjari} et~al.,}{{Saydjari}
  et~al.}{2023}]{saydjari23}
{Saydjari} A.~K.,  et~al., 2023, \mn@doi [\apjs] {10.3847/1538-4365/aca594},
  \href {https://ui.adsabs.harvard.edu/abs/2023ApJS..264...28S} {264, 28}

\bibitem[\protect\citeauthoryear{{Schapera} et~al.,}{{Schapera}
  et~al.}{2022}]{schapera22}
{Schapera} N.,  et~al., 2022, \mn@doi [Research Notes of the American
  Astronomical Society] {10.3847/2515-5172/ac9141}, \href
  {https://ui.adsabs.harvard.edu/abs/2022RNAAS...6..189S} {6, 189}

\bibitem[\protect\citeauthoryear{{Schechter}, {Mateo}  \& {Saha}}{{Schechter}
  et~al.}{1993}]{dophot1}
{Schechter} P.~L.,  {Mateo} M.,   {Saha} A.,  1993, \mn@doi [\pasp]
  {10.1086/133316}, \href
  {https://ui.adsabs.harvard.edu/abs/1993PASP..105.1342S} {105, 1342}

\bibitem[\protect\citeauthoryear{{Schlafly} et~al.,}{{Schlafly}
  et~al.}{2018}]{schlafly18}
{Schlafly} E.~F.,  et~al., 2018, \mn@doi [\apjs] {10.3847/1538-4365/aaa3e2},
  \href {https://ui.adsabs.harvard.edu/abs/2018ApJS..234...39S} {234, 39}

\bibitem[\protect\citeauthoryear{{Schlegel}, {Finkbeiner}  \&
  {Davis}}{{Schlegel} et~al.}{1998}]{sfd98}
{Schlegel} D.~J.,  {Finkbeiner} D.~P.,   {Davis} M.,  1998, \mn@doi [\apj]
  {10.1086/305772}, \href
  {https://ui.adsabs.harvard.edu/abs/1998ApJ...500..525S} {500, 525}

\bibitem[\protect\citeauthoryear{{Schmidt}, {West}, {Hawley}  \&
  {Pineda}}{{Schmidt} et~al.}{2010}]{schmidt10}
{Schmidt} S.~J.,  {West} A.~A.,  {Hawley} S.~L.,   {Pineda} J.~S.,  2010,
  \mn@doi [\aj] {10.1088/0004-6256/139/5/1808}, \href
  {https://ui.adsabs.harvard.edu/abs/2010AJ....139.1808S} {139, 1808}

\bibitem[\protect\citeauthoryear{{Schneider}, {Greco}, {Cushing},
  {Kirkpatrick}, {Mainzer}, {Gelino}, {Fajardo-Acosta}  \& {Bauer}}{{Schneider}
  et~al.}{2016}]{schneider16}
{Schneider} A.~C.,  {Greco} J.,  {Cushing} M.~C.,  {Kirkpatrick} J.~D.,
  {Mainzer} A.,  {Gelino} C.~R.,  {Fajardo-Acosta} S.~B.,   {Bauer} J.,  2016,
  \mn@doi [\apj] {10.3847/0004-637X/817/2/112}, \href
  {https://ui.adsabs.harvard.edu/abs/2016ApJ...817..112S} {817, 112}

\bibitem[\protect\citeauthoryear{{Schneider}, {Munn}, {Vrba}, {Bruursema},
  {Dahm}, {Williams}, {Liu}  \& {Dorland}}{{Schneider}
  et~al.}{2023}]{schneider23}
{Schneider} A.~C.,  {Munn} J.~A.,  {Vrba} F.~J.,  {Bruursema} J.,  {Dahm}
  S.~E.,  {Williams} S.~J.,  {Liu} M.~C.,   {Dorland} B.~N.,  2023, \mn@doi
  [\aj] {10.3847/1538-3881/ace9bf}, \href
  {https://ui.adsabs.harvard.edu/abs/2023AJ....166..103S} {166, 103}

\bibitem[\protect\citeauthoryear{{Seifahrt}, {Reiners}, {Almaghrbi}  \&
  {Basri}}{{Seifahrt} et~al.}{2010}]{seifahrt10}
{Seifahrt} A.,  {Reiners} A.,  {Almaghrbi} K.~A.~M.,   {Basri} G.,  2010,
  \mn@doi [\aap] {10.1051/0004-6361/200913368}, \href
  {https://ui.adsabs.harvard.edu/abs/2010A&A...512A..37S} {512, A37}

\bibitem[\protect\citeauthoryear{{Skrutskie} et~al.,}{{Skrutskie}
  et~al.}{2006}]{2mass}
{Skrutskie} M.~F.,  et~al., 2006, \mn@doi [\aj] {10.1086/498708}, \href
  {https://ui.adsabs.harvard.edu/abs/2006AJ....131.1163S} {131, 1163}

\bibitem[\protect\citeauthoryear{{Smart} et~al.,}{{Smart}
  et~al.}{2021}]{smart21}
{Smart} R.~L.,  et~al., 2021, \mn@doi [\aap] {10.1051/0004-6361/202039498},
  \href {https://ui.adsabs.harvard.edu/abs/2021A&A...649A...6G} {649, A6}

\bibitem[\protect\citeauthoryear{{Smith} et~al.,}{{Smith}
  et~al.}{2014}]{smith14}
{Smith} L.,  et~al., 2014, \mn@doi [\mnras] {10.1093/mnras/stu1295}, \href
  {https://ui.adsabs.harvard.edu/abs/2014MNRAS.443.2327S} {443, 2327}

\bibitem[\protect\citeauthoryear{{Smith} et~al.,}{{Smith}
  et~al.}{2015}]{smith15}
{Smith} L.~C.,  et~al., 2015, \mn@doi [\mnras] {10.1093/mnras/stv2290}, \href
  {https://ui.adsabs.harvard.edu/abs/2015MNRAS.454.4476S} {454, 4476}

\bibitem[\protect\citeauthoryear{{Smith} et~al.,}{{Smith} et~al.}{2018}]{LCS18}
{Smith} L.~C.,  et~al., 2018, \mn@doi [\mnras] {10.1093/mnras/stx2789}, \href
  {http://adsabs.harvard.edu/abs/2018MNRAS.474.1826S} {474, 1826}

\bibitem[\protect\citeauthoryear{{Smith} et~al.,}{{Smith} et~al.}{2021}]{wit08}
{Smith} L.~C.,  et~al., 2021, \mn@doi [\mnras] {10.1093/mnras/stab1211}, \href
  {https://ui.adsabs.harvard.edu/abs/2021MNRAS.505.1992S} {505, 1992}

\bibitem[\protect\citeauthoryear{{Sormani} et~al.,}{{Sormani}
  et~al.}{2022}]{sormani22}
{Sormani} M.~C.,  et~al., 2022, \mn@doi [\mnras] {10.1093/mnras/stac639}, \href
  {https://ui.adsabs.harvard.edu/abs/2022MNRAS.512.1857S} {512, 1857}

\bibitem[\protect\citeauthoryear{{Stassun} \& {Torres}}{{Stassun} \&
  {Torres}}{2021}]{stassun21}
{Stassun} K.~G.,  {Torres} G.,  2021, \mn@doi [\apjl]
  {10.3847/2041-8213/abdaad}, \href
  {https://ui.adsabs.harvard.edu/abs/2021ApJ...907L..33S} {907, L33}

\bibitem[\protect\citeauthoryear{{Stetson}}{{Stetson}}{1987}]{daophot}
{Stetson} P.~B.,  1987, \mn@doi [\pasp] {10.1086/131977}, \href
  {https://ui.adsabs.harvard.edu/abs/1987PASP...99..191S} {99, 191}

\bibitem[\protect\citeauthoryear{{Stetson}}{{Stetson}}{1996}]{stetsonJK}
{Stetson} P.~B.,  1996, \mn@doi [\pasp] {10.1086/133808}, \href
  {https://ui.adsabs.harvard.edu/abs/1996PASP..108..851S} {108, 851}

\bibitem[\protect\citeauthoryear{{Sutherland} et~al.,}{{Sutherland}
  et~al.}{2015}]{vista}
{Sutherland} W.,  et~al., 2015, \mn@doi [\aap] {10.1051/0004-6361/201424973},
  \href {https://ui.adsabs.harvard.edu/abs/2015A&A...575A..25S} {575, A25}

\bibitem[\protect\citeauthoryear{{Taylor}}{{Taylor}}{2005}]{topcat}
{Taylor} M.~B.,  2005, in {Shopbell} P.,  {Britton} M.,   {Ebert} R.,  eds,
  Astronomical Society of the Pacific Conference Series Vol. 347, Astronomical
  Data Analysis Software and Systems XIV. p.~29

\bibitem[\protect\citeauthoryear{{Terzan}, {Bernard}, {Fresneau}  \&
  {Ju}}{{Terzan} et~al.}{1980}]{terzan80}
{Terzan} A.,  {Bernard} A.,  {Fresneau} A.,   {Ju} K.~H.,  1980, Academie des
  Sciences Paris Comptes Rendus Serie B Sciences Physiques, \href
  {https://ui.adsabs.harvard.edu/abs/1980CRASB.290..321T} {290, 321}

\bibitem[\protect\citeauthoryear{{Vos}, {Faherty}, {Gagn{\'e}}, {Marley},
  {Metchev}, {Gizis}, {Rice}  \& {Cruz}}{{Vos} et~al.}{2022}]{vos22}
{Vos} J.~M.,  {Faherty} J.~K.,  {Gagn{\'e}} J.,  {Marley} M.,  {Metchev} S.,
  {Gizis} J.,  {Rice} E.~L.,   {Cruz} K.,  2022, \mn@doi [\apj]
  {10.3847/1538-4357/ac4502}, \href
  {https://ui.adsabs.harvard.edu/abs/2022ApJ...924...68V} {924, 68}

\bibitem[\protect\citeauthoryear{{Welch} \& {Stetson}}{{Welch} \&
  {Stetson}}{1993}]{stetsonI}
{Welch} D.~L.,  {Stetson} P.~B.,  1993, \mn@doi [\aj] {10.1086/116556}, \href
  {https://ui.adsabs.harvard.edu/abs/1993AJ....105.1813W} {105, 1813}

\bibitem[\protect\citeauthoryear{{Wenger} et~al.,}{{Wenger}
  et~al.}{2000}]{wenger00}
{Wenger} M.,  et~al., 2000, \mn@doi [\aaps] {10.1051/aas:2000332}, \href
  {https://ui.adsabs.harvard.edu/abs/2000A&AS..143....9W} {143, 9}

\bibitem[\protect\citeauthoryear{{Werner} et~al.,}{{Werner}
  et~al.}{2004}]{werner04}
{Werner} M.~W.,  et~al., 2004, \mn@doi [\apjs] {10.1086/422992}, \href
  {https://ui.adsabs.harvard.edu/abs/2004ApJS..154....1W} {154, 1}

\bibitem[\protect\citeauthoryear{{Whitney} et~al.,}{{Whitney}
  et~al.}{2011}]{whitney11}
{Whitney} B.,  et~al., 2011, {Deep GLIMPSE: Exploring the Far Side of the
  Galaxy}, Spitzer Proposal ID \#80074

\bibitem[\protect\citeauthoryear{{Wright} et~al.,}{{Wright}
  et~al.}{2010}]{wright10}
{Wright} E.~L.,  et~al., 2010, \mn@doi [\aj] {10.1088/0004-6256/140/6/1868},
  \href {https://ui.adsabs.harvard.edu/abs/2010AJ....140.1868W} {140, 1868}

\bibitem[\protect\citeauthoryear{{Zhang} et~al.,}{{Zhang}
  et~al.}{2017}]{zhang17}
{Zhang} Z.~H.,  et~al., 2017, \mn@doi [\mnras] {10.1093/mnras/stw2438}, \href
  {https://ui.adsabs.harvard.edu/abs/2017MNRAS.464.3040Z} {464, 3040}

\bibitem[\protect\citeauthoryear{{Zhang} et~al.,}{{Zhang}
  et~al.}{2018}]{zhang18}
{Zhang} Z.~H.,  et~al., 2018, \mn@doi [\mnras] {10.1093/mnras/sty2054}, \href
  {https://ui.adsabs.harvard.edu/abs/2018MNRAS.480.5447Z} {480, 5447}

\bibitem[\protect\citeauthoryear{von Neumann}{von Neumann}{1941}]{eta1}
von Neumann J.,  1941, \mn@doi [The Annals of Mathematical Statistics]
  {10.1214/aoms/1177731677}, 12, 367

\bibitem[\protect\citeauthoryear{von Neumann}{von Neumann}{1942}]{eta2}
von Neumann J.,  1942, \mn@doi [The Annals of Mathematical Statistics]
  {10.1214/aoms/1177731645}, 13, 86

\makeatother
\end{thebibliography}



\appendix

\section{Details of VIRAC2$\beta{}$}\label{app:virac2b}

For the astrometric calibration (Section \ref{sec:astrocalib}): VIRAC2$\beta{}$ used Gaia DR2 as a reference catalogue, where VIRAC2 used Gaia eDR3. The earlier version used a fixed 8 degree Chebyshev polynomial, where the final one used varying degrees. The version of \textsc{DoPhot} used for VIRAC2$\beta{}$ did not output centroid uncertainties, so these were fitted against the scatter in the residuals to the astrometric calibration as an approximate function of magnitude. For VIRAC2 we modified \textsc{DoPhot} such that it output centroid uncertainties, which were then calibrated against Gaia eDR3.

For the main pipeline: since the centroid uncertainties were fairly unreliable estimates, uncertainties from 5-parameter astrometric fits were scaled to set the reduced $\chi{}^2$ to unity. Additionally, we employed no residual-over-error selection threshold during source position matching. For the final version of the pipeline, the centroid uncertainties (which at this stage were output by \textsc{DoPhot} and scaled to match the scatter seen during astrometric calibration) were much more reliable, hence we found it unnecessary to perform the scaling on the 5-parameter astrometric fit uncertainties. We also found it was now useful to use a residual-over-error selection threshold to clean the positional matching.

For VIRAC2$\beta{}$, the default proper motion was taken as the mean of local field stars from VIRAC version 1.1 (see \citealp{LCS18}, and \citealp{sanders19} section 2.1). For VIRAC2, the default proper motion was taken from VIRAC2$\beta$ data.

\section{Some failure modes in parallax solutions}\label{sec:app_failures}

Here we describe two cases where our parallax-based search for new nearby sources (see section \ref{sec:50_pc_search}) produced initially promising candidates, based on visual inspection of images to confirm high proper motion, but the VIRAC2 parallax was soon found to be incorrect.

(1) VIRAC2 source 13071355014500 was confirmed as a high proper motion star,
with the images in fact showing it to be a triple system comprising
two relatively bright, blended sources and a fainter source $\sim$3\arcsec~away. However, the VIRAC2 entries for this system
showed that while all three sources have very similar proper motions,
the other two components have smaller VIRAC2 parallaxes: $6.2\pm1.0$~mas and $6.7\pm2.2$~mas for sources 13071355002815 and 13071355010344, compared to $32.0\pm3.6$~mas for 13071355014500. The two smaller parallaxes are confirmed by Gaia DR3 ($\sim$5.6~mas in both cases), though Gaia does not list a parallax for 13071355014500.
We deduce that the parallax fit was corrupted by blending in the case of 13071355014500, which is the fainter component of the blend.

(2) VIRAC2 source 13431779012483 showed a clear motion of 0.5\arcsec~between
2010 and 2018 but we noted that the image profile of this
source was symmetric in some images but slightly elongated in others. On inspecting plots of RA vs. time and Dec vs. time we saw that the overall trend in each case was not well fit by a straight line: there were signs of curvature and a distinct jump between 2012 and 2013. This source has an entry as a Long Period Variable (LPV) star in the VIVACE catalogue of candidate periodic variable VIRAC2 sources \citep{molnar22}.
We then noticed that the $K_s$ light curve was correlated with the position vs time plots, i.e. brightness was related to position. This, along with the elongation of the source in some images, made clear that source 13431779012483 is a blend of an LPV and another star. It is possible that there is a significant proper motion for one of the pair, to account for the overall trend in position vs. time, but the high VIRAC2 parallax ($\varpi \approx 46$~mas) is an artefact of the blending and variability. This candidate had ${\rm UWE} = 1.16$, somewhat higher than most bona fide nearby stars.

\clearpage
\section{Table schemata}\label{app:schema}

\begin{table*}
    \centering
    \caption{Field names, units and descriptions for the source table. An example row is also given, that of the early L dwarf $\beta$ Circini B \citep{smith15}. The \textit{sourceid} field links to the time series table (see Table \ref{tab:timeseries_schema}). This format is the same for both the main and reject tables.}
\label{tab:sources_schema}
\begin{tabular}{|l|l|l|c|}
\hline
  \multicolumn{1}{|l|}{Name} &
  \multicolumn{1}{l|}{Units} &
  \multicolumn{1}{l|}{Description} &
  \multicolumn{1}{c|}{Example row} \\
\hline
  sourceid &  & unique source identifier & 15869033004249\\
  astfit\_epochs &  & number of epochs used for astrometric solution & 180\\
  astfit\_params &  & number of astrometric solution parameters & 5\\
  duplicate &  & flag indicating a likely duplicate entry & 0\\
  ref\_epoch & yr & astrometric reference epoch & 2014.0\\
  ra & deg & right ascension & 229.33928126317744\\
  ra\_error & mas & uncertainty on right ascension & 0.48293653887707855\\
  de & deg & declination & -58.85879996158586\\
  de\_error & mas & uncertainty on declination & 0.7933554229463711\\
  parallax & mas & trigonometric parallax & 31.564542172410846\\
  parallax\_error & mas & uncertainty on trigonometric parallax & 0.9661602760068156\\
  pmra & mas/yr & proper motion in right ascension times cos(dec) & -94.70174133405408\\
  pmra\_error & mas/yr & uncertainty on proper motion in right ascension times cos(dec) & 0.34828527350530797\\
  pmde & mas/yr & proper motion in declination & -134.84857408202345\\
  pmde\_error & mas/yr & uncertainty on proper motion in declination & 0.34549854257712753\\
  ra\_de\_corr &  & correlation between ra and de & 0.029712955\\
  ra\_parallax\_corr &  & correlation between ra and parallax & 0.33010614\\
  ra\_pmra\_corr &  & correlation between ra and pmra & 0.2891742\\
  ra\_pmde\_corr &  & correlation between ra and pmde & 0.011853172\\
  de\_parallax\_corr &  & correlation between de and parallax & 0.09001031\\
  de\_pmra\_corr &  & correlation between de and pmra & 0.0033594724\\
  de\_pmde\_corr &  & correlation between de and pmde & -0.032637153\\
  parallax\_pmra\_corr &  & correlation between parallax and pmra & 0.0373232\\
  parallax\_pmde\_corr &  & correlation between parallax and pmde & 0.035907157\\
  pmra\_pmde\_corr &  & correlation between pmra and pmde & 0.0013401698\\
  chisq &  & chi squared of astrometric fit & 336.16397\\
  uwe &  & unit weight error of astrometric fit & 0.9731088\\
  phot\_z\_mean\_mag & mag & mean $Z$ band magnitude & 16.75006\\
  phot\_z\_std\_mag & mag & standard deviation of $Z$ band magnitude & 0.021935267\\
  phot\_z\_n\_epochs &  & number of $Z$ band epochs contributing to statistics & 8\\
  z\_n\_obs &  & approximate number of $Z$ band observations & 9\\
  z\_n\_det &  & number of $Z$ band detections & 9\\
  z\_n\_amb &  & number of $Z$ band detections shared with another source & 0\\
  phot\_y\_mean\_mag & mag & mean $Y$ band magnitude & 15.495929\\
  phot\_y\_std\_mag & mag & standard deviation of $Y$ band magnitude & 0.016722715\\
  phot\_y\_n\_epochs &  & number of $Y$ band epochs contributing to statistics & 10\\
  y\_n\_obs &  & approximate number of $Y$ band observations & 11\\
  y\_n\_det &  & number of $Y$ band detections & 11\\
  y\_n\_amb &  & number of $Y$ band detections shared with another source & 0\\
  phot\_j\_mean\_mag & mag & mean $J$ band magnitude & 14.492317\\
  phot\_j\_std\_mag & mag & standard deviation of $J$ band magnitude & 0.011483114\\
  phot\_j\_n\_epochs &  & number of $J$ band epochs contributing to statistics & 5\\
  j\_n\_obs &  & approximate number of $J$ band observations & 6\\
  j\_n\_det &  & number of $J$ band detections & 6\\
  j\_n\_amb &  & number of $J$ band detections shared with another source & 0\\
  phot\_h\_mean\_mag & mag & mean $H$ band magnitude & 13.750394\\
  phot\_h\_std\_mag & mag & standard deviation of $H$ band magnitude & 0.014054991\\
  phot\_h\_n\_epochs &  & number of $H$ band epochs contributing to statistics & 5\\
  h\_n\_obs &  & approximate number of $H$ band observations & 6\\
  h\_n\_det &  & number of $H$ band detections & 6\\
  h\_n\_amb &  & number of $H$ band detections shared with another source & 0\\
  phot\_ks\_mean\_mag & mag & mean $K_s$ band magnitude & 13.21982\\
  phot\_ks\_std\_mag & mag & standard deviation of $K_s$ band magnitude & 0.019139782\\
  phot\_ks\_n\_epochs &  & number of $K_s$ band epochs contributing to statistics & 141\\
  ks\_n\_obs &  & approximate number of $K_s$ band observations & 184\\
  ks\_n\_det &  & number of $K_s$ band detections & 180\\
  ks\_n\_amb &  & number of $K_s$ band detections shared with another source & 0\\
\hline\end{tabular}
\end{table*}

\begin{table*}
    \centering
\caption*{continued from previous page}
\begin{tabular}{|l|l|l|c|}
\hline
  \multicolumn{1}{|l|}{Name} &
  \multicolumn{1}{l|}{Units} &
  \multicolumn{1}{l|}{Description} &
  \multicolumn{1}{c|}{Example row} \\
\hline
  ks\_first\_epoch & MJD & epoch of first $K_s$ band detection & 55260.36375826\\
  ks\_last\_epoch & MJD & epoch of last $K_s$ band detection & 58717.06490433\\
  ks\_skew &  & skewness of $K_s$ band magnitudes & 0.6108212\\
  ks\_p0 & mag & 0th percentile (i.e. min) of $K_s$ band magnitudes & 13.172082\\
  ks\_p1 & mag & 1st percentile of $K_s$ band magnitudes & 13.1780205\\
  ks\_p2 & mag & 2nd percentile of $K_s$ band magnitudes & 13.185369\\
  ks\_p4 & mag & 4th percentile of $K_s$ band magnitudes & 13.191636\\
  ks\_p5 & mag & 5th percentile of $K_s$ band magnitudes & 13.193609\\
  ks\_p8 & mag & 8th percentile of $K_s$ band magnitudes & 13.197648\\
  ks\_p16 & mag & 16th percentile of $K_s$ band magnitudes & 13.20169\\
  ks\_p25 & mag & 25th percentile of $K_s$ band magnitudes & 13.206783\\
  ks\_p32 & mag & 32nd percentile of $K_s$ band magnitudes & 13.210841\\
  ks\_p50 & mag & 50th percentile (i.e. median) of $K_s$ band magnitudes & 13.217017\\
  ks\_p68 & mag & 68th percentile of $K_s$ band magnitudes & 13.2265835\\
  ks\_p75 & mag & 75th percentile of $K_s$ band magnitudes & 13.231348\\
  ks\_p84 & mag & 84th percentile of $K_s$ band magnitudes & 13.238225\\
  ks\_p92 & mag & 92nd percentile of $K_s$ band magnitudes & 13.245653\\
  ks\_p95 & mag & 95th percentile of $K_s$ band magnitudes & 13.2479105\\
  ks\_p96 & mag & 96th percentile of $K_s$ band magnitudes & 13.249347\\
  ks\_p98 & mag & 98th percentile of $K_s$ band magnitudes & 13.267736\\
  ks\_p99 & mag & 99th percentile of $K_s$ band magnitudes & 13.27843\\
  ks\_p100 & mag & 100th percentile (i.e. max) of $K_s$ band magnitudes & 13.288555\\
  ks\_mad & mag & median absolute deviation from the median $K_s$ band magnitude & 0.012143135\\
  ks\_med\_err & mag & median uncertainty of $K_s$ band magnitudes & 0.0201695\\
  ks\_Stetson\_I &  & Stetson I index for $K_s$ band magnitudes & 0.21573663035284185\\
  ks\_Stetson\_J &  & Stetson J index for $K_s$ band magnitudes & 0.13496087363194595\\
  ks\_Stetson\_K &  & Stetson K index for $K_s$ band magnitudes & 0.8183742496792374\\
  ks\_Stetson\_group\_count &  & number of observation groups used for Stetson indices & 61\\
  ks\_eta &  & von Neumann eta index & 1.5645293017153363\\
  ks\_eta\_f &  & modified von Neumann eta index & 1051447886.8728735\\
\hline\end{tabular}
\end{table*}

\begin{table*}
\centering
\caption{Field names, units and descriptions for the time series table. An example row is also given, that of the first time series element for the example source shown in Table \ref{tab:sources_schema}. The \textit{sourceid} field links to the source table (see Table \ref{tab:sources_schema}), and the \textit{catid} field links to the observation table (see Table \ref{tab:observations_schema}). This format is the same for both the main and reject tables.}\label{tab:timeseries_schema}
\begin{tabular}{|l|l|l|c|}
\hline
  \multicolumn{1}{|l|}{Name} &
  \multicolumn{1}{l|}{Units} &
  \multicolumn{1}{l|}{Description}  &
  \multicolumn{1}{c|}{Example row} \\
\hline
  sourceid &  & unique source identifier  & 15869033004249\\
  catid &  & unique observation identifier  & 120741\\
  mjdobs & MJD & epoch of observation  & 55260.36375826\\
  filter &  & observation bandpass name & $K_s$\\
  seeing & arcsec & observation seeing  & 0.624116289\\
  ra & deg & right ascension  & 229.33949898506464\\
  de & deg & declination  & -58.85866266424065\\
  era & mas & error on right ascension  & 7.892984765147925\\
  edec & mas & error on declination  & 9.436603552353597\\
  mag & mag & magnitude  & 13.245622\\
  emag & mag & error on magnitude  & 0.021564407\\
  phot\_flag &  & photometric error flag  & 0\\
  x & pixel & detector $X$ position  & 1152.587\\
  y & pixel & detector $Y$ position  & 918.431\\
  ex & pixel & error on detector $X$ position  & 0.023\\
  ey & pixel & error on detector $Y$ position  & 0.02\\
  cnf\_ctr &  & CASU confidence value of centroid pixel  & 95\\
  chi &  & dophot chi of detection  & 2.52\\
  objtype &  & dophot object type  & 1\\
  ext &  & VIRCAM detector number  & 13\\
  ast\_res\_chisq &  & chi squared of astrometric residual  & 4.3247724\\
  ambiguous\_match &  & flag indicating shared detection  & 0\\
\hline\end{tabular}
\end{table*}

\begin{table*}
\centering
\caption{Field names, units and descriptions for the observation index table. An example row is also given, that of the observation corresponding to the time series element shown in Table \ref{tab:timeseries_schema}. The \textit{catid} field links to the time series table (see Table \ref{tab:timeseries_schema}).}\label{tab:observations_schema}
\begin{tabular}{|l|l|l|c|}
\hline
  \multicolumn{1}{|l|}{Name} &
  \multicolumn{1}{l|}{Units} &
  \multicolumn{1}{l|}{Description}  &
  \multicolumn{1}{c|}{Example row} \\
\hline
  catid &  & unique observation identifier & 120741\\
  filename &  & FITS filename of the image & v20100304\_00780\_st.fits.fz\\
  tile &  & VVV tile name & d018\\
  ob &  & OB name & d018v-1\\
  filter &  & filter name & $K_s$\\
  ra & deg & right ascension & 228.858112\\
  de & deg & declination & -59.48647\\
  l & deg & galactic longitude & 320.302109749199\\
  b & deg & galactic latitude & -1.56127378157683\\
  exptime & s & exposure time & 4.0\\
  mjdobs & MJD & MJD of observation & 55260.36375826\\
  airmass &  & airmass & 1.222\\
  skylevel &  & sky level (CASU) & 4981.22\\
  skynoise &  & sky noise (CASU) & 45.515\\
  elliptic &  & ellipticity (CASU) & 0.10192925\\
  seeing & arcsec & seeing & 0.624116289\\
\hline\end{tabular}
\end{table*}

\clearpage
\section{UCD finder charts}\label{app:ucd_finders}
\begin{figure*}
  \begin{center}
    \includegraphics[width=\textwidth,keepaspectratio]{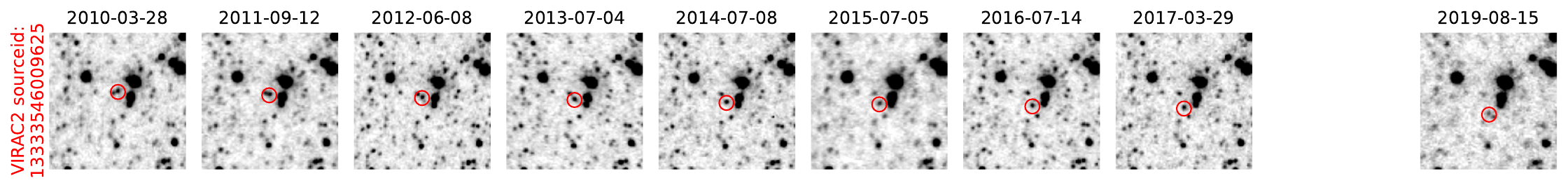}\\
    \includegraphics[width=\textwidth,keepaspectratio]{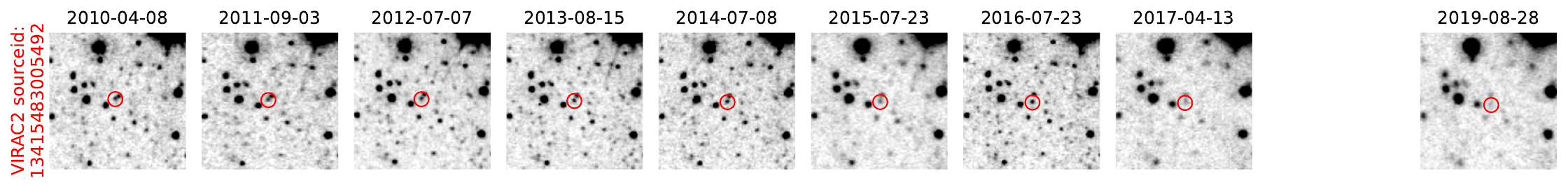}\\
    \includegraphics[width=\textwidth,keepaspectratio]{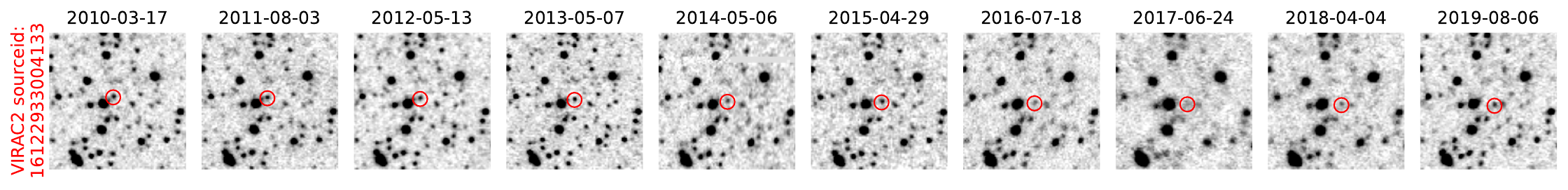}\\
    \includegraphics[width=\textwidth,keepaspectratio]{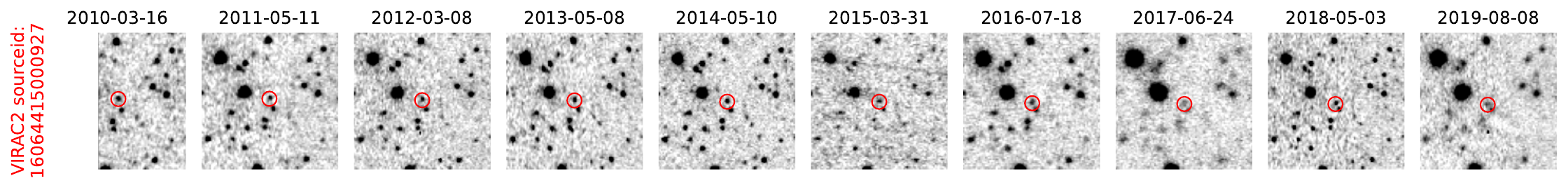}\\
    \includegraphics[width=\textwidth,keepaspectratio]{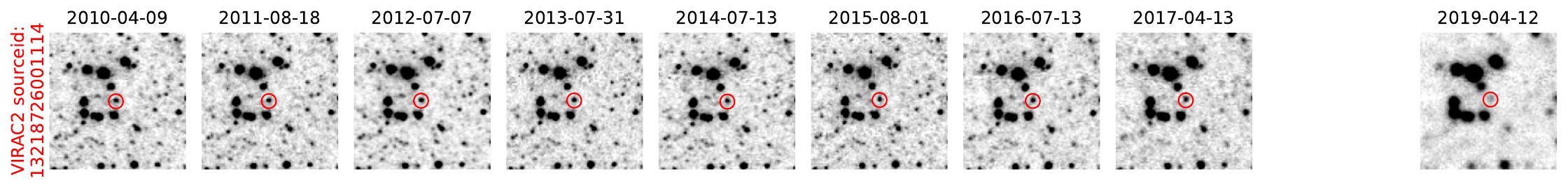}\\
    \includegraphics[width=\textwidth,keepaspectratio]{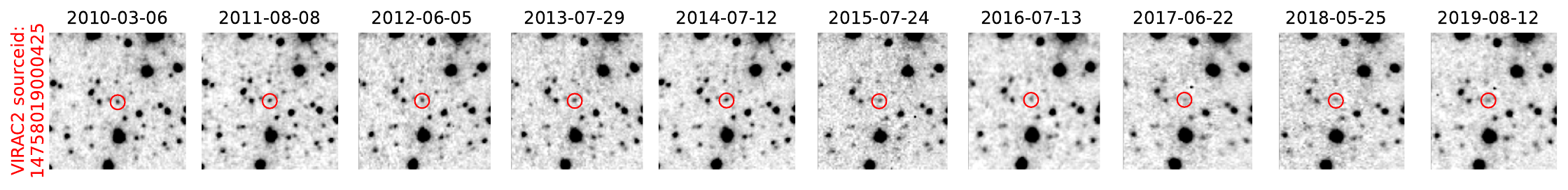}\\
    \includegraphics[width=\textwidth,keepaspectratio]{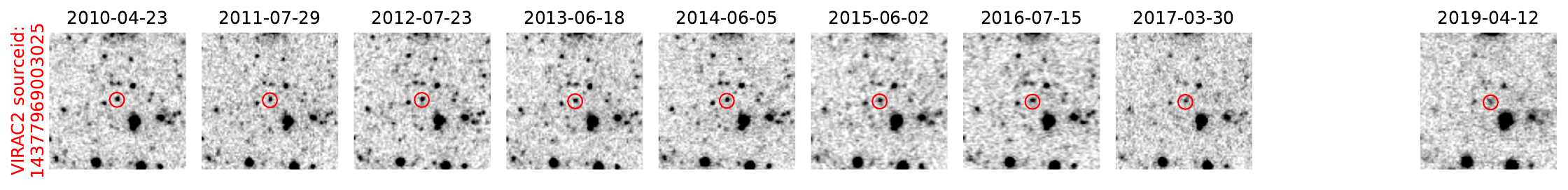}\\
    \includegraphics[width=\textwidth,keepaspectratio]{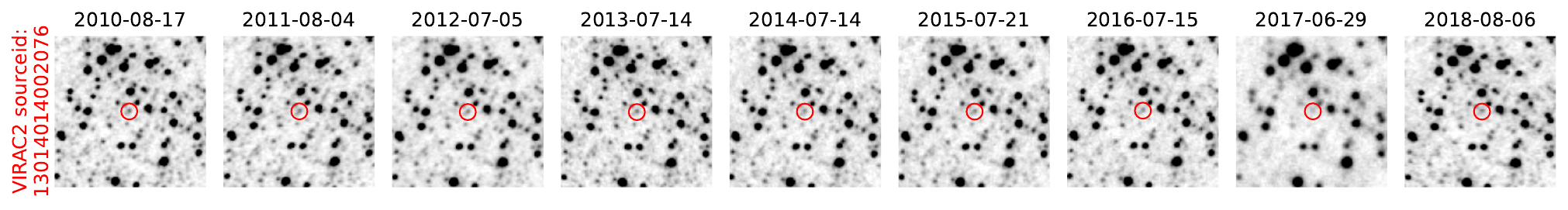}
    \caption{Finder charts, one per calendar year, for eight of the more interesting UCD candidates from Section \ref{sec:100pc}. The charts are centered on the VIRAC2 2014.0 position of each target, they are 30~\arcsec{} in size, north is up and east is to the left. From top to bottom, the targets are: VVV~J1814-2654, VVV~1820-2742, VVV~1253-6339, VVV~J1210-6227, VVV~J1728-2543, VVV~J1705-4245, VVV~J1754-3813, and VVV~J1759-2340. The red circle in each panel indicates the location of the target at that epoch, taking the VIRAC2 proper motion and parallax into account.}
    \label{fig:ucd_finders}
  \end{center}
\end{figure*}


\bsp	
\label{lastpage}
\end{document}